\documentclass[a4paper, 12pt]{article}

\usepackage{array}
\usepackage{fullpage}
\usepackage{authblk}
\usepackage{placeins}
\usepackage{lineno}
\usepackage{setspace}
\usepackage{tikz}
\usepackage{pdflscape}
\usepackage{todonotes}
\usepackage{url}
\usepackage{natbib}
\usepackage{graphicx}
\usepackage{subcaption}
\usepackage[toc,page]{appendix}

\usepackage{booktabs}
\usepackage{tabularx}
\usepackage{threeparttable}

\begin{document}

\title{Ground-motion simulation of multi-segment ruptures on the Alpine-Marlborough fault system}
\author[1]{Brendon A. Bradley\thanks{Corresponding author: Private Bag 4800, Christchurch 8140, New Zealand. Email: brendon.bradley@canterbury.ac.nz}}
\author[1]{Jake Faulkner}
\author[1]{Felipe Kuncar}
\author[1]{Cesar A. Pajaro}
\affil[1]{Department of Civil and Environmental Engineering, University of Canterbury, New Zealand}
\maketitle
\thispagestyle{empty}

\begin{abstract}
	The Alpine Fault (AF) is the major plate boundary crustal structure in New Zealand (NZ), producing large-to-great ($M=7-8+$) earthquakes approximately every 300-500 years, and has long been a focal point in the consideration of earthquake resilience in NZ.
	This study advances prior ground-motion modelling work on AF ruptures by considering multi-segment ruptures that initiate on the AF, and rupture northward onto structures in the Marlborough Fault Zone (MFZ), ultimately resulting in ground shaking of consequential amplitudes for the upper South Island and Lower North Island, including Wellington city.
	Through an analysis of potential rupture geometries from the 2022 NZ National Seismic Hazard Model (NSHM), three multi-segment geometries that include the Wairau, Clarence, and Hope fault structures in the MFZ are obtained, as well as associated scenario rupture rates (i.e., recurrence intervals).
	Hybrid broadband ground-motion simulations using a multi-segment kinematic source description and a 3D NZ crustal velocity model are performed.
	Three rupture realizations per rupture geometry are also considered to enable comparison of the relative effects of kinematic rupture randomness to that of the different rupture geometries.
	In order to provide meaningful comparisons with the prior ground-motion simulations used for public outreach in the AF8 programme, we also simulated the segmented geometry for a M7.9 Alpine Fiordland-to-Kelly rupture from~\citet{bradleyGroundMotionSimulations2017}, as well as adopting the same southern hypocentre location in all rupture scenarios, thus enabling an examination of differences due to eight years of advancements in kinematic rupture and crustal velocity modelling.
	Results illustrate strong northward rupture directivity and sedimentary basin response, with the largest Wellington amplitudes resulting from the Alpine-Wairau and Alpine-Clarence scenarios -- peak ground accelerations of approximately 0.1~g, comparable to those observed in Wellington during the 2016 Kaik\=oura earthquake -- and, in aggregate, these Alpine-MFZ ruptures have a likelihood similar to the conventionally-considered Wellington Fault and Hikurangi subduction sources.
\end{abstract}

\setstretch{1.25}

\section{Introduction}

The Alpine Fault (AF) is the major plate boundary fault in the South Island of New Zealand (NZ) and accounts for 70–75\% of the plate motion~\citep{wallaceBalancingPlateMotion2007}.
Its recurrence interval of large magnitude earthquakes has been extensively studied~\citep{rhoadesEstimatesTimevaryingHazard2003,sutherlandQuaternarySlipRate2006,berrymanLateHoloceneRupture2012,depascaleNewOnfaultEvidence2012,ghisettiDeformationTopBasement2014,howarth2016,howarth2021}, with recent estimates indicating a 62\%\footnote{Based on a 75\% probability of rupture in 50 years, and an 82\% probability that such a rupture will be $M\geq8$.} probability of a M8+ event in the next 50 years~\citep{howarth2021}.

The relatively high-likelihood of a great earthquake on the Alpine fault, and its anticipated impacts, have resulted in it being a focal rupture scenario to use for scientific modelling, coordinated response planning, and community engagement, designed to build resilience.
Most notably, the Alpine Fault magnitude 8 (AF8) programme~\citep{orchiston2018} has been the interdisciplinary `boundary organisation' through which research, policy and practice have been intertwined toward this goal over the past decade.

As part of the AF8 programme, \citet{bradleyGroundMotionSimulations2017} performed kinematic ground-motion simulations of a 411-km long rupture of the Fiord-to-Kelly (F2K) planar segment of the Alpine fault (rupture trace geometry based on the 2010 NZ NSHM).
The 3D wave propagation simulations were performed to a maximum frequency of $f=0.25$~Hz within the Canterbury basin (0.5~Hz elsewhere), %
using a 3D velocity model of the South Island region that included an explicit sedimentary basin model only for the Canterbury region~\citep{thomson2020}.
They considered the effect of hypocentre position through three -- northern, central, and southern -- hypocentral locations along the rupture plane.
Owing to the focal nature of the AF8 scenario, the results of the simulations from Bradley et al. have been widely used for examining urban, rural, and distributed infrastructure resilience in both research and practitioner settings~\citep{robinsonCoseismicLandslidingEstimates06,harris2024,aghababaei2021,davies2021,becker2020,geyinmertcan2020}.

Although the AF8 scenario has been, and continues to be, valuable for preparedness activities, a notable aspect is that the northern termination of the scenario rupture near the mid-latitudes of the South Island results in trivial ground-motion amplitudes at the top of the South Island and lower North Island, including the capital city, Wellington.
This presents a missed opportunity, particularly because of the heightened focus on Wellington due to its high absolute seismic hazard, and observed vulnerability to even the moderate levels of ground-motion amplitudes which occurred in the 2016 Kaikoura earthquake~\citep[e.g.,][]{bradley2017,bradley2018,cubrinovski2020}.

While the AF8 rupture geometry, from the 2010 NZ NSHM, is a planar segmented representation of a great earthquake on the AF, it is increasingly recognised that it is imperative to consider non-segmented and multi-segment earthquake ruptures in seismic source models used with probabilistic seismic hazard analysis~\citep{fieldEstimatingEarthquakeRuptureRates2011,gerstenberger2024,milner2023}.
Recent large-magnitude NZ earthquakes, such as the 2010 $M_W$ 7.1 Darfield and 2016 $M_W$ 7.8 Kaik\=oura events, have exhibited multi-segment ruptures, reinforcing the importance of accounting for this feature in ground motion studies in NZ.
The current 2022 NZ NSHM also adopted a non-segmented multi-segment seismic source model.
In this modelling paradigm, there are numerous fault structures that are modelled to `link' with the AF to produce large magnitude earthquakes.
Many such ruptures include further northward rupture propagation into the Marlborough Fault Zone (MFZ), which partitions the crustal plate motion from the AF across several major fault structures, most notably the Wairau, Awatere, Clarence and Hope Faults~\citep[e.g.,][]{zinke2019,hatem2020}.

While seismic source models have adopted multi-segment rupture geometries, the consideration of multi-segment ruptures in a ground-motion prediction context is not well developed.
For example, conventional ground-motion prediction using empirical ground-motion models (GMMs), such as those from the NGA-W2 and NGA-Sub projects~\citep{bozorgnia2022,bozorgniaNGAWest2ResearchProject2014}, are based on a single planar representation of the seismic source.
Similarly, past physics-based ground-motion simulations with kinematic rupture models have also been either single segment representations~\citep[e.g.,][]{dregerIntroductionFocusSection2015,GP2010} or multi-segment models heuristically-developed from past events~\citep[e.g.,][]{bradley2017}.
More recently, \citet{bradley2025b} developed a generalizable multi-segment kinematic rupture generator that is compatible with typical national seismic hazard model source descriptions, and adopted here.

In this study, we adopt leading-edge techniques and models for the hybrid broadband ground-motion simulation of multi-segment ruptures on the AF that extend northward into the MFZ, resulting in ground shaking of consequential amplitudes for the upper South Island and Lower North Island, including Wellington city.
In the subsequent sections of the paper, we first outline the development of the rupture geometries that were obtained from the 2022 NZ NSHM, and the multi-segment kinematic rupture generation.
Second, the simulation methodology and the adopted 3D crustal model of the South Island and lower North Island of NZ are presented, with a focus on the advancements in its representation of shallow sedimentary basin structure relative to the prior simulation study of \citet{bradleyGroundMotionSimulations2017}, followed by the simulated ground motions, which are discussed with respect to salient rupture and wave propagation physics, and their amplitudes over the simulation domain in comparison with conventional empirical GMMs.
Third, the simulations are compared with the prior simulations of \citet{bradleyGroundMotionSimulations2017}, including a reproduction of the 2017 scenario that separates the effects of the updated source and crustal modelling.
Lastly, the ground motions in the Wellington city region are examined in detail, because of its concentrated asset exposure.

\section{Multi-segment Alpine-Marlborough Fault rupture development}
\subsection{Rupture geometries}
\label{sec:rupture-geometries}

In order to restrict the solution space of possible multi-segment ruptures encompassing the AF and MFZ crustal sources, we applied three criteria as constraints:
\begin{enumerate}
	\item Rupture geometries that are present in the seismic source model of the 2022 NZ NSHM;
	\item Rupture includes the equivalent F2K segment of the Alpine fault;
	\item The nearest point of the rupture to Wellington city is less than 100 km.
\end{enumerate}
The first point ensures that we can assign annual rates to the resulting ruptures from the 2022 NZ NSHM, which greatly aids in talking about the aggregate likelihood of the scenarios that are simulated.
The second point enables the ruptures to be a multi-segment `extension' to the Alpine F2K `base' rupture and resulting ground-motion simulations from~\citet{bradleyGroundMotionSimulations2017}, enabling useful comparisons to that prior work that has been utilized in impact analysis and planning exercises.
The third point ensures that the ruptures produce ground-motions of engineering relevance to Wellington, given the identified limitations of the 2017 scenarios, as previously discussed.

\begin{figure}
	\centering
	\includegraphics[width=\linewidth]{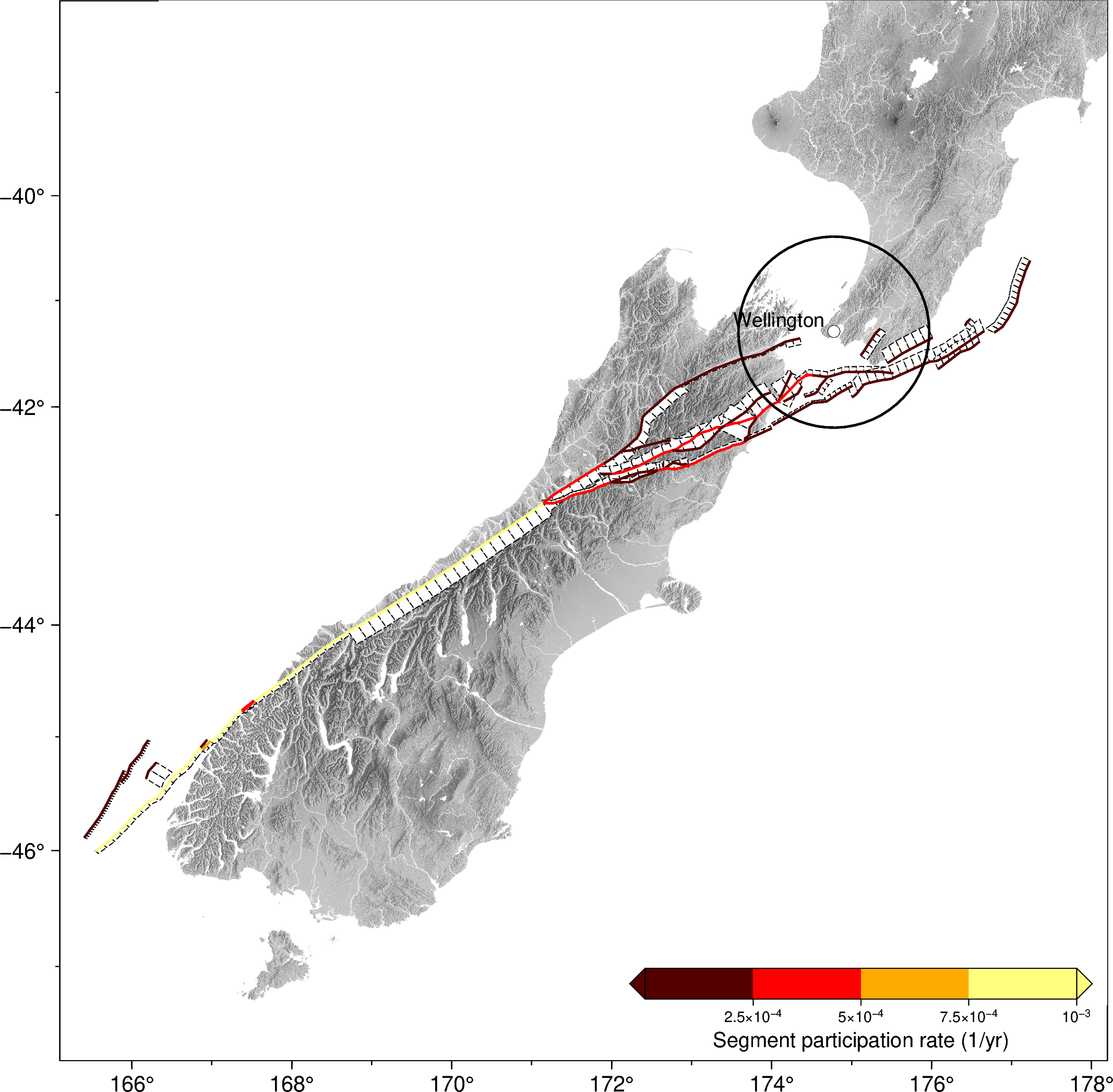}
	\caption{120 plausible rupture scenarios from the 2022 NZ NSHM source model that rupture the Alpine fault (segments from 'Caswell' to 'Kaniere') and include a segment within 100 km of Wellington city. The colors on the top edge of the rupture surfaces indicate the 'participation rate', i.e., the annual rate of rupture including that segment.}
	\label{fig:RuptureExplorer}
\end{figure}

The `Rupture Explorer' tool from the NZ NSHM (\url{https://nshm.gns.cri.nz/}) was utilized to extract scenarios from the 2022 NZ NSHM source model.
Because the Alpine F2K segment is not a formal definition in the 2022 NZ NSHM (it being used in the 2010 NSHM), the closest equivalent was to include all of the ('Caswell', 'Caswell-South George', 'George Seaward', 'George Landward', 'George to Jacksons', 'Jacksons to Kaniere') segments\footnote{The constraints for 'George Seaward', 'George Landward' were applied as "OR" constraints, since these segments are sub-parallel and no ruptures include both structures.} that are defined in the 2022 source model.
There were 694 unique geometries that initially satisfied criteria 1 and 2, and a total of 120 that additionally satisfied criterion 3.
Figure~\ref{fig:RuptureExplorer} illustrates the geometries of these 120 scenarios, with participation factors indicated by the colors on the top edge of the fault surfaces.
Most notably, rupture geometries include fault structures that are part of the broader Wairau, Clarence, or Hope fault systems,  including the offshore extensions into Cook Strait or upper plate faults of the Hikurangi subduction zone, as well as several offshore Fiordland faults.
Absent are rupture geometries that include the Awatere Fault (the other major fault structure in the MFZ) -- while there are geometries in the 2022 NZ NSHM that include the Awatere Fault with the northern segments of the AF, there are no geometries that extend sufficiently southward along the AF to meet the third criterion imposed.

Clustering was then performed to group the 120 unique geometries into Wairau, Clarence, and Hope-type ruptures, because the non-segmented rupture set contains many ruptures that share very similar geometries.
Table~\ref{tab:RuptureStatistics} illustrates the breakdown of the 120 plausible rupture geometries based on the clustering by the predominant Wairau, Clarence, or Hope fault systems that they include.  A minority of the geometries included ruptures within both the Hope and Clarence systems, and in such cases the discrete classification was based on where the larger rupture area occurred.
According to the inversion fault model component of the 2022 NZ NSHM, the annual rate of all 120 ruptures is $\lambda_{Rup}=7.97\times10^{-4}$/year\footnote{The rates from the inversion fault model component of the 2022 NZ NSHM are multiplied by 0.8 for combination with the distributed source model~\citep{gerstenberger2024}, but this factor is not applied here.}, or a 1,255 year recurrence interval, and the magnitudes of the events are $M_w=8.5\pm0.1$\footnote{As clarified in Section~\ref{sec:sim-method}, the simulations use a different moment magnitude ($M_w=8.28$--$8.31$) than that directly specified by the 2022 NZ NSHM, owing to the different magnitude--area scaling relations adopted.}.
The aggregate recurrence intervals of the ruptures including the Clarence and Hope fault structures are on the order of 3000 years, while those on the Wairau fault are on the order of 6000 years.
This is consistent with the slip rate partitioning on these sub-parallel structures of the MFZ, with the highest slip rate on the Hope system, and lowest on the Wairau system~\citep{vandissen2024}.

\begin{table}[h]
	\centering
	\caption{Statistics of ruptures from the inversion fault model-component of the 2022 NZ NSHM meeting the three constraint criteria. $\lambda_{Rup}$ and RI are the aggregation of all $N_{Rup}$ ruptures for each of the three MFZ fault system types.}
	\begin{tabular}{l c c c }
		\hline
		MFZ fault & $N_{Rup}$ & Rate, $\lambda_{Rup}$ (1/yr) & Recurrence interval, RI (yr) \\ \hline
		Wairau    & 35        & $1.53\times10^{-4}$          & 6553                         \\
		Clarence  & 27        & $3.14\times10^{-4}$          & 3182                         \\
		Hope      & 58        & $3.27\times10^{-4}$          & 3061                         \\
		Total     & 120       & $7.97\times10^{-4}$          & 1255                         \\ \hline
	\end{tabular}
	\label{tab:RuptureStatistics}
\end{table}

\begin{figure}
	\centering
	\includegraphics[width=0.5\linewidth]{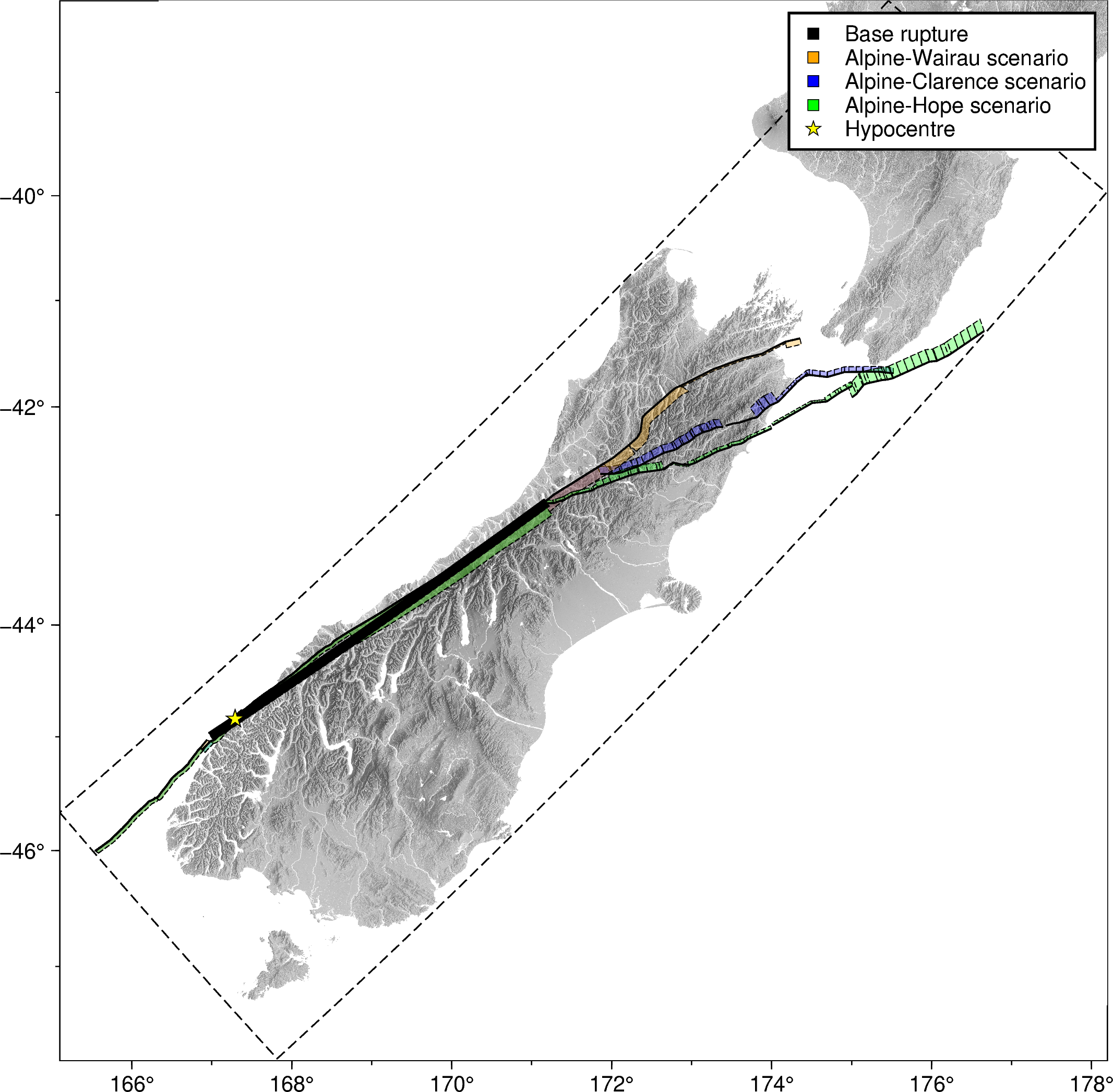}
	\caption{Geometries of the three selected multi-segment ruptures of the Alpine fault and Wairau, Clarence, and Hope fault systems.  The hypocentre location is indicated with a star, and is located toward the southern end of the Alpine fault. The simulation domain is also indicated with a dashed black line.}
	\label{fig:SelectedRupturesInDomain}
\end{figure}

For each fault system-based cluster, the rupture geometry with the highest individual rupture rate was selected for the purpose of subsequent ground-motion simulations.
Figure~\ref{fig:SelectedRupturesInDomain} illustrates the geometries of the three ruptures, as well as the hypocentre locations and also the spatial domain of the simulations.
For reference, the long-term recurrence interval of major rupture events (M$>$7) on the Alpine fault is $\approx$300 years.
Therefore, these three M8.5 scenarios occur on the order of 1-in-10 to 1-in-20 times for major Alpine fault ruptures.
It is also insightful to compare against rupture on the Hutt Valley section of the Wellington fault (historically emphasised due to the surface trace passing through the city itself) and major ruptures on the Hikurangi subduction interface.
Ruptures on the Hutt Valley section of the Wellington fault (which range from $M=7.2-8.3$) have an aggregate recurrence interval of 1,500 years ($\lambda_{Rup}=6.5e-4$) from the 2022 NZ NSHM; and M8.5+ ruptures on the Hikurangi subduction interface, which are within 100km of Wellington, have an aggregate recurrence interval of 930 years, indicating these Alpine-MFZ ruptures, in aggregate, have a similar likelihood to these two more widely considered rupture sources.

\subsection{Kinematic rupture generation}
\begin{figure}[h]
	\centering
	\includegraphics[width=0.5\linewidth]{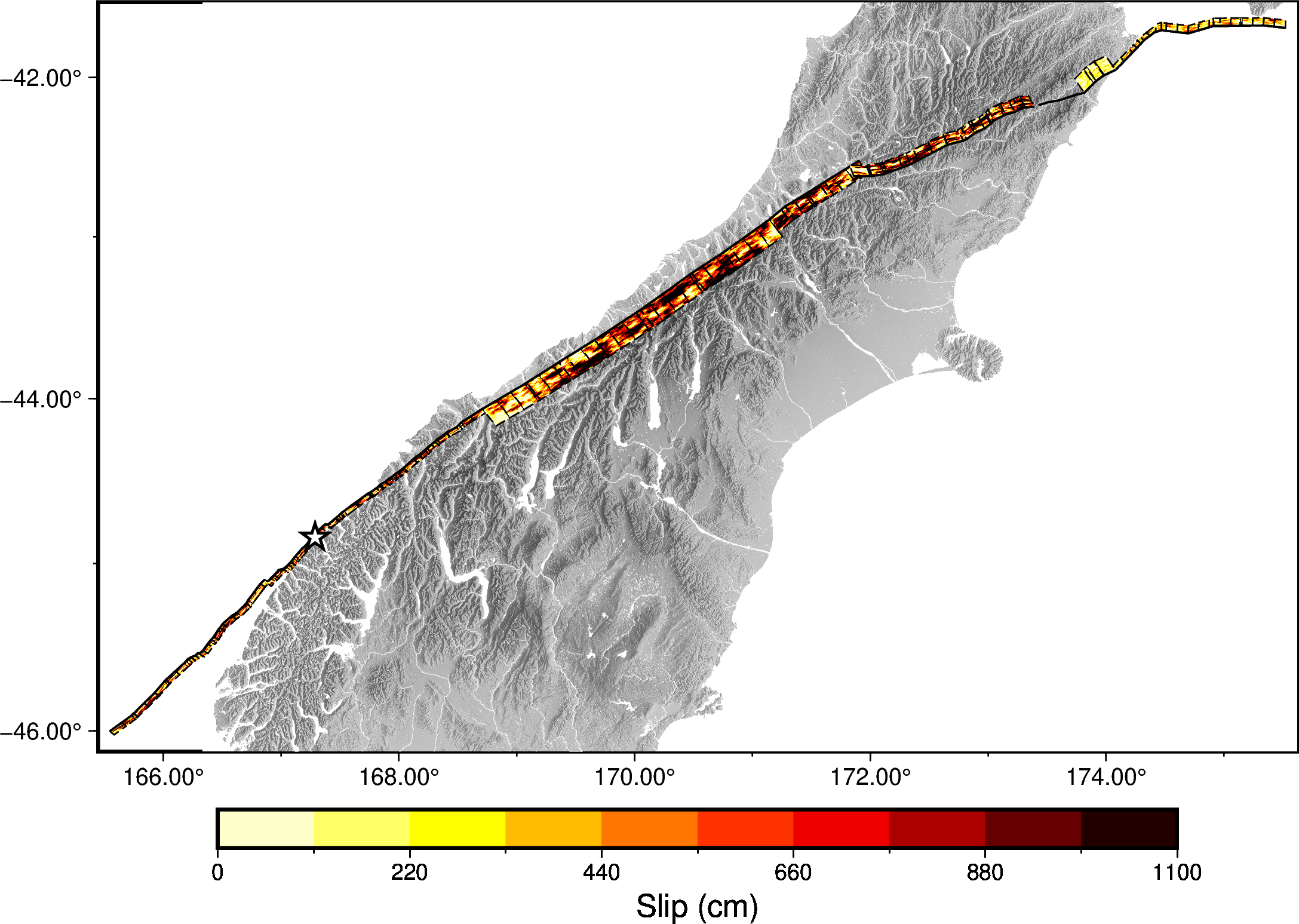}
	\caption{Illustration of the kinematic slip amplitude distribution of realization R1 on the Alpine-Clarence rupture geometry. The star indicates the hypocentre.}
	\label{fig:KinematicRuptureRealizationClarence}
\end{figure}

For each of the rupture geometries that were considered in Figure~\ref{fig:SelectedRupturesInDomain}, we developed multi-segment kinematic rupture realizations using the general multi-segment method of~\citet{bradley2025b}.
This method builds on the single segment rupture generator of~\citet{GP2015}, where moment is apportioned between the segments in relation to the segment-specific rigidity-weighted area, and rupture jumping between segments is based on kinematic triggering considerations.
Figure~\ref{fig:KinematicRuptureRealizationClarence} provides an illustrative example of the slip distribution from a kinematic rupture realization (R1) on the Alpine-Clarence geometry.

Three kinematic realizations were considered for each rupture geometry.
This was intended to provide a superficial examination of the effect of rupture realization variability on the resulting simulated amplitudes, in comparison to the effect of the different rupture geometries.

\section{Simulated ground motions}
\subsection{Simulation method and computational details}
\label{sec:sim-method}
Ground-motion simulation was undertaken using the~\citet{GP2015} methodology, a ‘hybrid’ approach in which the low frequency (LF) waveforms are comprehensively computed by solving the elastodynamic equation in the 3D crustal model domain, while the high frequency (HF) waveforms utilise a phenomenological simplified physics approach.
This simulation method has been validated for crustal earthquakes in New Zealand using multiple observations from past events~\citep{bradley2017,razafindrakoto2018,lee2022,pajaro2025,Lee2021}, providing confidence in the reliability of the simulation results for future scenario earthquakes.
The LF-HF transition frequency was $f_t=1Hz$, based on a 100m spatial discretization with a minimum shear-wave velocity of $V_{S,min}=500$m/s.
The LF simulation was based on the staggered-grid finite difference scheme of~\citet{gravesSimulatingSeismicWave1996}.
This adopts a flat free surface, such that topography was considered using the squashed-tapered approach~\citep{thomson2020}.
The HF component of the simulation is generated using a stochastic representation of seismic source radiation, combined with a Green's function calculated for a one-dimensional $V_S$ profile.
The profile represents a NZ basin and has a $V_{S,min}=500$~m/s.
A $V_{S30}$-based site amplification factor based on the CB14 model~\citep{CB2014} was applied to the HF simulation component to account for shallow site effects following prior validation studies~\citep{Lee2020,kuncar2025b}.
The moment magnitude of each scenario was obtained from its total rupture area using the \citet{leonard2010} magnitude--area scaling relation for strike-slip faulting, giving $M_w=8.28$, 8.31, and 8.31 for the Alpine-Wairau, Alpine-Clarence, and Alpine-Hope ruptures, respectively.
These are approximately 0.2 magnitude units lower than the values of $M_w\approx8.5$ for the same rupture areas in the 2022 NZ NSHM inversion fault model (Section~\ref{sec:rupture-geometries}), which reflects the different magnitude--area scaling relations adopted in the 2022 NZ NSHM.
For the base rupture, the $M_w=7.9$ of~\citet{bradleyGroundMotionSimulations2017} was retained to enable a direct comparison.
Computational details of the simulations are provided in Electronic Supplement~\ref{sec:ESupp_C}.

In order to understand the broader implications of the simulated ground motion, such as their use in probabilistic seismic hazard analysis~\citep{baker2021}, in the subsequently presented results we also compare the simulated ground motion pSA amplitudes with mean predictions from the 2022 NZ NSHM logic tree~\citep{bradley2024}.
In this logic tree, a combined weight of 0.67 is given to backbone models developed specifically for NZ, while the remaining 0.33 is distributed among the other candidate models.
The backbone GMMs are the A22 \citep{Atkinson2022}, weighted at 0.42, and the S22 \citep{Stafford2022} models, weighted at 0.58.
The remaining five models are ASK14 \citep{ASK14}, CY14 \citep{ChiouYoungs2014}, CB14 \citep{CB2014}, BSSA14 \citep{Boore2014}, and B13 \citep{Bradley2013}, each equally weighted.

\subsection{3D crustal model}

\begin{figure}[htbp]
	\centering
	\includegraphics[width=0.5\linewidth]{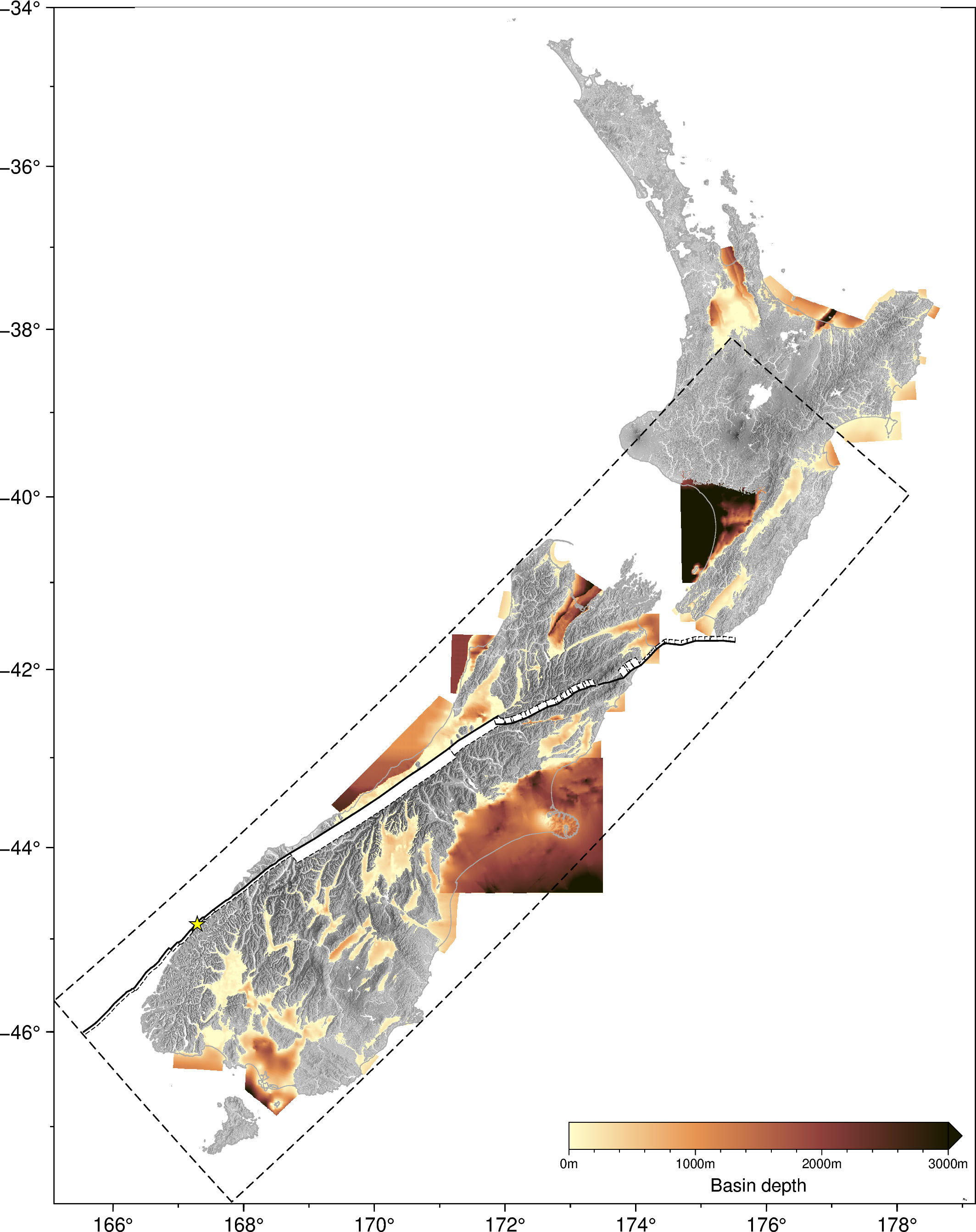}
	\caption{Sediment depth in the 45 explicitly modelled basins (and the Banks Peninsula volcanics) in the adopted 3D crustal model. The fault geometry and hypocentre location for the Alpine-Clarence scenario are illustrated as well as the simulation domain.}
	\label{fig:BasinsModelledDepthsWairauRuptureExample}
\end{figure}

The 3D crustal model used in the ground-motion simulations was obtained through v2.09 of the NZ Velocity Model (NZVM)~\citep{thomson2020,tiwari2025}. %
The major difference with~\citet{bradleyGroundMotionSimulations2017} is the extent of explicit sedimentary basin modelling -- in that study only the Canterbury basin was explicitly modelled, based on the work of~\citet{lee2017a}.
In contrast, the adopted v2.09 of the NZVM includes 45 explicitly modelled sedimentary basins (in addition to the Banks Peninsula volcanics), as documented in~\citet{tiwari2025}.
Figure~\ref{fig:BasinsModelledDepthsWairauRuptureExample} illustrates the sediment depths in these 45 modelled basins, together with the Alpine-Clarence rupture geometry, which reflects essentially all large bodies of sedimentary deposits in the simulation domain of interest.
Another minor difference in crustal modelling is that NZVM v2.09 uses the deeper crustal structure from~\citet{eberhart-phillips_new_2020}, as compared to a prior version~\citep{eberhart-phillipsEstablishingVersatile3D2010} used in~\citet{bradleyGroundMotionSimulations2017}.

\subsection{General features of rupture directivity and basin-directivity coupling.}

\begin{figure}[htbp]
	\centering
	\includegraphics[width=\linewidth]{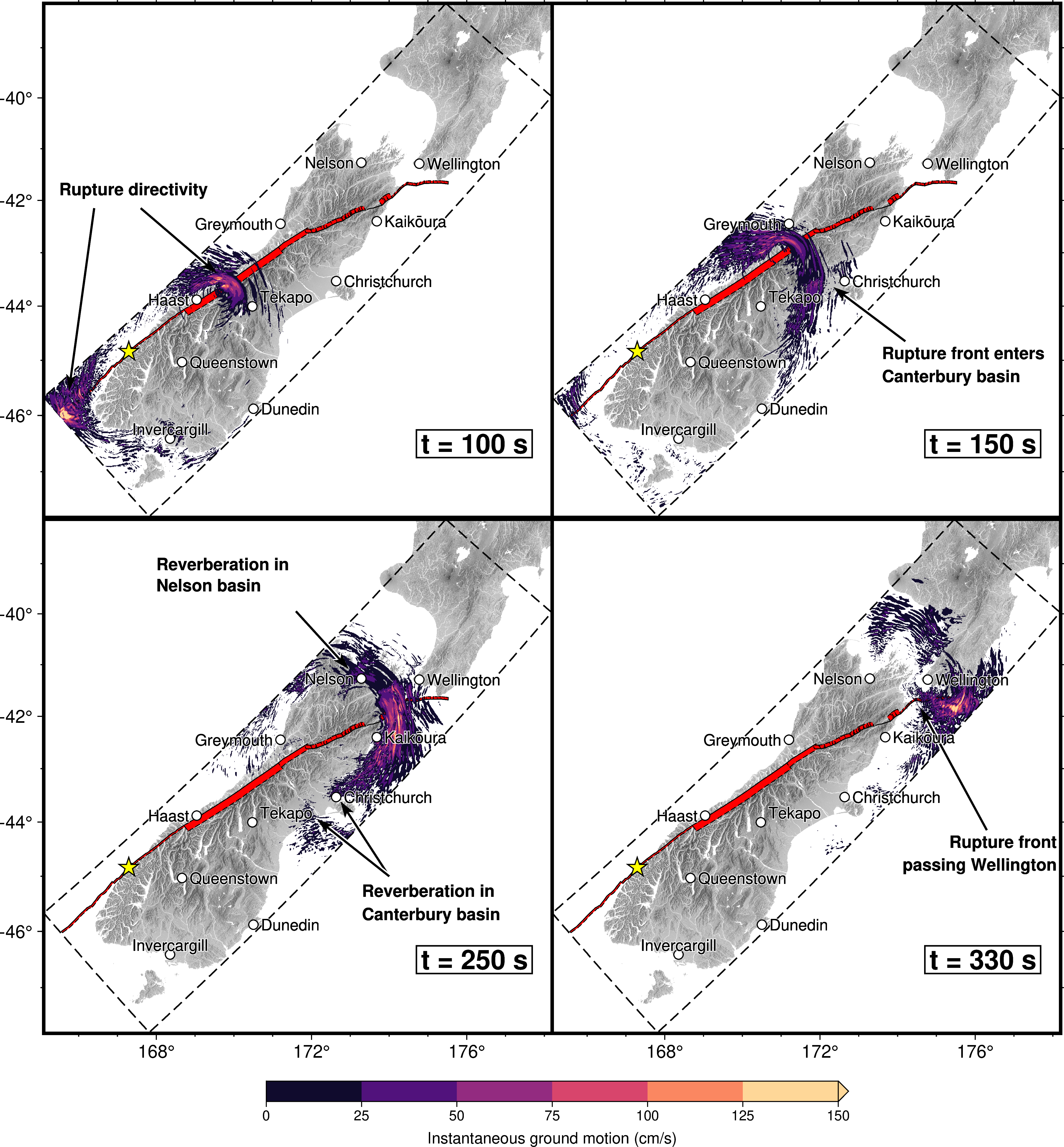}
	\caption{Illustration of the particle velocity seismic wavefield resulting from the Alpine-Clarence rupture (realization R1) at four time instants ($t=100$, 150, 250, and 330~s after rupture initiation). Areas of notable rupture directivity and wavefield reverberations in explicitly modelled sedimentary basins are annotated. Velocities below 10~cm/s are not shown. The dashed line indicates the simulation domain and the star the hypocentre.}
	\label{fig:WavefieldExampleClarence}
\end{figure}

Figure~\ref{fig:WavefieldExampleClarence} provides an example of the high-level rupture and wave propagation features via four time snapshots of the Alpine-Clarence rupture (realization R1).
The most notable features are the significant rupture directivity that occurs with the northward-propagating rupture front, and also the wave reverberations in the explicitly modelled sedimentary basins (see Figure~\ref{fig:BasinsModelledDepthsWairauRuptureExample} for locations).
As will be subsequently illustrated, the amplifications in the sedimentary basins vary appreciably with oscillator vibration period, principally as a function of sediment depth, but also the basin/rock interface geometry.
Given the magnitude of the ruptures, it is also notable that the rupture duration extends to approximately 295--395 seconds (approximately 295, 320--330, and 385--395~s for the Alpine-Wairau, Alpine-Clarence, and Alpine-Hope ruptures, respectively), which results in long duration ground-motions at specific station locations, which is obviously important when considering the ability of structures to withstand a large number of equivalent cycles of shaking.

\subsection{Detailed examination of simulation features}

\subsubsection{Spatial distribution of general simulation features}

\begin{figure}[htbp]
	\centering
	\includegraphics[width=\linewidth]{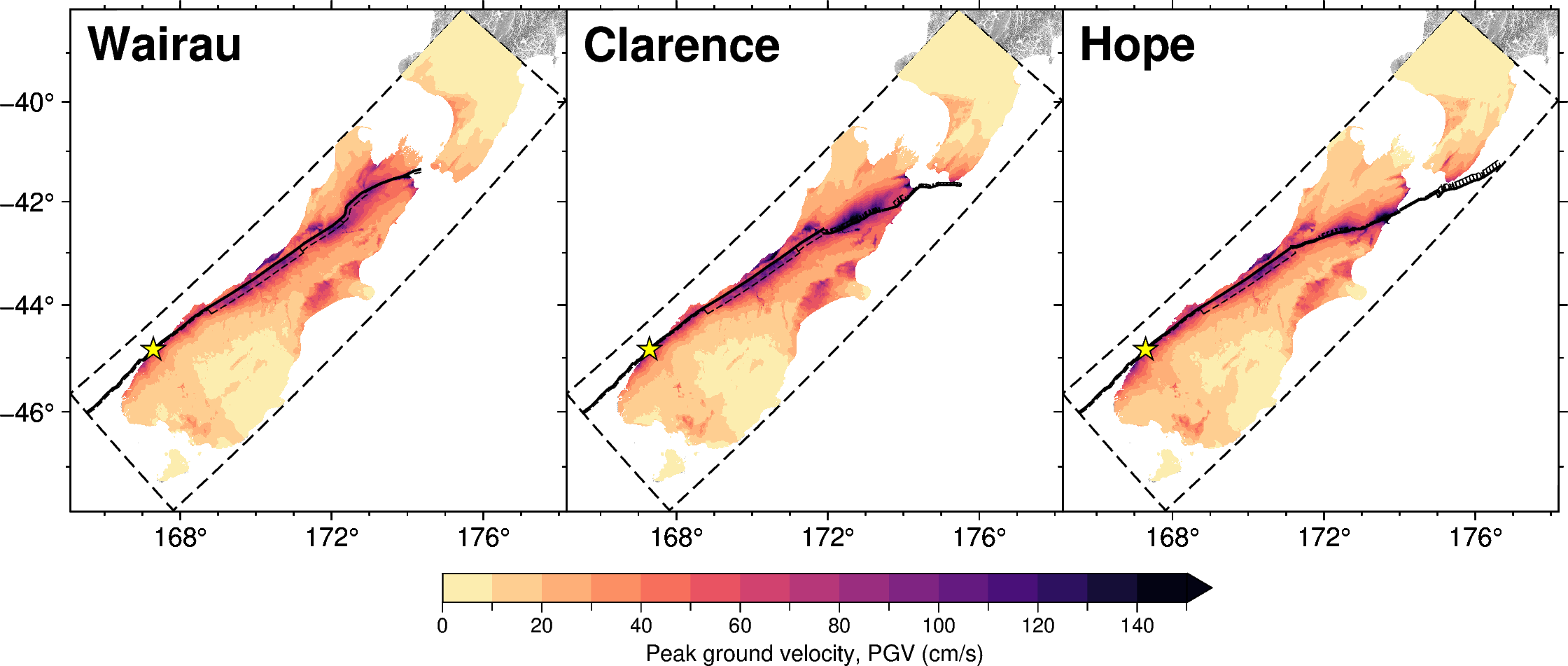}
	\caption{PGV (RotD50) over the simulation region for the Alpine-Wairau, Alpine-Clarence, and Alpine-Hope ruptures. At each site the value shown is the geometric mean of the three realizations (R1--R3), i.e., the per-site geometric mean rather than a single realization. The fault trace, rupture surface projection, hypocentre (star), and simulation domain (dashed) are also shown.}
	\label{fig:PgvMaps}
\end{figure}

Figure~\ref{fig:PgvMaps} illustrates the peak ground velocity (PGV) over the simulated domain from the three rupture geometry cases.
These results highlight a general trend of PGV attenuation as a function of distance from the causative rupture, overlaid with the rupture directivity and sedimentary basin amplification effects noted earlier.
For example, larger PGV amplitudes are evident toward the northern portion of the simulation domain, for a given source-to-site distance.
The larger PGV amplitudes are also evident in the locations of the Grey Valley, Canterbury, Southland, Mackenzie, seaward and inland Kaikoura and Hanmer, Nelson, Marlborough, and Palmerston North basins.
There are also more localised site amplifications in smaller-scale basins that are less visible at the scale of Figure~\ref{fig:PgvMaps}.

Comparing the amplitudes at specific locations for the three different rupture cases, it is expectedly seen that the largest amplitudes in the Nelson and Northwestern South Island occur as a result of the Alpine-Wairau rupture, because of its closer rupture proximity to this region.
Similarly, the largest shaking in the Kaikoura, Northern Canterbury, and Southeastern portion of the lower North Island occurs from the Alpine-Hope rupture.
As will be discussed subsequently, the shaking in the Wellington city region is generally largest from the Alpine-Wairau and Alpine-Clarence ruptures, both of which have a similar source-to-site distance to Wellington city, with the shaking from the Alpine-Hope rupture notably weaker.

\begin{figure}[htbp]
	\centering
	\includegraphics[width=\linewidth]{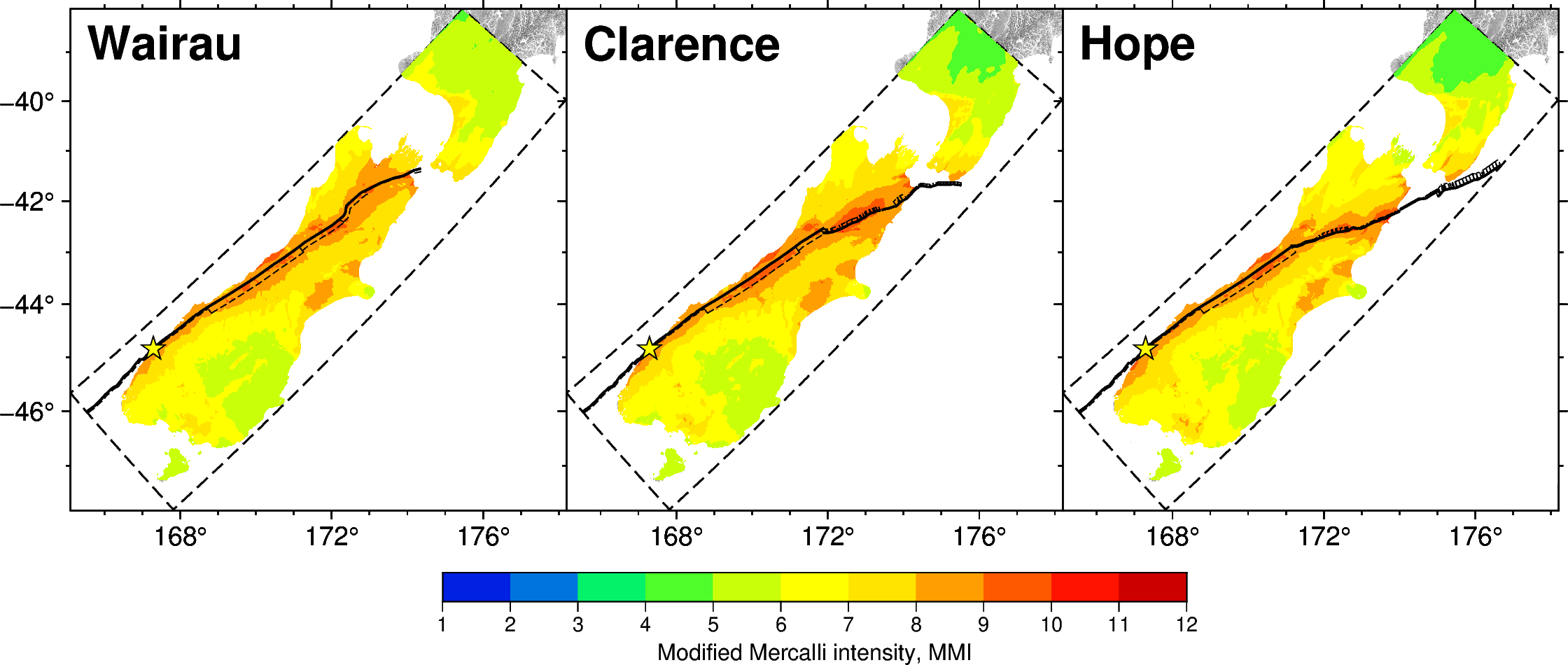}
	\caption{MMI over the simulation region for the Alpine-Wairau, Alpine-Clarence, and Alpine-Hope ruptures, computed with the~\citet{wordenProbabilisticRelationshipsGroundMotion2012} relation from the per-site geometric mean PGV of realizations R1--R3 shown in Figure~\ref{fig:PgvMaps}.}
	\label{fig:MmiMaps}
\end{figure}

Figure~\ref{fig:MmiMaps} illustrates the modified Mercalli intensity (MMI) over the simulated domain from the three rupture geometry cases.
The MMI values were computed from PGV based on the~\citet{wordenProbabilisticRelationshipsGroundMotion2012} model, and therefore exhibit the same general spatial features as discussed with respect to PGV.

\subsubsection{Scaling with source-to-site distance}

\begin{figure}[htbp]
	\centering
	\includegraphics[width=\linewidth]{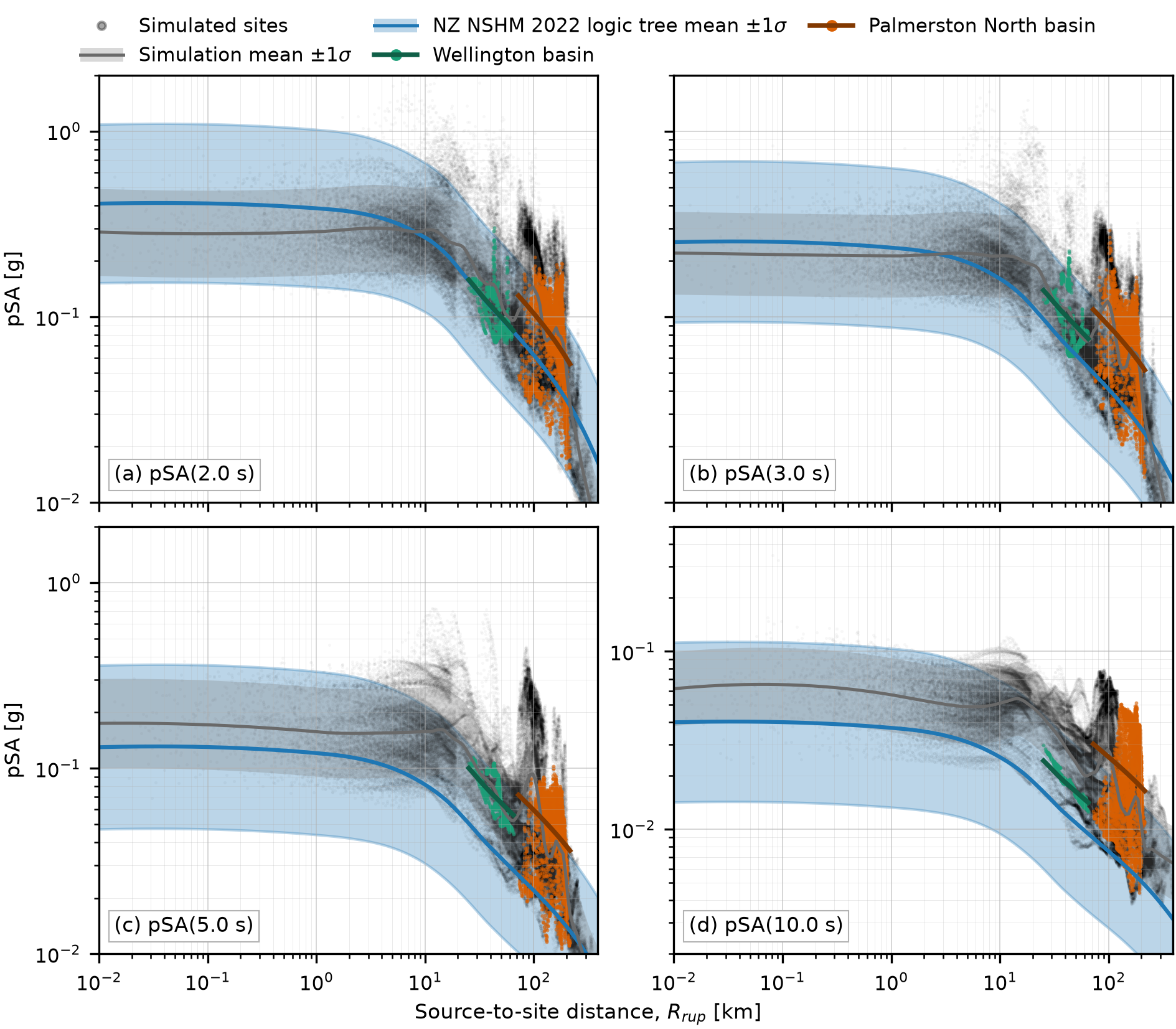}
	\caption{Response spectral amplitudes for four vibration periods, (a) pSA(2.0~s), (b) pSA(3.0~s), (c) pSA(5.0~s), and (d) pSA(10.0~s), as a function of source-to-site distance for the Alpine-Clarence rupture. Points represent simulated amplitudes, taken at each site as the geometric mean of realizations R1--R3. The grey line and band are a local regression (LOESS) of the simulated amplitudes against distance and its $\pm1$ standard deviation (approximately the $16^{th}$--$84^{th}$ percentile range). The blue line and band are the mean and $\pm1$ total standard deviation of the NZ NSHM 2022 logic tree. Simulated amplitudes in the Wellington and Palmerston North basins are highlighted, and the coloured lines show the empirical model shifted by the mean residual of the sites in each basin, over the distance range of those sites. Panels (a)--(c) share a vertical scale; panel (d) has its own, lower by a factor of about five.}
	\label{fig:SaAmplitudesWithDistance}
\end{figure}

Figure~\ref{fig:SaAmplitudesWithDistance} illustrates the variation in simulated ground-motion amplitudes for several vibration periods with nearest source-to-site distance to the causative rupture.
For pSA(2.0~s) (Figure~\ref{fig:SaAmplitudesWithDistance}a), there is a general agreement between the trend of the simulations (grey line) and the empirical logic tree mean (blue line) within approximately 30~km of the rupture (ratios of approximately 0.75--1.3), with the simulated amplitudes increasingly exceeding the empirical prediction at larger distances (by a factor of approximately 2.3 at 100~km).
As the vibration period increases, the mean simulated amplitudes are generally larger than the empirical mean prediction, e.g., at pSA(10~s) by a factor of approximately 1.5--2 within 10~km of the rupture, increasing with distance.
The specific amplitudes in the Wellington and Palmerston North basins are also annotated in these figures.
In all four vibration periods shown there is a stronger saturation of ground motion amplitudes with distance in the near-fault region (i.e., $R_{rup} < 10$km), including some degree of over-saturation.
It is seen that there is generally greater amplifications modelled in the Palmerston North basin, for all shown vibration periods, but there is also significant variation in the simulated amplitudes within a given sedimentary basin -- e.g., for pSA(2.0) at $R_{rup} < 60$km, the simulated Wellington basin amplitudes range from 0.07 to 0.30~g (i.e., a factor of approximately four).

\begin{figure}[htbp]
	\centering
	\includegraphics[width=\linewidth]{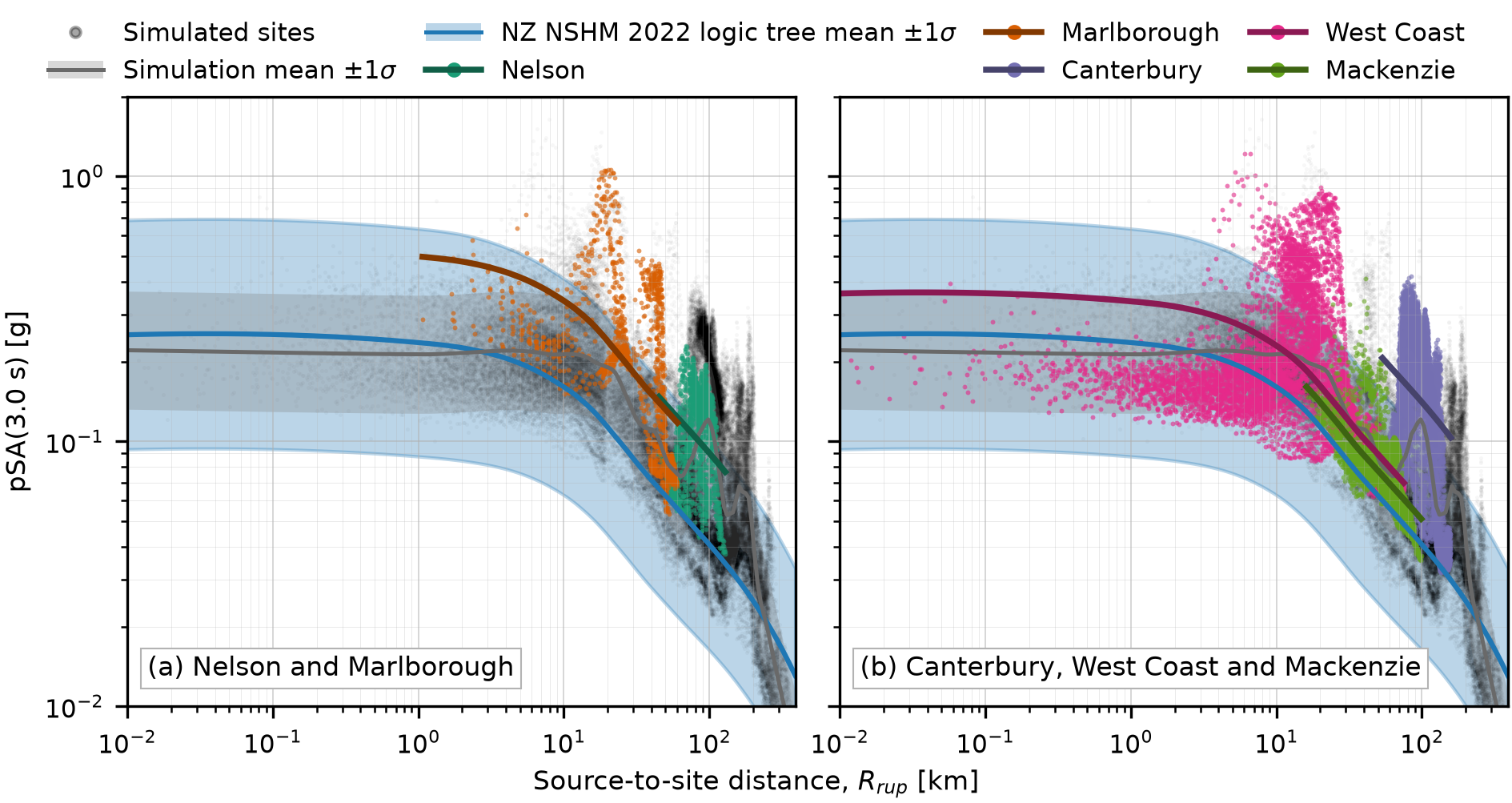}
	\caption{pSA(3.0~s) amplitudes as a function of source-to-site distance for the Alpine-Clarence rupture, highlighting sites in (a) the Nelson and Marlborough basins and (b) the Canterbury, West Coast, and Mackenzie basins. Points, lines and bands are as in Figure~\ref{fig:SaAmplitudesWithDistance}.}
	\label{fig:SaAmplitudesWithDistance_OtherBasins}
\end{figure}

Figure~\ref{fig:SaAmplitudesWithDistance_OtherBasins} illustrates the pSA(3.0) amplitudes, similar to Figure~\ref{fig:SaAmplitudesWithDistance}b, but highlighting five other sedimentary basin locations in the South Island -- Nelson and Marlborough in the Northern South Island (Figure~\ref{fig:SaAmplitudesWithDistance_OtherBasins}a) and Canterbury, West Coast, and Mackenzie (Figure~\ref{fig:SaAmplitudesWithDistance_OtherBasins}b) in the central South Island.
In the Nelson, Marlborough, Canterbury, and West Coast basins the within basin amplitudes are on average larger than the simulation-average amplitudes for the associated source-to-site distances (by factors of approximately 1.05--1.6), whereas in the Mackenzie basin they are on average lower (by a factor of approximately 0.8).
There are also consistent differences in the average amplifications in each of the basins, as well as significant intra-basin amplitude variations.
Such effects are well understood to result from the different sediment depths and basin/rock interface geometry beneath the sites, and at the lateral edges of the basins. %

\section{Comparison with the 2017 AF8 simulations}
\label{sec:af8-comparison}

\subsection{Comparison to the `base' AF8 scenario}
\label{sec:af8-base}
\begin{figure}[htbp]
	\centering
	\includegraphics[width=\linewidth]{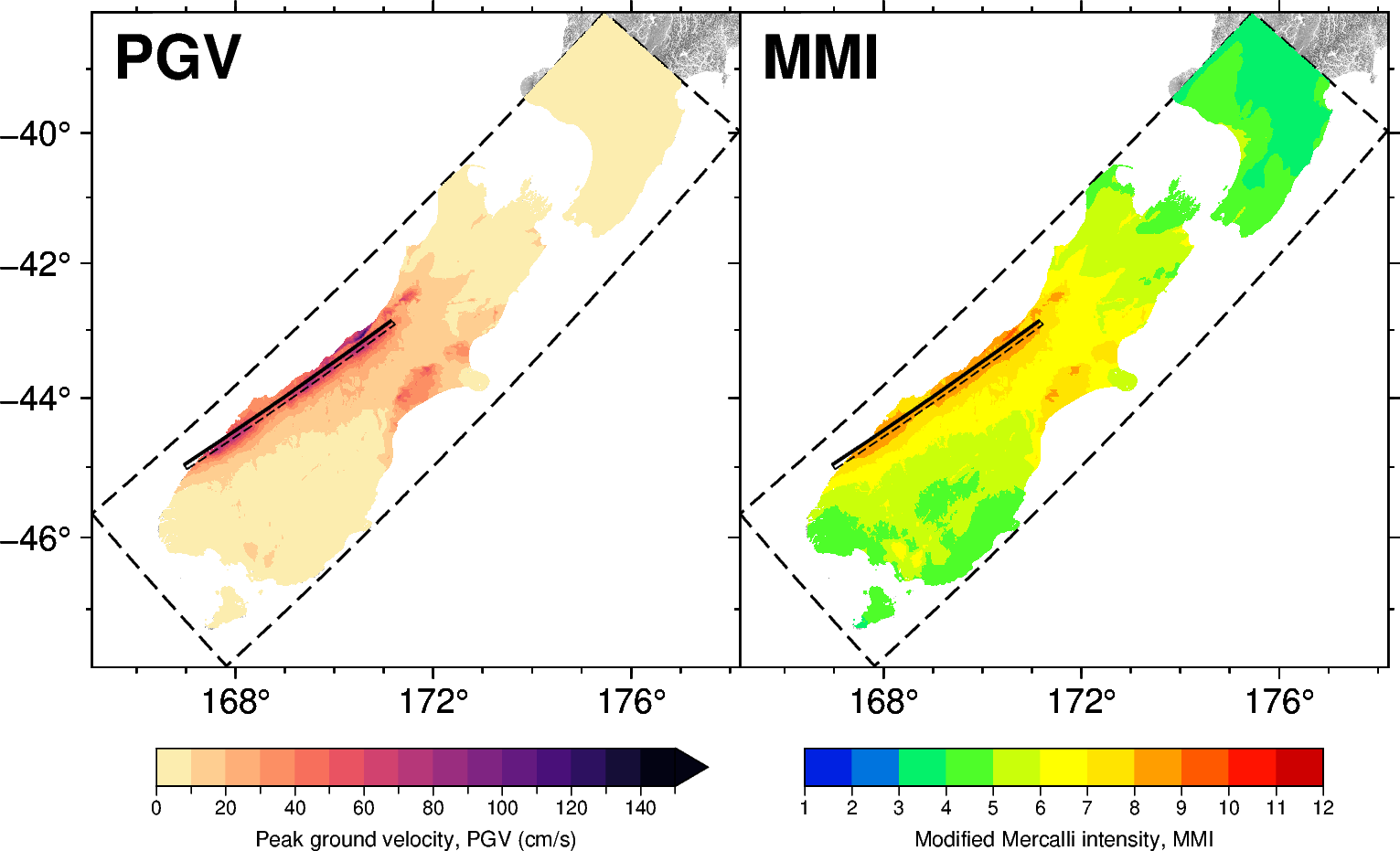}
	\caption{PGV (RotD50) and MMI over the simulation region for the `base' AF8 rupture definition, as the per-site geometric mean of realizations R1--R3. The PGV colour scale is the same as in Figure~\ref{fig:PgvMaps}, and MMI is computed from PGV with the~\citet{wordenProbabilisticRelationshipsGroundMotion2012} relation.}
	\label{fig:baseScenarioMaps}
\end{figure}

Figure~\ref{fig:baseScenarioMaps} illustrates the PGV and MMI distributions over the simulated domain for the `base' AF8 rupture scenario, as defined by~\citet{bradleyGroundMotionSimulations2017}.
As previously noted, while the same rupture geometry and hypocentre position was adopted from the prior 2017 study, calculations were re-performed to account for the advances in kinematic rupture generation and 3D velocity modelling (i.e., the addition of numerous explicitly-modelled sedimentary basins).
As compared to the Alpine-MFZ simulations shown in Figure~\ref{fig:PgvMaps}, the base AF8 scenario expectedly produces appreciably weaker shaking in the Northern South Island and Wellington regions, due to the more southerly termination of rupture.
However, examination of the shaking intensity in the Otago and Canterbury regions also illustrates reduced shaking intensities.
Further examination illustrated that this result is most likely attributed to the smaller average slip displacements that result from the magnitude-area scaling relationships when applied to this base AF8 rupture ($M_w$7.9), compared to the much longer AF-MFZ ($M_w$8.3) ruptures.

\begin{figure}[htbp]
	\centering
	\includegraphics[width=\linewidth]{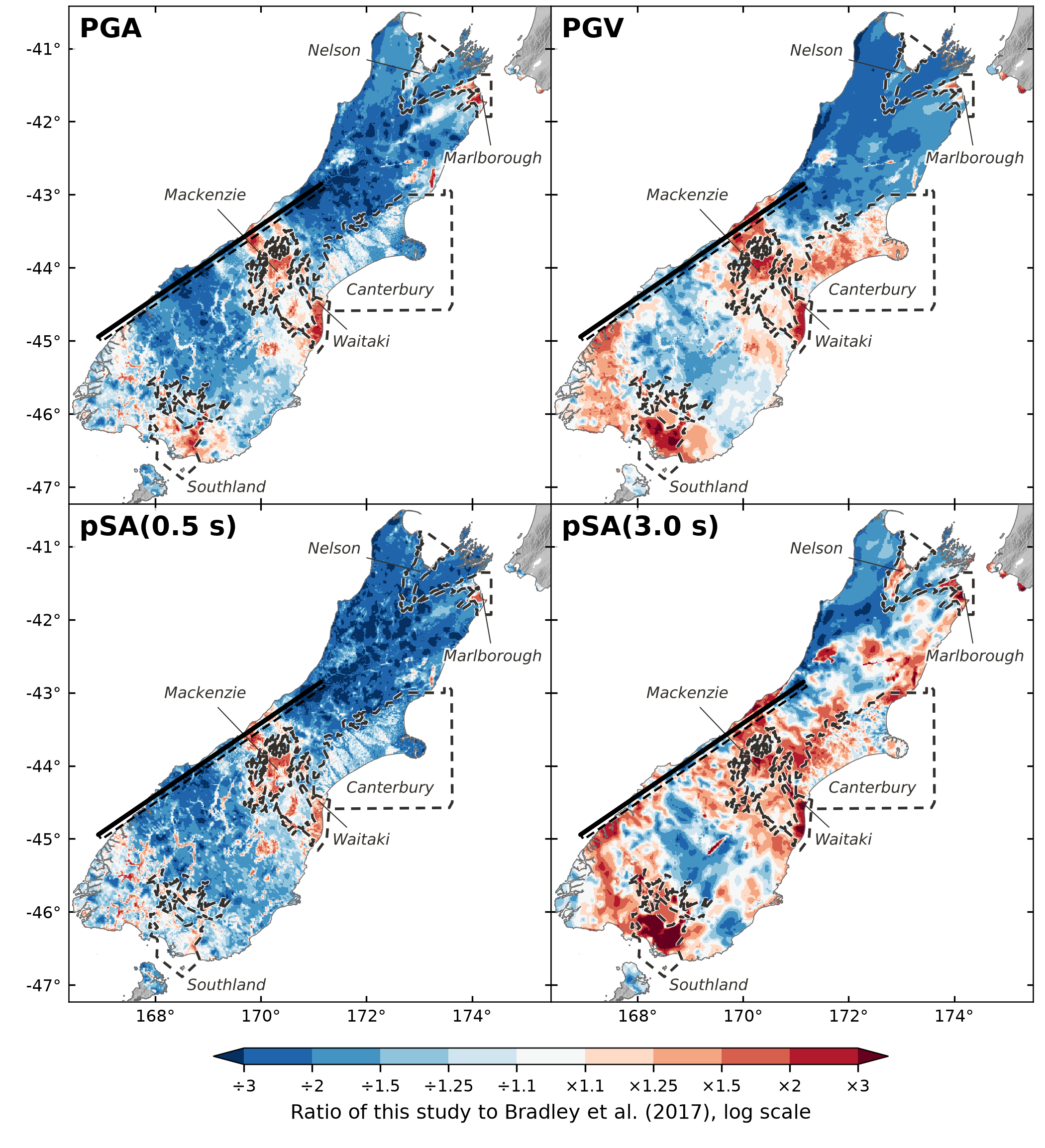}
	\caption{Ratio of this study's base rupture intensity measures to those of
		the AF8 scenario of \citet{bradleyGroundMotionSimulations2017} (southern
		hypocentre) for PGA, PGV, pSA(0.5~s) and pSA(3.0~s), on a logarithmic colour
		scale. Values above one indicate stronger motion in this study. This study is
		the geometric mean of realisations R1--R3, and the AF8 values are interpolated
		onto the stations of this study; stations more than 8~km from an AF8 station
		are omitted. The basins of interest are outlined and named: Nelson,
		Marlborough, Canterbury, Mackenzie, Waitaki and Southland. The AF8 fault trace is shown
		bold and its surface projection dashed. The same ratio with all modelled
		basins outlined is given in Figure~\ref{fig:af8-ratio}, and all modelled
		basins are named in Figure~\ref{fig:basins-named}.}
	\label{fig:af8-basins}
\end{figure}

Figure~\ref{fig:af8-basins} shows the ratio of the intensity measures of this study's base rupture to those of the AF8 scenario, with the basins of interest for this comparison outlined.
Supporting results are provided in Electronic Supplement~\ref{sec:ESupp_A}: the same ratio with all modelled basins outlined (Figure~\ref{fig:af8-ratio}) together with a map naming every modelled basin (Figure~\ref{fig:basins-named}), close-up views of the basins of interest (Figure~\ref{fig:af8-zooms}), the equivalent comparison in terms of MMI (Figure~\ref{fig:af8-mmi}), and the geometric-mean ratio within each basin of interest (Table~\ref{tab:af8-basin-ratios}).
Of the highlighted basins, only Canterbury was modelled in the AF8 study, and of the basins south of the northern deficit it shows the least amplification relative to the AF8 scenario.
The remaining southern basins---Mackenzie, Waitaki and Southland---show significant amplification, with geometric-mean ratios of up to approximately 1.5 for PGV and 1.9 for pSA(3.0~s).
Nelson and Marlborough follow the general trend of lower intensities in the northern portion of the rupture at short periods, but at pSA(3.0~s) the amplification in these basins approximately offsets the deficit in source strength (geometric-mean ratios of 0.97 and 1.09, respectively).

\begin{figure}[htbp]
	\centering
	\includegraphics[width=\linewidth]{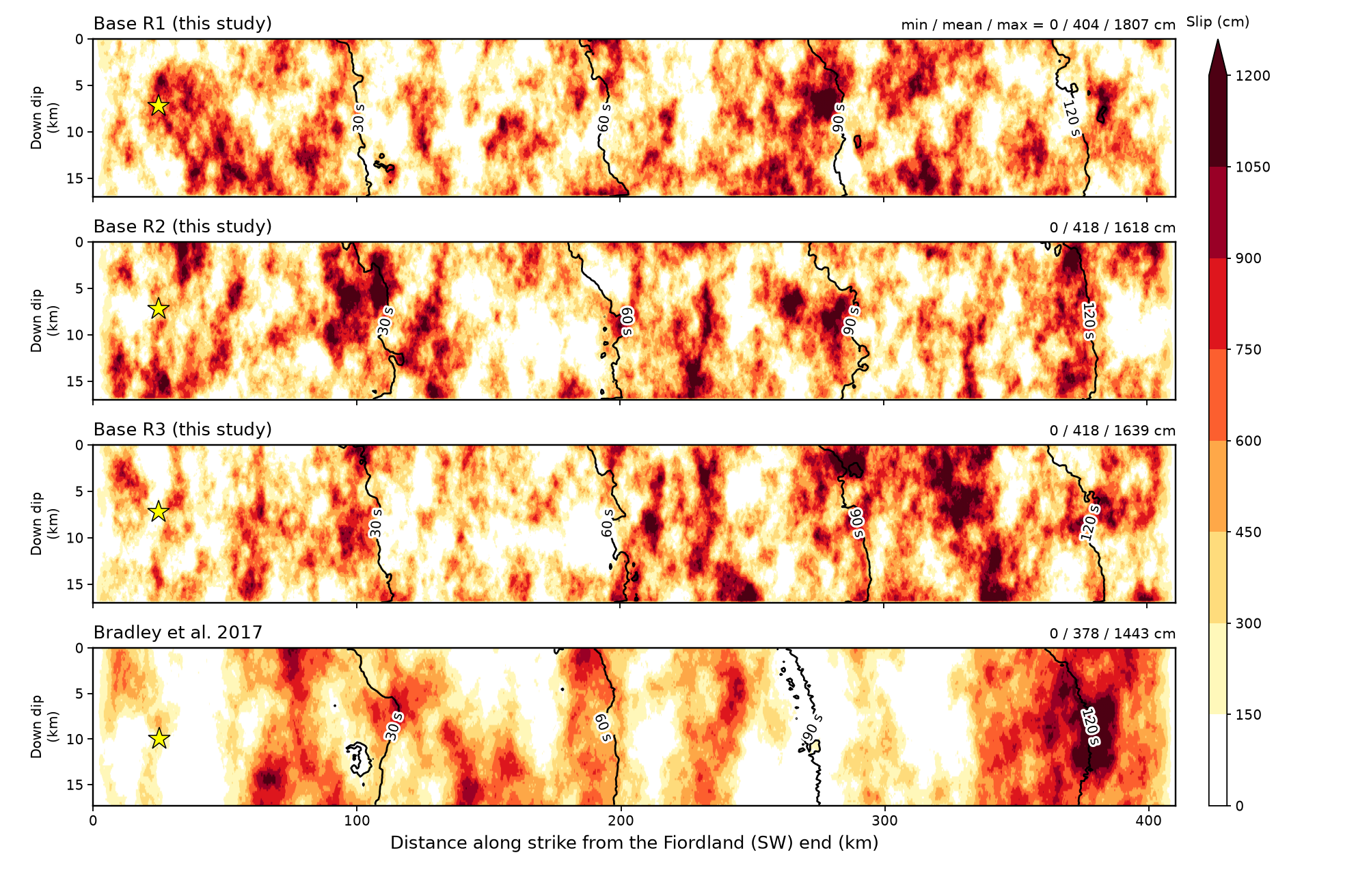}
	\caption{Slip on the three base realisations of this study and on the AF8
		rupture of \citet{bradleyGroundMotionSimulations2017}, on one colour scale. Contours are isochrones of rupture onset at
		30~s intervals; stars mark the hypocentres. The fault is exaggerated
		$3.5\times$ down dip.}
	\label{fig:af8-slip}
\end{figure}

\begin{figure}[htbp]
	\centering
	\includegraphics[width=\linewidth]{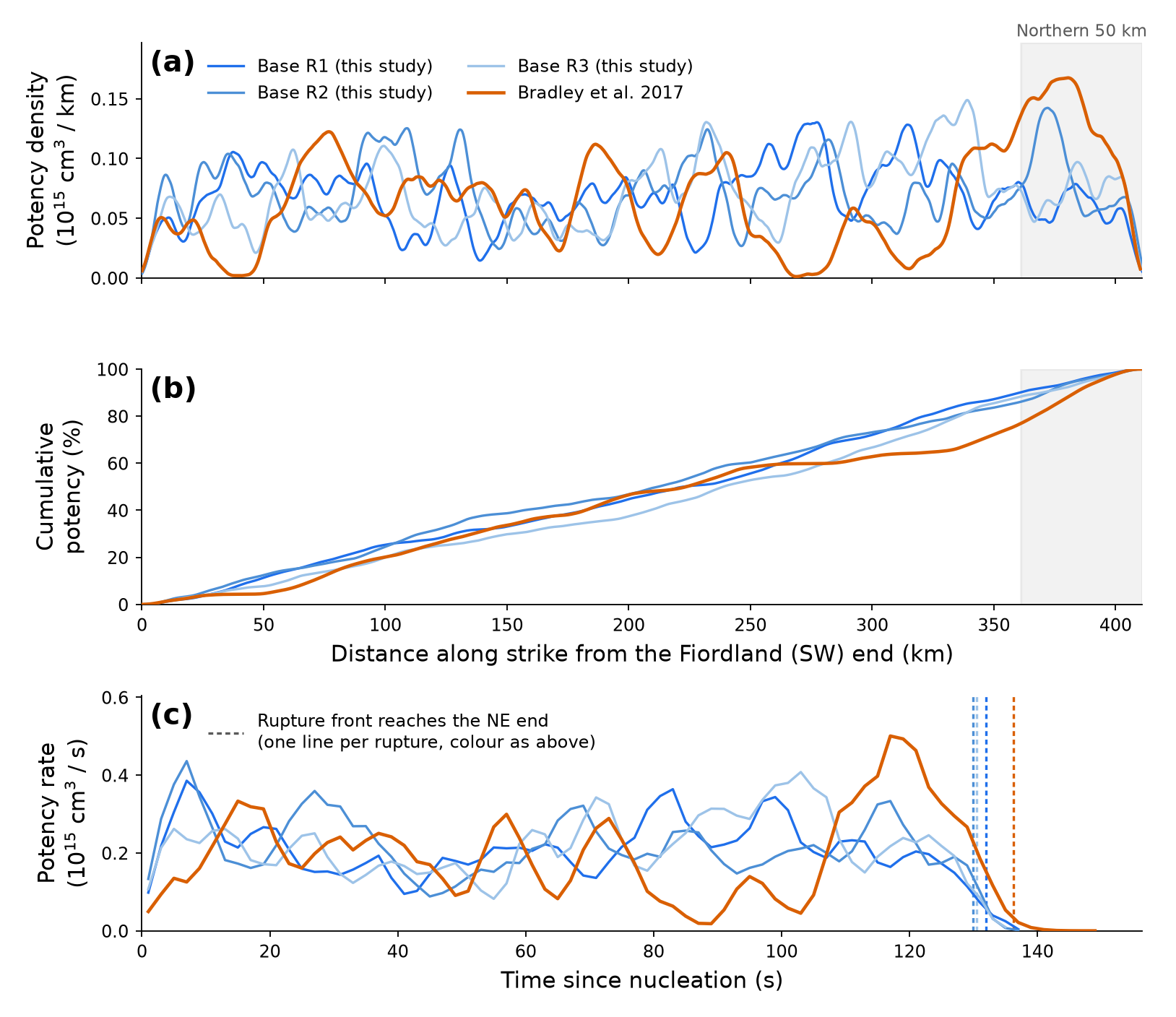}
	\caption{Potency ($=$ slip $\times$ area) of the three base realisations of
		this study and the AF8 rupture of \citet{bradleyGroundMotionSimulations2017}:
		(a) density along strike in 5~km windows, (b) cumulative fraction from the
		Fiordland end, and (c) potency binned by rupture onset time. Dashed vertical
		lines in (c) mark when each rupture front reaches the north-eastern end of the
		fault (130.0--132.0~s for this study's realisations and 136.3~s for the AF8
		rupture). The AF8 rupture puts 23.5\% of its potency in the northern 50~km,
		against 9.9--13.9\% for this study's realisations.}
	\label{fig:af8-potency}
\end{figure}

To explain the generally lower ground motion in the northern South Island, Figure~\ref{fig:af8-slip} compares the slip distribution of this study's three base realisations against the AF8 rupture, and Figure~\ref{fig:af8-potency} shows the corresponding potency release ($=$ slip $\times$ area).
Two differences are apparent.
First, the AF8 slip distribution is noticeably `smoother' than those of this study, indicative of less small-scale spatial complexity in the slip, which is consistent with the coarser subfault discretisation and earlier version of the rupture generator used in 2017.
Second, the AF8 rupture was unusual in concentrating a large fraction of its potency at its northern extent: 23.5\% of its potency is released in the northern 50~km of the rupture (approximately 12\% of its length), compared with 9.9--13.9\% for the three realisations of this study, i.e., close to the fraction expected for a uniform distribution of slip.
Because this potency is released at the end of the northward-propagating rupture, where forward directivity is strongest and the source is closest to the northern South Island, it yielded stronger shaking in the northern South Island in the AF8 scenario than in any of the realisations simulated here.
This highlights a limitation of drawing conclusions from individual scenario realisations: the regional distribution of shaking in the AF8 scenario is, in part, a consequence of the particular stochastic slip distribution of its single realisation, which motivates the consideration of multiple realisations per rupture geometry in this study.

\subsection{Reproducing the 2017 scenario}

\begin{figure}[htbp]
	\centering
	\includegraphics[width=\textwidth]{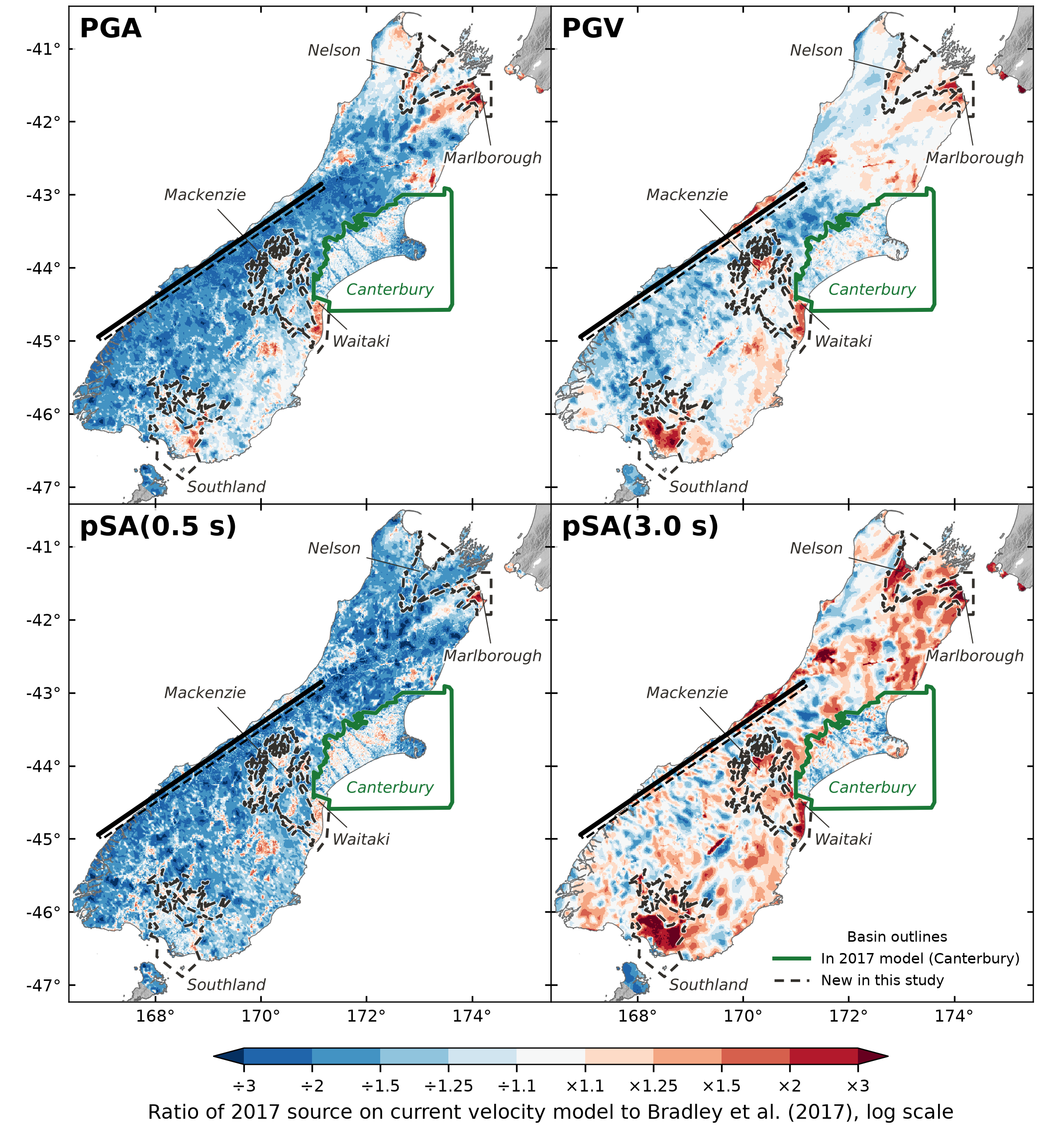}
	\caption{Ratio of simulated intensity measures from the reproduction of the
		\citet{bradleyGroundMotionSimulations2017} southern hypocentre scenario (the 2017 source simulated
		on the current velocity model) to the archived 2017 results, for PGA, PGV,
		and pSA at 0.5s and 3s, shown on a log scale. The 2017 values are
		interpolated from the 2017 station grid onto the current stations;
		stations more than 8 km from a 2017 station are omitted. The 2017 pSA
		values are the geometric mean of the horizontal components and the
		reproduction values are RotD50. Outlines mark the basins discussed
		in the text (Nelson, Marlborough, Canterbury, Mackenzie, Waitaki, and
		Southland). Canterbury (solid green) was the only one of these basins
		present in the velocity model of the 2017 study; the other basins (dashed)
		are new in the velocity model of this study. The solid black line is the fault trace and the dashed black
		outline the surface projection of the 2017 rupture, which is common to
		both runs.
	}
	\label{fig:repro_ratio}
\end{figure}

\begin{figure}[htbp]
	\centering
	\includegraphics[width=\textwidth]{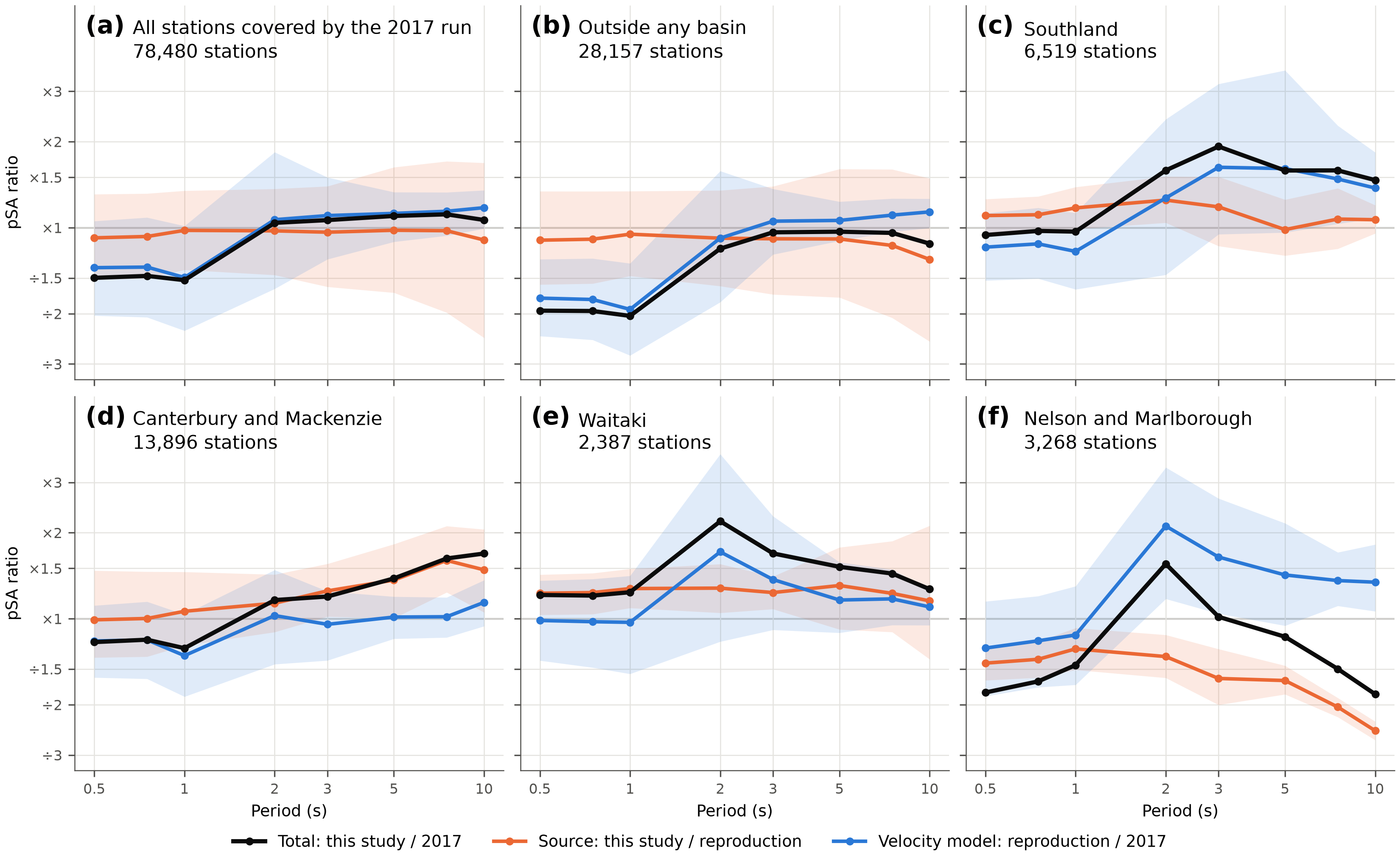}
	\caption{Decomposition of the change in pSA between the \citet{bradleyGroundMotionSimulations2017}
		scenario and this study into velocity-model and source terms, from 0.5s to
		10s. The velocity-model term (blue; reproduction / 2017) compares the 2017
		source simulated on the current velocity model with the archived 2017 results.
		The source term (orange; this study / reproduction) compares the new source
		with the 2017 source on the current velocity model. On the log scale the
		two sum to the total change (black; this study / 2017). Lines are means
		across the stations in each group and shading spans the 16th to 84th
		percentiles. The number of stations in each group is given in each panel.
	}
	\label{fig:decomposition}
\end{figure}

To isolate the effect of the updated velocity model, we reran the southern
hypocentre source of \citet{bradleyGroundMotionSimulations2017}
on the current velocity model and station grid.
The reference results for this comparison are the archived broadband outputs of the 2017 simulations (as distributed for scenario planning via SeisFinder); it is noted that \citet{bradleyGroundMotionSimulations2017} itself presented only the LF-dominated PGV and MMI measures, and hence the short-period intensity measures compared here derive from these archived outputs rather than from published values.
Because the source is unchanged, the ratio between this reproduction and the archived 2017 results
(Figure~\ref{fig:repro_ratio}) reflects the velocity model together with any
other change in the simulation workflow since 2017. Figure~\ref{fig:decomposition}
summarises this velocity-model term by station group, alongside the source
term that makes up the remainder of the total change.
The source term is the ratio of this study's base rupture to the reproduction,
so that, on a logarithmic scale, the velocity-model and source terms sum exactly
to the total change between the archived 2017 results and this study (this
additivity holds for the station-group means, but not for the percentile bands).
It is emphasised that the velocity-model term includes every difference between
the archived 2017 run and the reproduction other than the source, including the
simulation code and its parameters, the station grid, and the horizontal
component definition (the geometric mean of the two horizontal components in
2017, and RotD50 in this study). Station groups are defined by the basin outlines of NZVM v2.09.

The reproduction yields markedly lower short-period motion almost everywhere:
the reduction averages (geometric mean over stations) a factor of
approximately 1.3 for PGA and 1.4--1.5 for pSA at 0.5--1s, with reductions of
a factor of 1.5 or more at approximately 37--49\% of stations. At longer periods the change reverses to
a modest average increase (\(\times\)1.07--1.17 at 2--10s). The abrupt step
between 1s and 2s, seen in every station group including sites outside
basins, suggests the short-period reduction reflects changes to the
high-frequency (HF) simulation method and its parameters rather than the 3D
velocity model itself.
In the reproduction, the HF component was recomputed from the 2017 rupture
with the same HF configuration as the other simulations in this study (HF
implementation v6.0.3 of the~\citet{GP2015} method, a stress parameter of
50~bar, $\kappa=0.045$~s, the 1D Canterbury crustal profile with
$V_{S,min}=500$~m/s, a 1~Hz LF--HF transition frequency, and the CB14
$V_{S30}$-based site factor). The HF configuration of the 2017 simulations
was not documented in~\citet{bradleyGroundMotionSimulations2017} and could not
be recovered from archived workflow records, and therefore the individual
differences between the two HF configurations (e.g., implementation version,
stress parameter, $\kappa$, crustal profile, and whether a $V_{S30}$-based
site factor was applied) cannot be itemised. The short-period reduction is
accordingly attributed to the HF methodology and its parameterisation in
aggregate, rather than to any single parameter.
Outside basins the short-period reduction is strongest (a factor of about
1.8--1.9), and a small long-period increase (\(\times\)1.05--1.14 at 3--10s)
remains, pointing to changes in the wider crustal model as well as the basins.

At long periods the response is concentrated in particular basins. Southland
shows the strongest and broadest amplification, \(\times\)1.6 on average at
3--5s, exceeding \(\times\)3 at approximately one in five stations and
reaching \(\times\)5--9 at the most strongly amplified stations. Waitaki and
Nelson--Marlborough both show a pronounced peak at 2s (\(\times\)1.7 and
\(\times\)2.1 respectively) that decays with period, consistent with basin
resonance. In contrast, the velocity model changes long-period motion little
in Canterbury and Mackenzie, where the ratio stays close to unity from 2s to
7.5s (\(\times\)1.14 at 10s).

\section{Ground-motion amplitudes in Wellington}
\label{sec:wellington}

As previously alluded to, one of the primary reasons for performing AF-MFZ scenario simulations, following the 2017 AF8 simulation work, was to produce scenarios that caused appreciable shaking in Wellington, due to its position as the capital city of NZ, with a recognized high seismic hazard and seismic vulnerabilities recently illustrated by the 2016 Kaik\={o}ura earthquake~\citep{cubrinovski2020}.
Such scenarios offer supplemental value for emergency planning beyond that of the traditional AF8 scenario, which does not cause any appreciable impacts to the North Island of NZ.

\subsection{Amplitudes from different scenarios}

\begin{figure}[htbp]
	\centering
	\includegraphics[width=\linewidth]{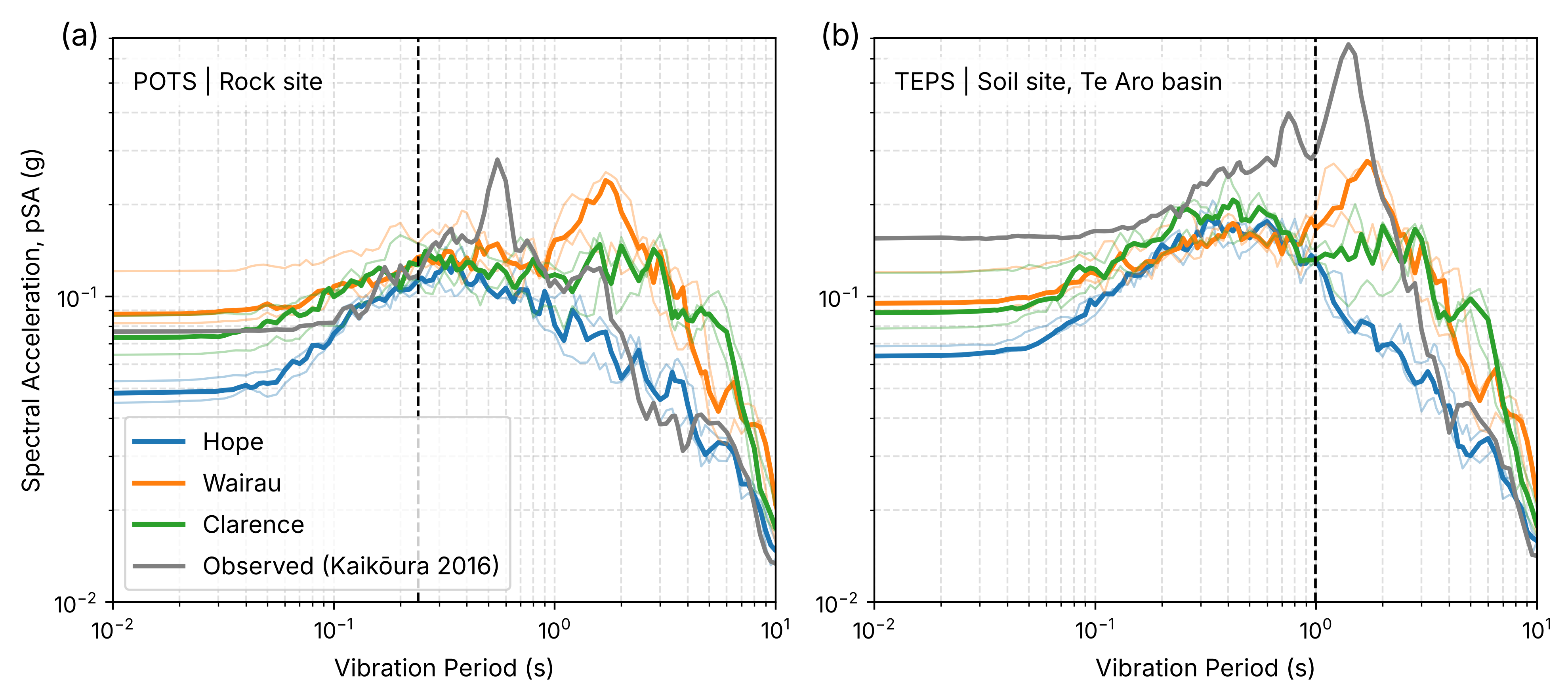}
	\caption{Simulated RotD50 response spectra at two Wellington sites (POTS and TEPS) for the three rupture scenarios (Hope, Wairau, Clarence). The simulations include $V_{S30}$-based site factor applied to the HF simulation component, with $V_{S30}$ values from \cite{wotherspoon_development_2024}. POTS is a rock site, and TEPS (Te Papa) is a soil site located in the Te Aro basin. For each scenario, the three simulated realizations (R1-R3) are shown as light lines and their geometric mean as a solid line. The observed response spectra from the 2016 Kaik\={o}ura earthquake are overlaid in gray. The vertical dashed black lines indicate the fundamental site period, $T_0$, from \cite{wotherspoon_development_2024}.}
	\label{fig:response_spectra_POTS_TEPS}
\end{figure}

Figure~\ref{fig:response_spectra_POTS_TEPS} illustrates the response spectral amplitudes from the three AF-MFZ rupture scenarios (considering the three realizations R1-R3 generated per scenario) for a typical rock site (POTS, Wellington Pottery Station) and a soft soil site (TEPS, Te Papa museum) in Wellington.
In addition, Table~\ref{tab:POTS_TEPS_comparison} shows the PGA and PGV values obtained at these two sites.
For both sites, the amplitudes are generally the lowest from the Alpine-Hope rupture, principally as a result of its larger source-to-site distance (see $R_{rup}$ values in Table~\ref{tab:POTS_TEPS_comparison}).
The amplitudes from Alpine-Wairau and Alpine-Clarence are of similar size, with the larger of the two changing as a function of vibration period and realization.

\begin{table}[htbp]
	\centering
	\caption{RotD50 peak ground acceleration (PGA) and peak ground velocity (PGV) on rock (POTS) and soil (TEPS) sites in central Wellington, for the 2016 Kaik\={o}ura earthquake (recorded values) and the AF-MFZ simulation scenarios. For the simulation scenarios the PGA and PGV values are given as the minimum-maximum range across the three realizations of each scenario. The source-to-site distance ($R_{rup}$) values are given as the range between the two sites.}
	\label{tab:POTS_TEPS_comparison}
	\begin{threeparttable}
		\begin{tabularx}{\linewidth}{@{} l c X X @{}}
			\toprule
			\textbf{Scenario}       & \textbf{$R_{rup}$ (km)} & \textbf{POTS (rock)}      & \textbf{TEPS (soil)}      \\
			\midrule
			2016 Kaik\={o}ura M7.87 & 66--68                  & 0.08~g, 17~cm/s           & 0.15~g, 39~cm/s           \\
			Alpine--Hope M8.31      & 52--54                  & 0.04--0.05~g, 12--14~cm/s & 0.06--0.07~g, 13--16~cm/s \\
			Alpine--Wairau M8.28    & 36                      & 0.08--0.12~g, 32--39~cm/s & 0.09--0.12~g, 37--44~cm/s \\
			Alpine--Clarence M8.31  & 41--43                  & 0.06--0.09~g, 29--38~cm/s & 0.08--0.12~g, 32--40~cm/s \\
			\bottomrule
		\end{tabularx}
	\end{threeparttable}
\end{table}

In addition to the AF-MFZ scenario simulation results,  Figure~\ref{fig:response_spectra_POTS_TEPS} and Table~\ref{tab:POTS_TEPS_comparison} also show the amplitudes recorded during the 2016 Kaik\={o}ura earthquake.
The 2016 Kaik\={o}ura earthquake produced ground motions with RotD50 PGA values of 0.076~g and 0.154~g at POTS and TEPS, respectively \citep{Pajaro2025b}.
Thus, the simulated PGA values for the Wairau and Clarence scenarios are similar to those from the 2016 earthquake at POTS, but smaller than observed at TEPS (Table~\ref{tab:POTS_TEPS_comparison}).
In terms of relative amplification between TEPS and POTS, the PGA amplification is 2.0 for the 2016 event and ranges from 1.0 to 1.6 across the simulated realizations.
Similarly, the PGV amplification is 2.3 for the 2016 event, while it is close to 1.1 for the simulated scenarios.
This relatively low amplification in the simulation, compared to the 2016 event, is also evident in Figure~\ref{fig:response_spectra_POTS_TEPS} in terms of spectral acceleration.

\begin{figure}[htbp]
	\centering
	\includegraphics[width=\linewidth]{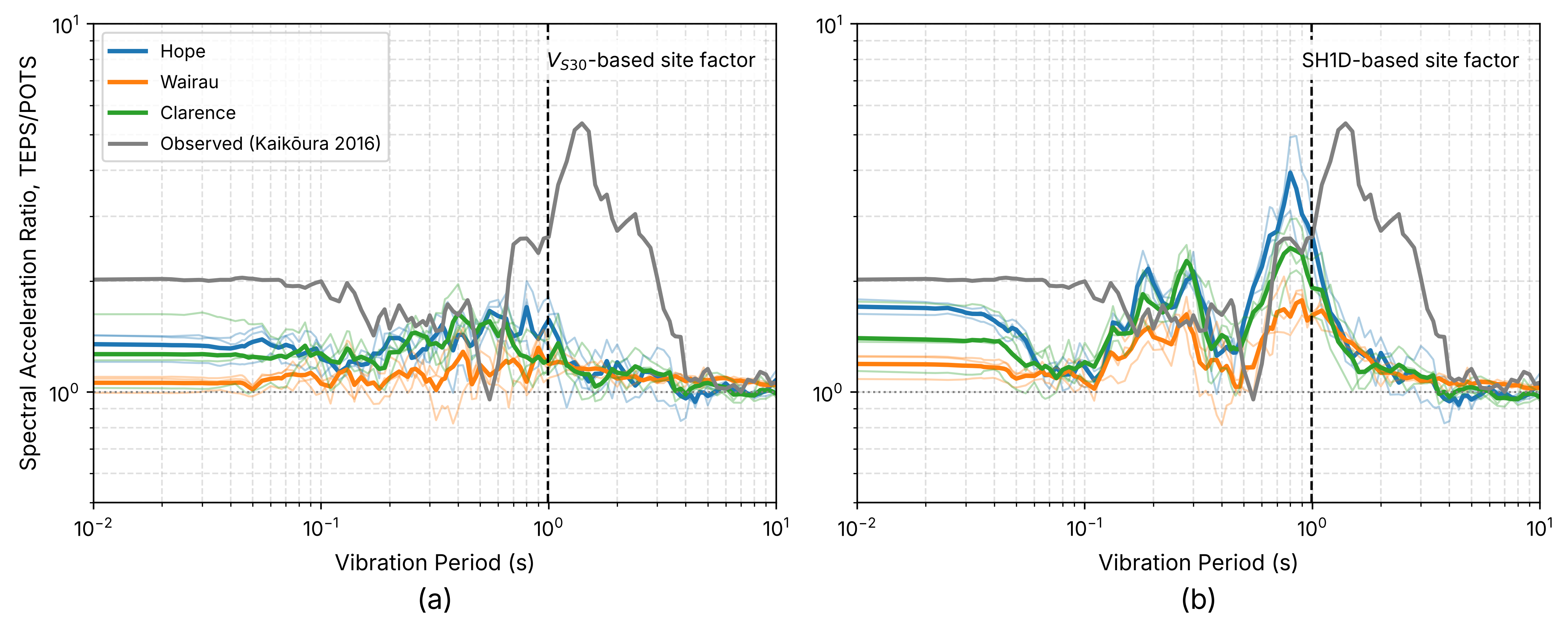}
	\caption{Spectral acceleration ratio between TEPS and POTS for the three rupture scenarios (Hope, Wairau, Clarence). For each scenario, the three simulated realizations (R1-R3) are shown as light lines and their geometric mean as a solid line. The observed spectral acceleration ratio from the 2016 Kaik\={o}ura earthquake is overlaid in gray. The vertical dashed black lines indicate the site period, $T_0$, at TEPS, from \cite{wotherspoon_development_2024}. Panel (a) shows the simulation results with the conventional $V_{S30}$-based site factor using $V_{S30}$ values from \cite{wotherspoon_development_2024}, and panel (b) shows the corresponding simulation results considering the SH1D-based site factor using $V_S$ profiles selected in \cite{kuncar2025b}.}
	\label{fig:response_spectra_ratio_POTS_TEPS_Vs30_SH1D}
\end{figure}

To better understand these trends, Figure~\ref{fig:response_spectra_ratio_POTS_TEPS_Vs30_SH1D}a shows the spectral acceleration ratio between TEPS and POTS for the different scenarios and realizations.
Given the short distance between TEPS and POTS, relative to $R_{rup}$ for these events, it can be assumed that source and path effects are similar for the two stations and that the spectral ratio is mainly influenced by local site effects \citep{Borcherdt1970}.
The figure shows that the simulations do not reproduce the site amplification observed in the 2016 event, particularly at the TEPS fundamental site period (indicated by the vertical dashed line), and at longer periods.
Similar results have been consistently reported in simulation-validation studies of small-magnitude \citep{kuncar2025b} and moderate-magnitude \citep{Lee2026} NZ earthquakes recorded at stations in central Wellington.
These previous validation studies have hypothesized that 3D ground motion simulations with a relatively coarse spatial resolution (such as the 100-m grid spacing used in these AF-MFZ simulations) and relatively high minimum $V_S$ (such as the 500~m/s used in this study) have important limitations for adequately representing longer-period site response in relatively shallow sedimentary basins like Wellington~\citep{stern_subsurface_2026}.

\subsubsection{Effect of shallow site response modelling}

A long-term solution to address the lack of site amplification discussed in the previous section is to increase the spatial resolution of the 3D simulations in the near surface in order to adequately represent wave propagation at smaller length scales and thus more appropriately discretize shallow sedimentary basins and their basin edges.
In the present work, we sought to better understand the degree to which the coarse spatial representation of the near-surface materials and the method used to account for shallow site effects affected ground-motion amplitudes in this study.
The standard method for incorporating shallow site effects in hybrid broadband ground-motion simulation consists of applying a $V_{S30}$-based site factor \citep[e.g.,][]{GP2010, kuncar2025}, as considered in this study (where it is only applied to the HF simulation component).
This method is practical for regional applications because $V_{S30}$ estimates are readily available at any location from regional maps or site-characterization databases~\citep[e.g.,][]{foster2019,wotherspoon_development_2024}.
When a site-specific $V_S$ profile or additional site-characterization data are available, a more sophisticated site adjustment can be performed \citep[e.g.,][]{kuncar2025}.
In particular, the validation study by \cite{kuncar2025b} showed that an adjustment method based on one-dimensional (1D) site-response analysis (``SH1D-based site factor” herein) can improve site-response predictions at sites in central Wellington near their fundamental site period, relative to the conventional $V_{S30}$-based approach.

\begin{figure}[htbp]
	\centering
	\includegraphics[width=\linewidth]{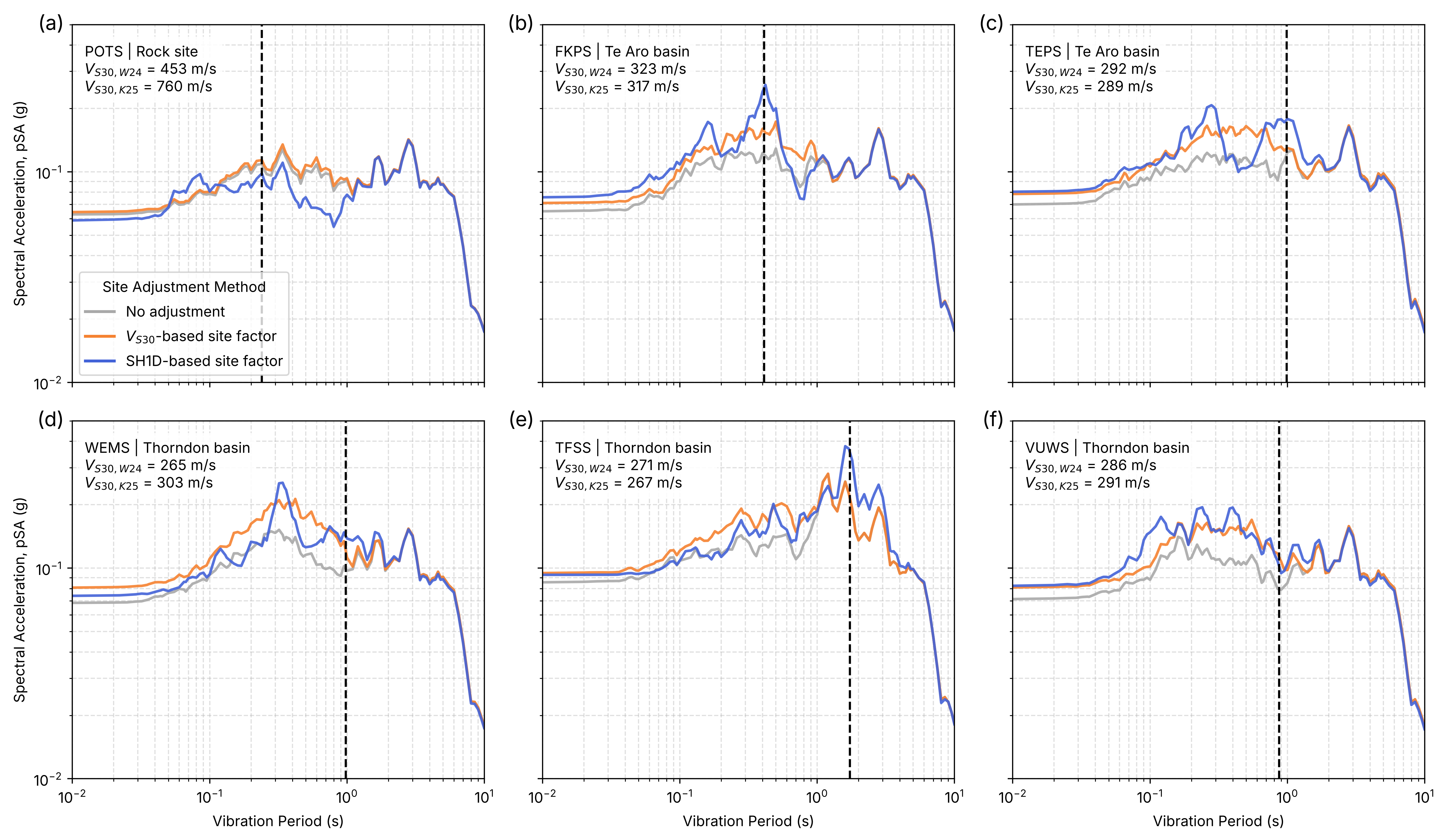}
	\caption{Comparison of simulated spectral accelerations for the Alpine–Clarence R1 rupture scenario at six Wellington sites: (a)~POTS, (b)~FKPS, (c)~TEPS, (d)~WEMS, (e)~TFSS, and (f)~VUWS. Results are shown for alternative approaches used to model shallow site effects. $V_{S30,W24}$ denotes the value from \cite{wotherspoon_development_2024} adopted in \cite{Pajaro2025b} and employed in the $V_{S30}$-based site adjustment. $V_{S30,K25}$ corresponds to the value derived from the site-specific $V_S$ profile adopted for the SH1D-based adjustment \citep{kuncar2025b}. The vertical dashed black line indicates the site period, $T_0$, from \cite{wotherspoon_development_2024}.}
	\label{fig:response_spectra_6sites_clarence_R1}
\end{figure}

Figure~\ref{fig:response_spectra_6sites_clarence_R1} shows the simulated ground motions at six strong-motion stations in central Wellington, for the Alpine-Clarence R1 scenario (results for the other rupture scenarios and realizations can be found in Electronic Supplement~\ref{sec:ESupp_A}).
The figure compares simulations performed without site adjustment, with the $V_{S30}$-based site factor, and with the SH1D-based site factor.
In each panel, the dashed vertical black line illustrates the fundamental site period from \citet{wotherspoon_development_2024}.

At several stations (e.g., FKPS, TEPS, and TFSS), the SH1D-based site factor yields higher response spectral amplitudes near the site period than both the conventional $V_{S30}$-based method and the case with no shallow site adjustment.
This behavior better aligns with observations from past events (e.g., Figure~\ref{fig:response_spectra_ratio_POTS_TEPS_Vs30_SH1D}).
The SH1D-based site factor also produces local peaks of higher amplitude at shorter vibration periods (e.g., FKPS, TEPS, WEMS, VUWS), arising from resonances associated with shallow impedance contrasts captured in the site-specific $V_S$ profiles.
Such features are not modeled by the $V_{S30}$-based approach, which represents the observed average site response over multiple sites sharing the same $V_{S30}$ value.

We note that the $V_{S30}$-based adjustment employed $V_{S30}$ values taken directly from the Site Characterization Database developed for the 2022 NZ National Seismic Hazard Model~\citep{wotherspoon_development_2024} and adopted in \cite{Pajaro2025b}, whereas the SH1D-based site factor used the preferred site-specific $V_S$ profile available at each location~\citep[e.g.,][]{cox_dynamic_2018} as selected by \cite{kuncar2025b} on the basis of all available data, including CPT, SPT, and borehole logs.
Figure~\ref{fig:response_spectra_6sites_clarence_R1} indicates, for each site, both the $V_{S30}$ value used for the $V_{S30}$-based adjustment ($V_{S30,W24}$) and the value derived from the $V_S$ profile used for the SH1D-based adjustment ($V_{S30,K25}$).
At FKPS, TEPS, TFSS, and VUWS the differences in $V_{S30}$ are minor and stem from alternative selections among available geophysical $V_S$ models.
At POTS the difference is substantial, with the SH1D-based method accounting for significantly stiffer conditions.
The stiffer representation of POTS is supported by the considerations discussed in \citet{kuncar2025b}, \citet{delatorre2024a}, and \citet{Kaiser2024}, and explains the marked reduction in spectral amplitudes in the $T=0.1-1$~s range observed in Figure~\ref{fig:response_spectra_6sites_clarence_R1}a when the SH1D-based site factor is applied.
At WEMS, the difference in $V_{S30}$ arises from using more recent site-investigation data in the SH1D-based site factor.
Hence, the differences observed between the two site-adjustment methods in Figure~\ref{fig:response_spectra_6sites_clarence_R1} result not only from the choice of the adjustment approach but also from a more refined selection of the site-characterization data.

Figure~\ref{fig:response_spectra_ratio_POTS_TEPS_Vs30_SH1D}b further shows the effect of the SH1D-based site factor and refined $V_S$ representation in terms of the spectral acceleration ratio between TEPS and POTS, consistent with the previously discussed Figure~\ref{fig:response_spectra_ratio_POTS_TEPS_Vs30_SH1D}a.
It shows that the more site-specific approach provides a better alignment with the observed spectral acceleration ratio, especially in the vicinity of TEPS's site period, but also at several other shorter vibration periods.
It also illustrates that this approach does not properly capture the strong amplification observed in the $T=1-3$~s range, which would require a more comprehensive treatment of 3D site effects.
While the SH1D-based site factor yields generally better results than the conventional $V_{S30}$-based approach, it can only be applied with a site-specific $V_S$ profile, hence it is not presently possible to use it for the entire simulation domain region.

\section{Conclusions}

This study performed leading-edge ground-motion simulations for great earthquakes on the Alpine Fault (AF) that rupture northward through the Marlborough Fault Zone (MFZ) to within 100 km of Wellington city.
Through an analysis of potential rupture geometries from the 2022 NZ National Seismic Hazard Model (NSHM), three multi-segment geometries that include the Wairau, Clarence, and Hope fault structures in the MFZ are obtained, as well as associated scenario rupture rates, which range from approximately 3000 - 6000 year recurrence intervals.

Hybrid broadband ground-motion simulations using a multi-segment kinematic source description and a 3D NZ crustal velocity model were performed.
Results illustrate that the strongest ground-motion amplitudes in Wellington resulted from the Alpine-Wairau and Alpine-Clarence ruptures, with those from the Alpine-Hope scenario being lower for all vibration periods.

The simulated ground motions exhibit strong northward rupture directivity, coupled with pronounced amplification and wavefield reverberation in the explicitly-modelled sedimentary basins.
The largest amplitudes in Nelson and the northwestern South Island result from the Alpine-Wairau rupture, and those in Kaikoura and northern Canterbury from the Alpine-Hope rupture, while shaking in Wellington is controlled by the Alpine-Wairau and Alpine-Clarence scenarios.
In central Wellington, the simulated peak ground accelerations of approximately 0.06--0.12~g are comparable to those observed at the same sites in the 2016 Kaik\=oura earthquake, however, with larger peak ground velocities, and resulting from rupture sources whose aggregate likelihood is comparable to that of the conventionally-considered Hutt Valley Wellington Fault and Hikurangi subduction sources.

Comparison with the 2017 AF8 simulations illustrates that the intervening modelling advancements produce two principal effects.
The velocity-model term, dominated by the explicit representation of 44 additional sedimentary basins, increases long-period motion within the newly-modelled basins (on average by factors of up to approximately 1.6--2.1 at 2--5~s in Southland, Waitaki, and Nelson--Marlborough), while short-period (0.5--1~s) motion is reduced by a factor of approximately 1.4--1.5 on average (1.8--1.9 outside basins), a reduction attributed primarily to differences in the high-frequency simulation methodology rather than to the velocity model itself.
The source term redistributes motion along strike: the single 2017 realisation concentrated 23.5\% of its potency in the northern 50~km of the rupture, as compared with 9.9--13.9\% for the three realisations of this study, which explains the stronger northern South Island shaking in the prior results and highlights the limitations of impact assessments based on individual scenario realisations.

With respect to shallow site response, the simulated spectral amplification between soil and rock sites in central Wellington is appreciably lower than that observed in the 2016 Kaik\=oura earthquake, consistent with prior validation findings for simulations with 100~m grid spacing and $V_{S,min}=500$~m/s in relatively shallow basins.
Site factors based on 1D site-response analysis improve predictions near the fundamental site period, however, they require a site-specific $V_S$ profile, and the $V_{S30}$-based approach therefore remains the default for regional-scale application.

Finally, it is noted that the presented simulations consider a single (southern) hypocentre location and three kinematic realisations per rupture geometry, with a LF--HF transition frequency of 1~Hz.
Future work should examine additional hypocentre locations and realisations, and higher spatial resolution in the Wellington region specifically, given the demonstrated sensitivity of long-period basin response to these modelling choices.
These scenarios, and their associated occurrence rates from the 2022 NZ NSHM, are intended to provide a quantitative basis for emergency planning and impact assessment through the AF8 programme and partner agencies, complementing the single-segment scenario that has underpinned such exercises to date.

\section{Acknowledgements}
This research was partially supported through the `It's Our Fault' research programme, QuakeCoRE: The New Zealand Centre for Earthquake Resilience, the New Zealand Natural Hazards Commission, and the New Zealand Marsden Fund.
Ground-motion simulation development and production calculations utilized the eResearch Infrastructure Platform hosted by the Crown company, Research and Education Advanced Network New Zealand (REANNZ) Ltd., and the Texas Advanced Computing Center (TACC) computing resources.
Anthropic's Claude model was used to assist with the drafting and editing of text, figure-generation code, and the organisational structuring of this paper; all content was reviewed and verified by the authors.
This is QuakeCoRE publication number 1189.

\section{Data Availability Statement}

The NZVM version 2.09 (\url{https://doi.org/10.5281/zenodo.19228330}) and the velocity modeling code used to build it (\url{https://doi.org/10.5281/zenodo.19228321}) are both archived at Zenodo.
The simulations were performed with the open-source simulation workflow \url{https://github.com/ucgmsim/workflow} (commit 7c08f7a), which implements the simulation domain and station grid generation, the multi-segment kinematic rupture generation of \citet{bradley2025b} (with the source geometry and magnitude scaling library \url{https://github.com/ucgmsim/source_modelling}), and the orchestration of the low-frequency, high-frequency, broadband, and intensity-measure calculations.
The Graves--Pitarka low-frequency and high-frequency simulation codes are available through the SCEC Broadband Platform (\url{https://github.com/SCECcode/bbp/wiki}). Code necessary to reproduce all figures, plus a complete intensity measure database is published on Zenodo (\url{https://doi.org/10.5281/zenodo.23586371}).
The kinematic source models (SRFs) for all simulation realizations (summarised in Electronic Supplement~B), the station set, and the computed intensity measures for every scenario and realization, together with the analysis and plotting scripts used to produce the figures in this paper, are available from the authors on request and will be deposited in a public archive prior to publication.
Observed ground motions from the 2016 Kaik\=oura earthquake at the Wellington strong-motion stations were obtained from the New Zealand Ground-Motion Database v4.3 \citep{Pajaro2025b} (\url{https://doi.org/10.17603/ds2-0gxc-j420}).
The archived broadband outputs of the \citet{bradleyGroundMotionSimulations2017} simulations used in the 2017-comparison section are available from the authors on request.
Due to their size, the full broadband waveform sets and source models are not hosted online; interested readers are encouraged to contact the authors to request specific datasets.

Computational details of the simulations are provided in Electronic Supplement~\ref{sec:ESupp_C}.

\bibliographystyle{SageH}
\bibliography{References_AlpineSimulations2025.bib}

\begin{thebibliography}{65}
\providecommand{\natexlab}[1]{#1}
\providecommand{\url}[1]{\texttt{#1}}
\providecommand{\urlprefix}{URL }
\expandafter\ifx\csname urlstyle\endcsname\relax
  \providecommand{\doi}[1]{DOI:\discretionary{}{}{}#1}\else
  \providecommand{\doi}{DOI:\discretionary{}{}{}\begingroup
  \urlstyle{rm}\Url}\fi

\bibitem[{Abrahamson et~al.(2014)Abrahamson, Silva and Kamai}]{ASK14}
Abrahamson NA, Silva WJ and Kamai R (2014) Summary of the {ASK14} {Ground}
  {Motion} {Relation} for {Active} {Crustal} {Regions}.
\newblock \emph{Earthquake Spectra} 30(3): 1025--1055.
\newblock \doi{10.1193/070913EQS198M}.
\newblock
  \urlprefix\url{https://journals.sagepub.com/doi/10.1193/070913EQS198M}.

\bibitem[{Aghababaei et~al.(2021)Aghababaei, Costello and
  Ranjitkar}]{aghababaei2021}
Aghababaei MTS, Costello SB and Ranjitkar P (2021) Measures to evaluate
  post-disaster trip resilience on road networks.
\newblock \emph{Journal of Transport Geography} 95: 103154.
\newblock \doi{10.1016/j.jtrangeo.2021.103154}.
\newblock
  \urlprefix\url{https://www.sciencedirect.com/science/article/pii/S0966692321002076}.

\bibitem[{Atkinson(2022)}]{Atkinson2022}
Atkinson GM (2022) Backbone ground-motion models for crustal, interface and
  slab earthquakes in {New} {Zealand} \doi{10.21420/QMJ6-P189}.
\newblock \urlprefix\url{https://shop.gns.cri.nz/sr_2022-11-pdf}.

\bibitem[{Baker et~al.(2021)Baker, Bradley and Stafford}]{baker2021}
Baker J, Bradley B and Stafford P (2021) \emph{Seismic Hazard and Risk
  Analysis}.
\newblock Cambridge University Press.
\newblock ISBN 978-1-108-42505-6.
\newblock
  \urlprefix\url{https://www.cambridge.org/core/books/seismic-hazard-and-risk-analysis/177B4BA01FC2600AA53A22B27598A06C}.

\bibitem[{Becker et~al.(2020)Becker, Potter, Prasanna, Tan, Payne, Holden,
  Horspool, Smith and Johnston}]{becker2020}
Becker JS, Potter SH, Prasanna R, Tan ML, Payne BA, Holden C, Horspool N, Smith
  R and Johnston DM (2020) Scoping the potential for earthquake early warning
  in aotearoa new zealand: A sectoral analysis of perceived benefits and
  challenges.
\newblock \emph{International Journal of Disaster Risk Reduction} 51: 101765.
\newblock \doi{10.1016/j.ijdrr.2020.101765}.
\newblock
  \urlprefix\url{https://www.sciencedirect.com/science/article/pii/S221242092031267X}.

\bibitem[{Berryman et~al.(2012)Berryman, Cooper, Norris, Villamor, Sutherland,
  Wright, Schermer, Langridge and Biasi}]{berrymanLateHoloceneRupture2012}
Berryman K, Cooper A, Norris R, Villamor P, Sutherland R, Wright T, Schermer E,
  Langridge R and Biasi G (2012) Late holocene rupture history of the alpine
  fault in south westland, new zealand.
\newblock \emph{Bulletin of the Seismological Society of America} 102(2):
  620--638.
\newblock \doi{10.1785/0120110177}.
\newblock \urlprefix\url{http://www.bssaonline.org/content/102/2/620.abstract}.

\bibitem[{Boore et~al.(2014)Boore, Stewart, Seyhan and Atkinson}]{Boore2014}
Boore DM, Stewart JP, Seyhan E and Atkinson GM (2014) {NGA}-{West2} {Equations}
  for {Predicting} {PGA}, {PGV}, and 5\% {Damped} {PSA} for {Shallow} {Crustal}
  {Earthquakes}.
\newblock \emph{Earthquake Spectra} 30(3): 1057--1085.
\newblock \doi{10.1193/070113EQS184M}.
\newblock
  \urlprefix\url{http://journals.sagepub.com/doi/10.1193/070113EQS184M}.

\bibitem[{Borcherdt(1970)}]{Borcherdt1970}
Borcherdt RD (1970) Effects of local geology on ground motion near {San}
  {Francisco} {Bay}*.
\newblock \emph{Bulletin of the Seismological Society of America} 60(1):
  29--61.
\newblock \doi{10.1785/BSSA0600010029}.
\newblock \urlprefix\url{https://doi.org/10.1785/BSSA0600010029}.

\bibitem[{Bozorgnia et~al.(2022)Bozorgnia, Abrahamson, Ahdi, Ancheta, Atik,
  Archuleta, Atkinson, Boore, Campbell, S-J~Chiou, Contreras, Darragh,
  Derakhshan, Donahue, Gregor, Gulerce, Idriss, Ji, Kishida, Kottke, Kuehn,
  Kwak, O-L~Kwok, Lin, Macedo, Mazzoni, Midorikawa, Muin, Parker, Rezaeian, Si,
  Silva, Stewart, Walling, Wooddell and Youngs}]{bozorgnia2022}
Bozorgnia Y, Abrahamson NA, Ahdi SK, Ancheta TD, Atik LA, Archuleta RJ,
  Atkinson GM, Boore DM, Campbell KW, S-J~Chiou B, Contreras V, Darragh RB,
  Derakhshan S, Donahue JL, Gregor N, Gulerce Z, Idriss I, Ji C, Kishida T,
  Kottke AR, Kuehn N, Kwak D, O-L~Kwok A, Lin P, Macedo J, Mazzoni S,
  Midorikawa S, Muin S, Parker GA, Rezaeian S, Si H, Silva WJ, Stewart JP,
  Walling M, Wooddell K and Youngs RR (2022) {NGA}-subduction research program.
\newblock \emph{Earthquake Spectra} 38(2): 783--798.
\newblock \doi{10.1177/87552930211056081}.
\newblock \urlprefix\url{https://doi.org/10.1177/87552930211056081}.

\bibitem[{Bozorgnia et~al.(2014)Bozorgnia, Abrahamson, Atik, Ancheta, Atkinson,
  Baker, Baltay, Boore, Campbell, Chiou, Darragh, Day, Donahue, Graves, Gregor,
  Hanks, Idriss, Kamai, Kishida, Kottke, Mahin, Rezaeian, Rowshandel, Seyhan,
  Shahi, Shantz, Silva, Spudich, Stewart, Watson-Lamprey, Wooddell and
  Youngs}]{bozorgniaNGAWest2ResearchProject2014}
Bozorgnia Y, Abrahamson NA, Atik LA, Ancheta TD, Atkinson GM, Baker JW, Baltay
  A, Boore DM, Campbell KW, Chiou BSJ, Darragh R, Day S, Donahue J, Graves RW,
  Gregor N, Hanks T, Idriss IM, Kamai R, Kishida T, Kottke A, Mahin SA,
  Rezaeian S, Rowshandel B, Seyhan E, Shahi S, Shantz T, Silva W, Spudich P,
  Stewart JP, Watson-Lamprey J, Wooddell K and Youngs R (2014) {NGA}-west2
  research project.
\newblock \emph{Earthquake Spectra} 30(3): 973--987.
\newblock \doi{10.1193/072113EQS209M}.
\newblock \urlprefix\url{http://dx.doi.org/10.1193/072113EQS209M}.

\bibitem[{Bradley et~al.(2017{\natexlab{a}})Bradley, Bae, Polak, Lee, Thomson
  and Tarbali}]{bradleyGroundMotionSimulations2017}
Bradley B, Bae S, Polak V, Lee R, Thomson E and Tarbali K (2017{\natexlab{a}})
  Ground motion simulations of great earthquakes on the alpine fault: Effect of
  hypocentre location and comparison with empirical modelling.
\newblock \emph{New Zealand Journal of Geology and Geophysics} 60(3): 188--198.
\newblock \doi{10.1080/00288306.2017.1297313}.

\bibitem[{Bradley and Faulkner(2025)}]{bradley2025b}
Bradley B and Faulkner J (2025) Extending kinematic rupture generators to
  multi-segment geometries.

\bibitem[{Bradley(2013)}]{Bradley2013}
Bradley BA (2013) A {New} {Zealand}-{Specific} {Pseudospectral} {Acceleration}
  {Ground}-{Motion} {Prediction} {Equation} for {Active} {Shallow} {Crustal}
  {Earthquakes} {Based} on {Foreign} {Models}.
\newblock \emph{Bulletin of the Seismological Society of America} 103(3):
  1801--1822.
\newblock \doi{10.1785/0120120021}.
\newblock
  \urlprefix\url{https://pubs.geoscienceworld.org/bssa/article/103/3/1801-1822/331663}.

\bibitem[{Bradley et~al.(2024)Bradley, Bora, Lee, Manea, Gerstenberger,
  Stafford, Atkinson, Weatherill, Hutchinson, de~la Torre, Hulsey and
  Kaiser}]{bradley2024}
Bradley BA, Bora SS, Lee RL, Manea EF, Gerstenberger MC, Stafford PJ, Atkinson
  GM, Weatherill G, Hutchinson J, de~la Torre CA, Hulsey AM and Kaiser AE
  (2024) The ground‐motion characterization model for the 2022 new zealand
  national seismic hazard model.
\newblock \emph{Bulletin of the Seismological Society of America} 114(1):
  329--349.
\newblock \doi{10.1785/0120230170}.
\newblock \urlprefix\url{https://doi.org/10.1785/0120230170}.

\bibitem[{Bradley et~al.(2017{\natexlab{b}})Bradley, Razafindrakoto and
  Polak}]{bradley2017}
Bradley BA, Razafindrakoto HNT and Polak V (2017{\natexlab{b}}) Ground‐motion
  observations from the 14 november 2016 mw7.8 kaikoura, new zealand,
  earthquake and insights from broadband simulations.
\newblock \emph{Seismological Research Letters} 88(3): 740--756.
\newblock \doi{10.1785/0220160225}.

\bibitem[{Bradley et~al.(2018)Bradley, Wotherspoon, Kaiser, Cox and
  Jeong}]{bradley2018}
Bradley BA, Wotherspoon LM, Kaiser AE, Cox BR and Jeong S (2018) Influence of
  site effects on observed ground motions in the wellington region from the
  mw 7.8 kaikōura, new zealand, earthquake.
\newblock \emph{Bulletin of the Seismological Society of America} 108(3B):
  1722--1735.
\newblock \doi{10.1785/0120170286}.
\newblock \urlprefix\url{https://doi.org/10.1785/0120170286}.

\bibitem[{Campbell and Bozorgnia(2014)}]{CB2014}
Campbell KW and Bozorgnia Y (2014) {NGA}-west2 ground motion model for the
  average horizontal components of {PGA}, {PGV}, and 5\% damped linear
  acceleration response spectra.
\newblock \emph{Earthquake Spectra} 30(3): 1087--1115.
\newblock \doi{10.1193/062913EQS175M}.

\bibitem[{Chiou and Youngs(2014)}]{ChiouYoungs2014}
Chiou BSJ and Youngs RR (2014) Update of the {Chiou} and {Youngs} {NGA} {Model}
  for the {Average} {Horizontal} {Component} of {Peak} {Ground} {Motion} and
  {Response} {Spectra}.
\newblock \emph{Earthquake Spectra} 30(3): 1117--1153.
\newblock \doi{10.1193/072813EQS219M}.
\newblock
  \urlprefix\url{https://journals.sagepub.com/doi/10.1193/072813EQS219M}.

\bibitem[{Cox and Vantassel(2018)}]{cox_dynamic_2018}
Cox B and Vantassel J (2018) Dynamic characterization of wellington, new
  zealand.
\newblock \doi{10.17603/DS24M6J}.
\newblock
  \urlprefix\url{https://www.designsafe-ci.org/data/browser/public/designsafe.storage.published/PRJ-2075}.

\bibitem[{Cubrinovski et~al.(2020)Cubrinovski, Bradley, Elwood, Johnston,
  Orchiston, Sullivan and Wotherspoon}]{cubrinovski2020}
Cubrinovski M, Bradley BA, Elwood KJ, Johnston D, Orchiston C, Sullivan T and
  Wotherspoon LM (2020) Wellington’s earthquake resilience: {Lessons} from
  the 2016 {Kaikōura} earthquake.
\newblock \emph{Earthquake Spectra} 36(3): 1448--1484.
\newblock \doi{10.1177/8755293020919426}.
\newblock \urlprefix\url{https://doi.org/10.1177/8755293020919426}.

\bibitem[{Davies et~al.(2021)Davies, Zorn, Wilson, Wotherspoon, Beavan, Davies
  and Hughes}]{davies2021}
Davies A, Zorn C, Wilson T, Wotherspoon L, Beavan S, Davies T and Hughes M
  (2021) Infrastructure failure propagations and recovery strategies from an
  alpine fault earthquake scenario: Establishing feedback loops between
  integrated modelling and participatory processes for impact reduction.
\newblock \emph{Bulletin of the New Zealand Society for Earthquake Engineering}
  54(2): 82--96.
\newblock \doi{10.5459/bnzsee.54.2.82-96}.
\newblock
  \urlprefix\url{https://bulletin.nzsee.org.nz/index.php/bnzsee/article/view/1538}.
\newblock Section: Articles.

\bibitem[{de~la Torre et~al.(2024)de~la Torre, Bradley, Kuncar, Lee,
  Wotherspoon and Kaiser}]{delatorre2024a}
de~la Torre CA, Bradley BA, Kuncar F, Lee RL, Wotherspoon LM and Kaiser AE
  (2024) Combining observed linear basin amplification factors with 1d
  nonlinear site-response analyses to predict site response for strong ground
  motions: Application to wellington, new zealand.
\newblock \emph{Earthquake Spectra} 40(1): 143--173.
\newblock \doi{10.1177/87552930231209726}.
\newblock \urlprefix\url{https://doi.org/10.1177/87552930231209726}.

\bibitem[{De~Pascale and Langridge(2012)}]{depascaleNewOnfaultEvidence2012}
De~Pascale G and Langridge R (2012) New on-fault evidence for a great
  earthquake in a.d. 1717, central alpine fault, new zealand.
\newblock \emph{Geology} \doi{10.1130/g33363.1}.
\newblock
  \urlprefix\url{http://geology.gsapubs.org/content/early/2012/06/21/G33363.1.abstract}.

\bibitem[{Dreger and Jordan(2015)}]{dregerIntroductionFocusSection2015}
Dreger DS and Jordan TH (2015) Introduction to the focus section on validation
  of the {SCEC} broadband platform v14.3 simulation methods.
\newblock \emph{Seismological Research Letters} 86(1): 15--16.
\newblock \doi{10.1785/0220140233}.
\newblock \urlprefix\url{http://srl.geoscienceworld.org/content/86/1/15.short}.

\bibitem[{Eberhart-Phillips et~al.(2020)Eberhart-Phillips, Bannister, Reyners
  and Henrys}]{eberhart-phillips_new_2020}
Eberhart-Phillips D, Bannister S, Reyners M and Henrys S (2020) New zealand
  wide model 2.2 seismic velocity and qs and qp models for new zealand.
\newblock \doi{10.5281/zenodo.3779523}.
\newblock \urlprefix\url{https://doi.org/10.5281/zenodo.3779523}.

\bibitem[{Eberhart-Phillips et~al.(2010)Eberhart-Phillips, Reyners, Bannister,
  Chadwick and Ellis}]{eberhart-phillipsEstablishingVersatile3D2010}
Eberhart-Phillips D, Reyners M, Bannister S, Chadwick M and Ellis S (2010)
  Establishing a versatile 3-d seismic velocity model for new zealand.
\newblock \emph{Seismological Research Letters} 81(6): 992--1000.
\newblock \doi{10.1785/gssrl.81.6.992}.
\newblock \urlprefix\url{http://srl.geoscienceworld.org}.

\bibitem[{Field and Page(2011)}]{fieldEstimatingEarthquakeRuptureRates2011}
Field EH and Page MT (2011) Estimating earthquake-rupture rates on a fault or
  fault system.
\newblock \emph{Bulletin of the Seismological Society of America} 101(1):
  79--92.
\newblock \doi{10.1785/0120100004}.
\newblock
  \urlprefix\url{http://www.bssaonline.org/cgi/content/abstract/101/1/79}.

\bibitem[{Foster et~al.(2019)Foster, Bradley, {McGann} and
  Wotherspoon}]{foster2019}
Foster KM, Bradley BA, {McGann} CR and Wotherspoon LM (2019) A {VS}30 map for
  new zealand based on geologic and terrain proxy variables and field
  measurements.
\newblock \emph{Earthquake Spectra} 35(4): 1865--1897.
\newblock \doi{10.1193/121118EQS281M}.

\bibitem[{Gerstenberger et~al.(2024)Gerstenberger, Bora, Bradley, {DiCaprio},
  Kaiser, Manea, Nicol, Rollins, Stirling, Thingbaijam, Van~Dissen, Abbott,
  Atkinson, Chamberlain, Christophersen, Clark, Coffey, de~la Torre, Ellis,
  Fraser, Graham, Griffin, Hamling, Hill, Howell, Hulsey, Hutchinson,
  Iturrieta, Johnson, Jurgens, Kirkman, Langridge, Lee, Litchfield, Maurer,
  Milner, Rastin, Rattenbury, Rhoades, Ristau, Schorlemmer, Seebeck, Shaw,
  Stafford, Stolte, Townend, Villamor, Wallace, Weatherill, Williams and
  Wotherspoon}]{gerstenberger2024}
Gerstenberger MC, Bora S, Bradley BA, {DiCaprio} C, Kaiser A, Manea EF, Nicol
  A, Rollins C, Stirling MW, Thingbaijam KKS, Van~Dissen RJ, Abbott ER,
  Atkinson GM, Chamberlain C, Christophersen A, Clark K, Coffey GL, de~la Torre
  CA, Ellis SM, Fraser J, Graham K, Griffin J, Hamling IJ, Hill MP, Howell A,
  Hulsey A, Hutchinson J, Iturrieta P, Johnson KM, Jurgens VO, Kirkman R,
  Langridge RM, Lee RL, Litchfield NJ, Maurer J, Milner KR, Rastin S,
  Rattenbury MS, Rhoades DA, Ristau J, Schorlemmer D, Seebeck H, Shaw BE,
  Stafford PJ, Stolte AC, Townend J, Villamor P, Wallace LM, Weatherill G,
  Williams CA and Wotherspoon LM (2024) The 2022 aotearoa new zealand national
  seismic hazard model: Process, overview, and results.
\newblock \emph{Bulletin of the Seismological Society of America} 114(1):
  7--36.
\newblock \doi{10.1785/0120230182}.
\newblock \urlprefix\url{https://doi.org/10.1785/0120230182}.

\bibitem[{Geyin and Maurer(2020)}]{geyinmertcan2020}
Geyin M and Maurer BW (2020) Fragility functions for liquefaction-induced
  ground failure.
\newblock \emph{Journal of Geotechnical and Geoenvironmental Engineering}
  146(12): 04020142.
\newblock \doi{10.1061/(ASCE)GT.1943-5606.0002416}.
\newblock \urlprefix\url{https://doi.org/10.1061/(ASCE)GT.1943-5606.0002416}.
\newblock Publisher: American Society of Civil Engineers.

\bibitem[{Ghisetti et~al.(2014)Ghisetti, Barnes and
  Sibson}]{ghisettiDeformationTopBasement2014}
Ghisetti FC, Barnes PM and Sibson RH (2014) Deformation of the top basement
  unconformity west of the alpine fault (south island, new zealand):
  seismotectonic implications.
\newblock \emph{New Zealand Journal of Geology and Geophysics} 57(3): 271--294.
\newblock \doi{10.1080/00288306.2013.876433}.
\newblock \urlprefix\url{http://dx.doi.org/10.1080/00288306.2013.876433}.

\bibitem[{Graves(1996)}]{gravesSimulatingSeismicWave1996}
Graves R (1996) Simulating seismic wave propagation in 3d elastic media using
  staggered-grid finite differences.
\newblock \emph{Bulletin of the Seismological Society of America} 86(4):
  1091--1106.
\newblock \doi{10.1785/BSSA0860041091}.

\bibitem[{Graves and Pitarka(2010)}]{GP2010}
Graves RW and Pitarka A (2010) Broadband ground-motion simulation using a
  hybrid approach.
\newblock \emph{Bulletin of the Seismological Society of America} 100(5A):
  2095--2123.
\newblock \doi{10.1785/0120100057}.

\bibitem[{Graves and Pitarka(2015)}]{GP2015}
Graves RW and Pitarka A (2015) Refinements to the graves and pitarka (2010)
  broadband ground‐motion simulation method.
\newblock \emph{Seismological Research Letters} 86(1): 75--80.
\newblock \doi{10.1785/0220140101}.

\bibitem[{Harris et~al.(2024)Harris, Robinson and Wilson}]{harris2024}
Harris OL, Robinson TR and Wilson TM (2024) Agent-based modelling of evacuation
  scenarios for a landslide-generated tsunami in milford sound, new zealand.
\newblock \emph{International Journal of Disaster Risk Reduction} 113: 104847.
\newblock \doi{10.1016/j.ijdrr.2024.104847}.
\newblock
  \urlprefix\url{https://www.sciencedirect.com/science/article/pii/S2212420924006095}.

\bibitem[{Hatem et~al.(2020)Hatem, Dolan, Zinke, Langridge, {McGuire}, Rhodes,
  Brown and Van~Dissen}]{hatem2020}
Hatem AE, Dolan JF, Zinke RW, Langridge RM, {McGuire} CP, Rhodes EJ, Brown N
  and Van~Dissen RJ (2020) Holocene to latest pleistocene incremental slip
  rates from the east-central hope fault (conway segment) at hossack station,
  marlborough fault system, south island, new zealand: Towards a dated path of
  earthquake slip along a plate boundary fault.
\newblock \emph{Geosphere} 16(6): 1558--1584.
\newblock \doi{10.1130/GES02263.1}.
\newblock \urlprefix\url{https://doi.org/10.1130/GES02263.1}.

\bibitem[{Howarth et~al.(2021)Howarth, Barth, Fitzsimons, Richards-Dinger,
  Clark, Biasi, Cochran, Langridge, Berryman and Sutherland}]{howarth2021}
Howarth JD, Barth NC, Fitzsimons SJ, Richards-Dinger K, Clark KJ, Biasi GP,
  Cochran UA, Langridge RM, Berryman KR and Sutherland R (2021) Spatiotemporal
  clustering of great earthquakes on a transform fault controlled by geometry.
\newblock \emph{Nature Geoscience} 14(5): 314--320.
\newblock \doi{10.1038/s41561-021-00721-4}.
\newblock \urlprefix\url{https://doi.org/10.1038/s41561-021-00721-4}.

\bibitem[{Howarth et~al.(2016)Howarth, Fitzsimons, Norris, Langridge and
  Vandergoes}]{howarth2016}
Howarth JD, Fitzsimons SJ, Norris RJ, Langridge R and Vandergoes MJ (2016) A
  2000 yr rupture history for the alpine fault derived from lake ellery, south
  island, new zealand.
\newblock \emph{{GSA} Bulletin} 128(3-4): 627--643.
\newblock \doi{10.1130/B31300.1}.
\newblock \urlprefix\url{https://doi.org/10.1130/B31300.1}.

\bibitem[{Kaiser et~al.(2024)Kaiser, Hill, De~La~Torre, Bora, Manea,
  Wotherspoon, Atkinson, Lee, Bradley, Hulsey, Stolte and
  Gerstenberger}]{Kaiser2024}
Kaiser AE, Hill MP, De~La~Torre C, Bora S, Manea E, Wotherspoon L, Atkinson GM,
  Lee R, Bradley B, Hulsey A, Stolte A and Gerstenberger M (2024) Overview of
  {Site} {Effects} and the {Application} of the 2022 {New} {Zealand} {NSHM} in
  the {Wellington} {Basin}, {New} {Zealand}.
\newblock \emph{Bulletin of the Seismological Society of America} 114(1):
  399--421.
\newblock \doi{10.1785/0120230189}.
\newblock
  \urlprefix\url{https://pubs.geoscienceworld.org/bssa/article/114/1/399/631792/Overview-of-Site-Effects-and-the-Application-of}.

\bibitem[{Kuncar et~al.(2025{\natexlab{a}})Kuncar, Bradley, de~la Torre,
  Rodriguez-Marek, Zhu and Lee}]{kuncar2025}
Kuncar F, Bradley BA, de~la Torre CA, Rodriguez-Marek A, Zhu C and Lee RL
  (2025{\natexlab{a}}) Methods to account for shallow site effects in hybrid
  broadband ground-motion simulations.
\newblock \emph{Earthquake Spectra} 41(2): 1272--1313.
\newblock \doi{10.1177/87552930241301059}.
\newblock \urlprefix\url{https://doi.org/10.1177/87552930241301059}.

\bibitem[{Kuncar et~al.(2025{\natexlab{b}})Kuncar, Bradley, de~la Torre,
  Rodriguez-Marek, Zhu and Lee}]{kuncar2025b}
Kuncar F, Bradley BA, de~la Torre CA, Rodriguez-Marek A, Zhu C and Lee RL
  (2025{\natexlab{b}}) Validating alternative methods to account for shallow
  site effects in hybrid broadband ground-motion simulation of small-magnitude
  earthquakes in new zealand.
\newblock \emph{Earthquake Spectra} 41(5): 3914--3948.
\newblock \doi{10.1177/87552930251377784}.
\newblock \urlprefix\url{https://doi.org/10.1177/87552930251377784}.
\newblock Publisher: {SAGE} Publications Ltd {STM}.

\bibitem[{Lee et~al.(2021)Lee, Bradley, Graves, Paterson and Motha}]{Lee2021}
Lee R, Bradley B, Graves R, Paterson J and Motha J (2021) Source modelling
  insights from ground motion simulation validation of moderate magnitude
  active shallow crustal earthquakes in new zealand.
\newblock In: \emph{{NZSEE} 2021 annual conference}. pp. 1--8.
\newblock Type: Conference Proceedings.

\bibitem[{Lee et~al.(2017)Lee, Bradley, Ghisetti and Thomson}]{lee2017a}
Lee RL, Bradley BA, Ghisetti FC and Thomson EM (2017) Development of a 3d
  velocity model of the canterbury, new zealand, region for broadband
  ground-motion simulation.
\newblock \emph{Bulletin of the Seismological Society of America} 107(5):
  2131--2150.
\newblock \doi{10.1785/0120160326}.
\newblock \urlprefix\url{http://dx.doi.org/10.1785/0120160326}.

\bibitem[{Lee et~al.(2026)Lee, Bradley, Hill, De~La~Torre, Kaiser and
  Wotherspoon}]{Lee2026}
Lee RL, Bradley BA, Hill MP, De~La~Torre CA, Kaiser AE and Wotherspoon LM
  (2026) {3D} {Ground}-{Motion} {Simulation}-{Based} {Site} {Amplification}
  {Considering} {Multiple} {Basin} {Geometries}: {A} {Wellington}, {New}
  {Zealand}, {Case} {Study}.
\newblock \emph{Bulletin of the Seismological Society of America} 116(5):
  1864--1883.
\newblock \doi{10.1785/0120260033}.
\newblock
  \urlprefix\url{https://pubs.geoscienceworld.org/bssa/article/doi/10.1785/0120260033/734189/3D-Ground-Motion-Simulation-Based-Site}.

\bibitem[{Lee et~al.(2020)Lee, Bradley, Stafford, Graves and
  Rodriguez-Marek}]{Lee2020}
Lee RL, Bradley BA, Stafford PJ, Graves RW and Rodriguez-Marek A (2020) Hybrid
  broadband ground motion simulation validation of small magnitude earthquakes
  in canterbury, new zealand.
\newblock \emph{Earthquake Spectra} 36(2): 673--699.
\newblock \doi{10.1177/8755293019891718}.

\bibitem[{Lee et~al.(2022)Lee, Bradley, Stafford, Graves and
  Rodriguez-Marek}]{lee2022}
Lee RL, Bradley BA, Stafford PJ, Graves RW and Rodriguez-Marek A (2022) Hybrid
  broadband ground-motion simulation validation of small magnitude active
  shallow crustal earthquakes in new zealand.
\newblock \emph{Earthquake Spectra} 38(4): 2548--2579.
\newblock \doi{10.1177/87552930221109297}.
\newblock \urlprefix\url{https://doi.org/10.1177/87552930221109297}.

\bibitem[{Leonard(2010)}]{leonard2010}
Leonard M (2010) Earthquake fault scaling: Self-consistent relating of rupture
  length, width, average displacement, and moment release.
\newblock \emph{Bulletin of the Seismological Society of America} 100(5A):
  1971--1988.
\newblock \doi{10.1785/0120090189}.

\bibitem[{Milner and Field(2024)}]{milner2023}
Milner KR and Field EH (2024) A comprehensive fault‐system inversion
  approach: Methods and application to {NSHM}23.
\newblock \emph{Bulletin of the Seismological Society of America} 114(1):
  486--522.
\newblock \doi{10.1785/0120230122}.
\newblock \urlprefix\url{https://doi.org/10.1785/0120230122}.

\bibitem[{Orchiston et~al.(2018)Orchiston, Mitchell, Wilson, Langridge, Davies,
  Bradley, Johnston, Davies, Becker and {McKay}}]{orchiston2018}
Orchiston C, Mitchell J, Wilson T, Langridge R, Davies T, Bradley B, Johnston
  D, Davies A, Becker J and {McKay} A (2018) Project {AF}8: developing a
  coordinated, multi-agency response plan for a future great alpine fault
  earthquake.
\newblock \emph{New Zealand Journal of Geology and Geophysics} 61(3): 389--402.
\newblock \doi{10.1080/00288306.2018.1455716}.
\newblock \urlprefix\url{https://doi.org/10.1080/00288306.2018.1455716}.
\newblock Publisher: Taylor \& Francis.

\bibitem[{Pajaro et~al.(2025{\natexlab{a}})Pajaro, Bradley, Ridden, Rampersad
  and Lee}]{Pajaro2025b}
Pajaro C, Bradley B, Ridden JN, Rampersad A and Lee R (2025{\natexlab{a}}) New
  {Zealand} {Ground}-{Motion} {Database} ({Version} 4.3).
\newblock \doi{10.17603/ds2-0gxc-j420}.

\bibitem[{Pajaro et~al.(2025{\natexlab{b}})Pajaro, Bradley and
  Lee}]{pajaro2025}
Pajaro CA, Bradley B and Lee R (2025{\natexlab{b}}) Validation of hybrid
  broadband ground-motion simulation for large-magnitude earthquakes in new
  zealand.
\newblock p.~9.
\newblock \urlprefix\url{https://tinyurl.com/PajaroLargeM}.

\bibitem[{Razafindrakoto et~al.(2018)Razafindrakoto, Bradley and
  Graves}]{razafindrakoto2018}
Razafindrakoto HNT, Bradley BA and Graves RW (2018) Broadband ground-motion
  simulation of the 2011 mw 6.2 christchurch, new zealand, earthquake.
\newblock \emph{Bulletin of the Seismological Society of America} 108(4):
  2130--2147.
\newblock \doi{10.1785/0120170388}.
\newblock \urlprefix\url{https://doi.org/10.1785/0120170388}.

\bibitem[{Rhoades and Van~Dissen(2003)}]{rhoadesEstimatesTimevaryingHazard2003}
Rhoades D and Van~Dissen R (2003) Estimates of the time-varying hazard of
  rupture of the alpine fault, new zealand, allowing for uncertainties.
\newblock \emph{New Zealand Journal of Geology and Geophysics} 46: 479--488.
\newblock \doi{10.1080/00288306.2003.9515023}.

\bibitem[{Robinson et~al.(2016)Robinson, Davies, Wilson and
  Orchiston}]{robinsonCoseismicLandslidingEstimates06}
Robinson TR, Davies TRH, Wilson TM and Orchiston C (2016) Coseismic landsliding
  estimates for an alpine fault earthquake and the consequences for erosion of
  the southern alps, new zealand.
\newblock \emph{Geomorphology} 263: 71--86.
\newblock \doi{10.1016/j.geomorph.2016.03.033}.
\newblock
  \urlprefix\url{http://www.sciencedirect.com/science/article/pii/S0169555X1630143X}.

\bibitem[{Stafford(2022)}]{Stafford2022}
Stafford PJ (2022) A model for the distribution of response spectral ordinates
  from {New} {Zealand} crustal earthquakes based upon adjustments to the
  {Chiou} and {Youngs} (2014) response spectral model \doi{10.21420/5098-0S19}.
\newblock \urlprefix\url{https://shop.gns.cri.nz/sr_2022-15-pdf/}.

\bibitem[{Stern et~al.(2026)Stern, Louie, Newton, Thorpe-Loversuch, Stronach
  and Stratford}]{stern_subsurface_2026}
Stern TA, Louie JN, Newton MO, Thorpe-Loversuch SB, Stronach AI and Stratford
  WR (2026) Subsurface {Expression} of the {Lambton} {Fault} and {Potential}
  for the {Basin} {Edge} {Effect} in the {Wellington} {Central} {Business}
  {District}.
\newblock \emph{New Zealand Journal of Geology and Geophysics} 69(1): e70034.
\newblock \doi{10.1002/jgo2.70034}.
\newblock \urlprefix\url{https://doi.org/10.1002/jgo2.70034}.

\bibitem[{Stirling et~al.(2012)Stirling, McVerry, Gerstenberger, Litchfield,
  Van~Dissen, Berryman, Barnes, Wallace, Villamor, Langridge, Lamarche, Nodder,
  Reyners, Bradley, Rhoades, Smith, Nicol, Pettinga, Clark and
  Jacobs}]{Stirling_2012}
Stirling M, McVerry G, Gerstenberger M, Litchfield N, Van~Dissen R, Berryman K,
  Barnes P, Wallace L, Villamor P, Langridge R, Lamarche G, Nodder S, Reyners
  M, Bradley B, Rhoades D, Smith W, Nicol A, Pettinga J, Clark K and Jacobs K
  (2012) National seismic hazard model for new zealand: 2010 update.
\newblock \emph{Bulletin of the Seismological Society of America} 102(4):
  1514–1542.
\newblock \doi{10.1785/0120110170}.
\newblock \urlprefix\url{http://dx.doi.org/10.1785/0120110170}.

\bibitem[{Sutherland et~al.(2006)Sutherland, Berryman and
  Norris}]{sutherlandQuaternarySlipRate2006}
Sutherland R, Berryman K and Norris R (2006) Quaternary slip rate and
  geomorphology of the alpine fault: Implications for kinematics and seismic
  hazard in southwest new zealand.
\newblock \emph{Geological Society of America Bulletin} 118: 464--474.
\newblock \doi{10.1130/B25627.1}.

\bibitem[{Thomson et~al.(2020)Thomson, Bradley and Lee}]{thomson2020}
Thomson EM, Bradley BA and Lee RL (2020) Methodology and computational
  implementation of a new zealand velocity model ({NZVM}2.0) for broadband
  ground motion simulation.
\newblock \emph{New Zealand Journal of Geology and Geophysics} 63(1): 110--127.
\newblock \doi{10.1080/00288306.2019.1636830}.
\newblock \urlprefix\url{https://doi.org/10.1080/00288306.2019.1636830}.

\bibitem[{Tiwari et~al.(2025)Tiwari, Bradley, Lee, Bae and de~la
  Torre}]{tiwari2025}
Tiwari A, Bradley B, Lee RL, Bae S and de~la Torre C (2025) New zealand-wide
  representation of sedimentary basin structures for use in 3d ground-motion
  simulations.

\bibitem[{Van~Dissen et~al.(2024)Van~Dissen, Johnson, Seebeck, Wallace,
  Rollins, Maurer, Gerstenberger, Williams, Hamling, Howell and
  {DiCaprio}}]{vandissen2024}
Van~Dissen RJ, Johnson KM, Seebeck H, Wallace LM, Rollins C, Maurer J,
  Gerstenberger MC, Williams CA, Hamling IJ, Howell A and {DiCaprio} CJ (2024)
  Upper plate and subduction interface deformation models in the 2022 revision
  of the aotearoa new zealand national seismic hazard model.
\newblock \emph{Bulletin of the Seismological Society of America} 114(1):
  37--56.
\newblock \doi{10.1785/0120230118}.
\newblock \urlprefix\url{https://doi.org/10.1785/0120230118}.

\bibitem[{Wallace et~al.(2007)Wallace, Beavan, {McCaffrey}, Berryman and
  Denys}]{wallaceBalancingPlateMotion2007}
Wallace LM, Beavan J, {McCaffrey} R, Berryman K and Denys P (2007) Balancing
  the plate motion budget in the south island, new zealand using {GPS},
  geological and seismological data.
\newblock \emph{Geophysical Journal International} 168(1): 332--352.
\newblock \doi{10.1111/j.1365-246X.2006.03183.x}.
\newblock \urlprefix\url{http://dx.doi.org/10.1111/j.1365-246X.2006.03183.x}.

\bibitem[{Worden et~al.(2012)Worden, Gerstenberger, Rhoades and
  Wald}]{wordenProbabilisticRelationshipsGroundMotion2012}
Worden CB, Gerstenberger MC, Rhoades DA and Wald DJ (2012) Probabilistic
  relationships between ground-motion parameters and modified mercalli
  intensity in california.
\newblock \emph{Bulletin of the Seismological Society of America} 102(1):
  204--221.
\newblock \doi{10.1785/0120110156}.
\newblock \urlprefix\url{http://www.bssaonline.org/content/102/1/204.abstract}.

\bibitem[{Wotherspoon et~al.(2024)Wotherspoon, Kaiser, Stolte and
  Manea}]{wotherspoon_development_2024}
Wotherspoon LM, Kaiser AE, Stolte AC and Manea EF (2024) Development of the
  site characterization database for the 2022 new zealand national seismic
  hazard model.
\newblock \emph{Seismological Research Letters} 95(1): 214--225.
\newblock \doi{10.1785/0220230219}.
\newblock \urlprefix\url{https://doi.org/10.1785/0220230219}.

\bibitem[{Zinke et~al.(2019)Zinke, Dolan, Rhodes, Van~Dissen, {McGuire}, Hatem,
  Brown and Langridge}]{zinke2019}
Zinke R, Dolan JF, Rhodes EJ, Van~Dissen R, {McGuire} CP, Hatem AE, Brown ND
  and Langridge RM (2019) Multimillennial incremental slip rate variability of
  the clarence fault at the tophouse road site, marlborough fault system, new
  zealand.
\newblock \emph{Geophysical Research Letters} 46(2): 717--725.
\newblock \doi{10.1029/2018GL080688}.
\newblock \urlprefix\url{https://doi.org/10.1029/2018GL080688}.
\newblock Publisher: John Wiley \& Sons, Ltd.

\end{thebibliography}

\newpage

\setcounter{secnumdepth}{3}

\newpage

\setcounter{secnumdepth}{3}

\begin{appendices}
	\setcounter{table}{0}
	\setcounter{figure}{0}
	\setcounter{equation}{0}
	\renewcommand{\thetable}{\Alph{section}.\arabic{table}}
	\renewcommand{\thefigure}{\Alph{section}.\arabic{figure}}
	\renewcommand{\theequation}{\Alph{section}.\arabic{equation}}
	\pagenumbering{roman}

	\begin{itemize}

		\item[--] \textbf{Electronic Supplement A: Additional Results}
		\item[--] \textbf{Electronic Supplement B: Simulation Source and Model Parameters}
		\item[--] \textbf{Electronic Supplement C: Computational Details}

	\end{itemize}

	\pagebreak{}

	\section{Electronic Supplement A: Additional Results}
	\label{sec:ESupp_A}
	\setcounter{figure}{0}
	\setcounter{table}{0}

\begin{figure}[h]
		\centering
		\includegraphics[width=\linewidth]{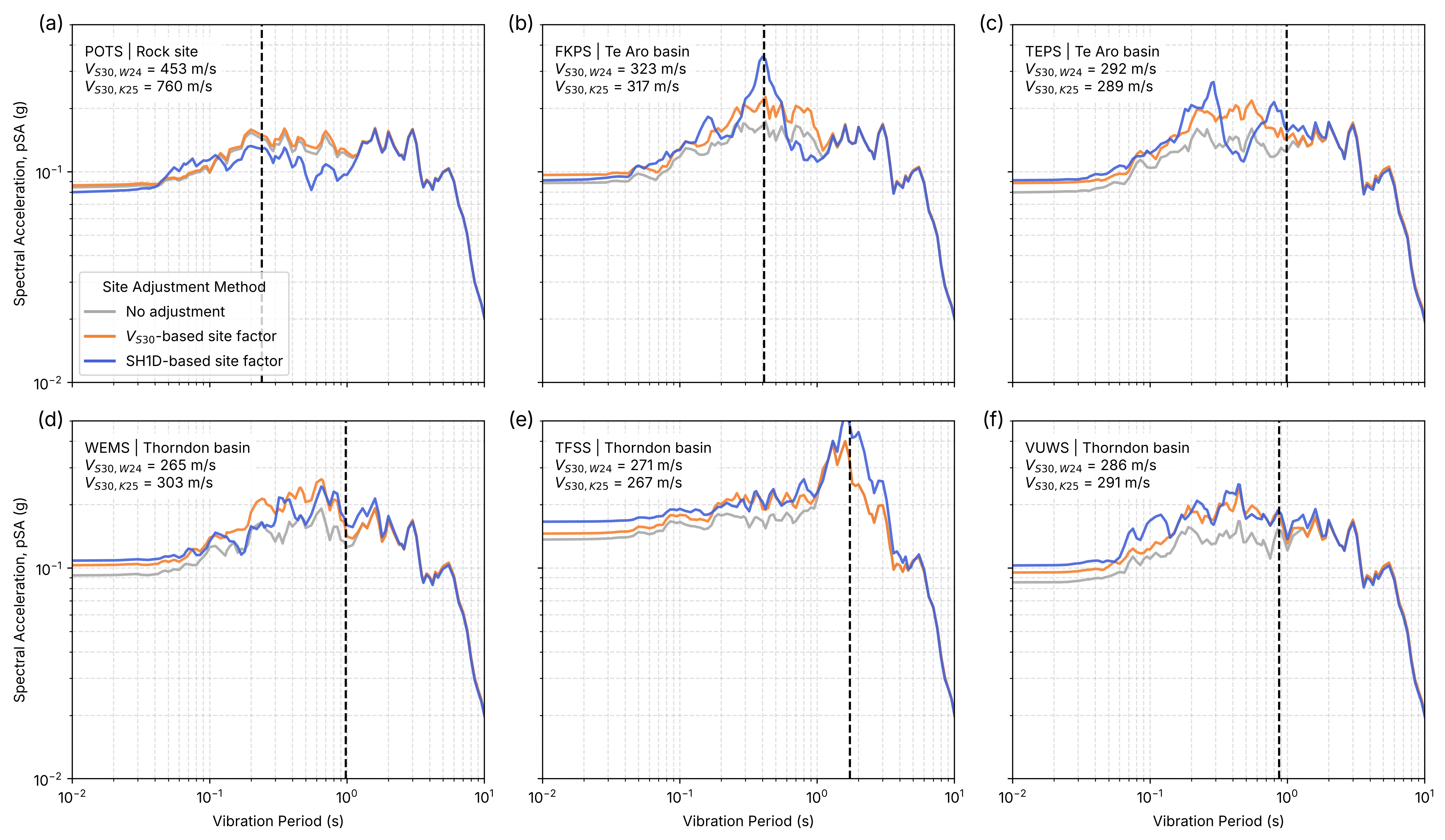}
		\caption{Comparison of simulated spectral accelerations for the Alpine–Clarence R2 rupture scenario at six Wellington sites: (a)~POTS, (b)~FKPS, (c)~TEPS, (d)~WEMS, (e)~TFSS, and (f)~VUWS. Results are shown for alternative approaches used to model shallow site effects. $V_{S30,W24}$ denotes the value from~\citet{wotherspoon_development_2024} employed in the $V_{S30}$-based site adjustment. $V_{S30,K25}$ corresponds to the value derived from the site-specific $V_S$ profile adopted for the SH1D-based adjustment \citep{kuncar2025b}. The vertical dashed black line indicates the site period, $T_0$, from~\citet{wotherspoon_development_2024}.}
		\label{fig:response_spectra_6sites_clarence_R2}
	\end{figure}

\begin{figure}[h]
		\centering
		\includegraphics[width=\linewidth]{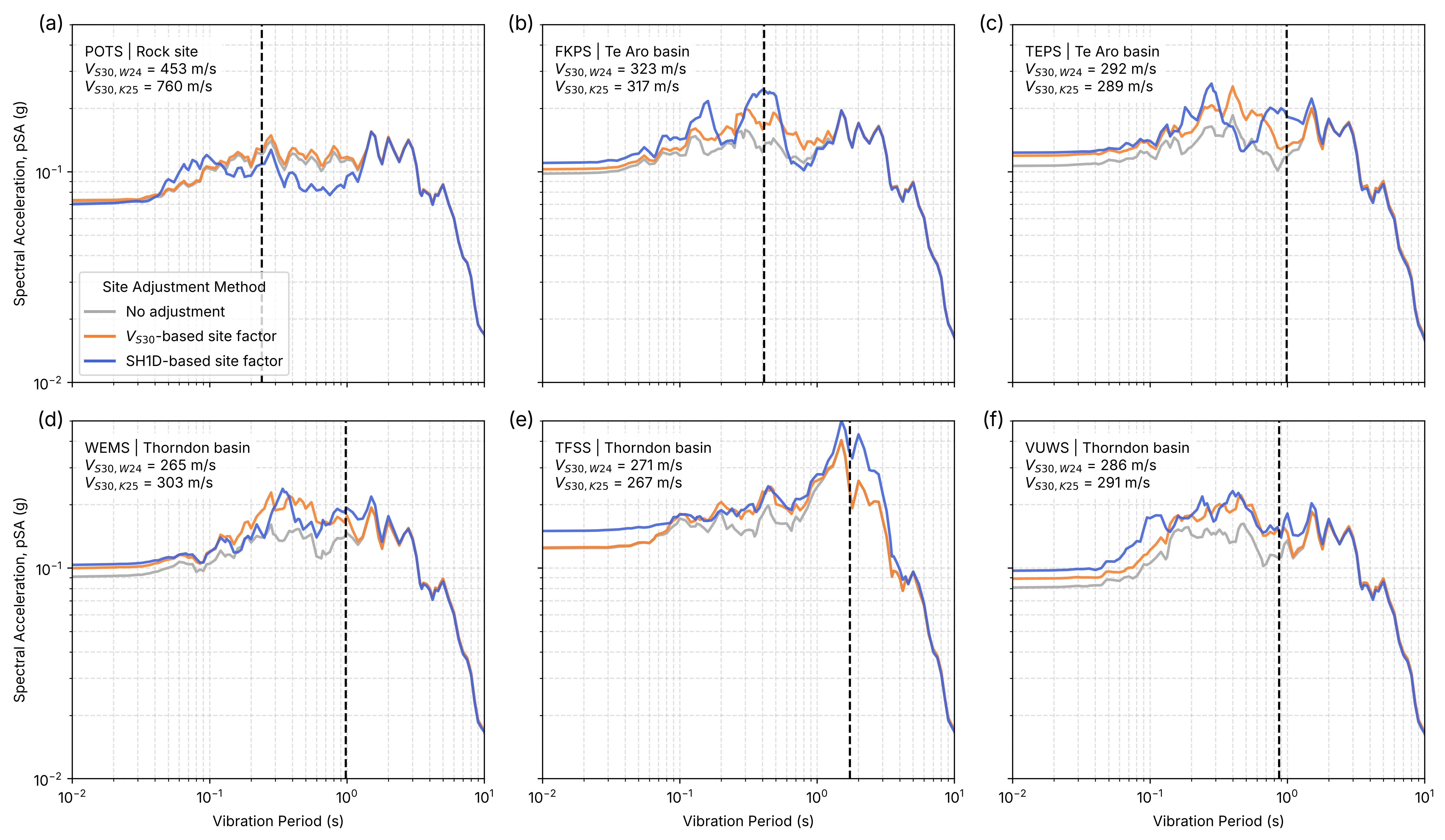}
		\caption{Comparison of simulated spectral accelerations for the Alpine–Clarence R3 rupture scenario at six Wellington sites: (a)~POTS, (b)~FKPS, (c)~TEPS, (d)~WEMS, (e)~TFSS, and (f)~VUWS. Results are shown for alternative approaches used to model shallow site effects. $V_{S30,W24}$ denotes the value from~\citet{wotherspoon_development_2024} employed in the $V_{S30}$-based site adjustment. $V_{S30,K25}$ corresponds to the value derived from the site-specific $V_S$ profile adopted for the SH1D-based adjustment \citep{kuncar2025b}. The vertical dashed black line indicates the site period, $T_0$, from~\citet{wotherspoon_development_2024}.}
		\label{fig:response_spectra_6sites_clarence_R3}
	\end{figure}

\begin{figure}[h]
		\centering
		\includegraphics[width=\linewidth]{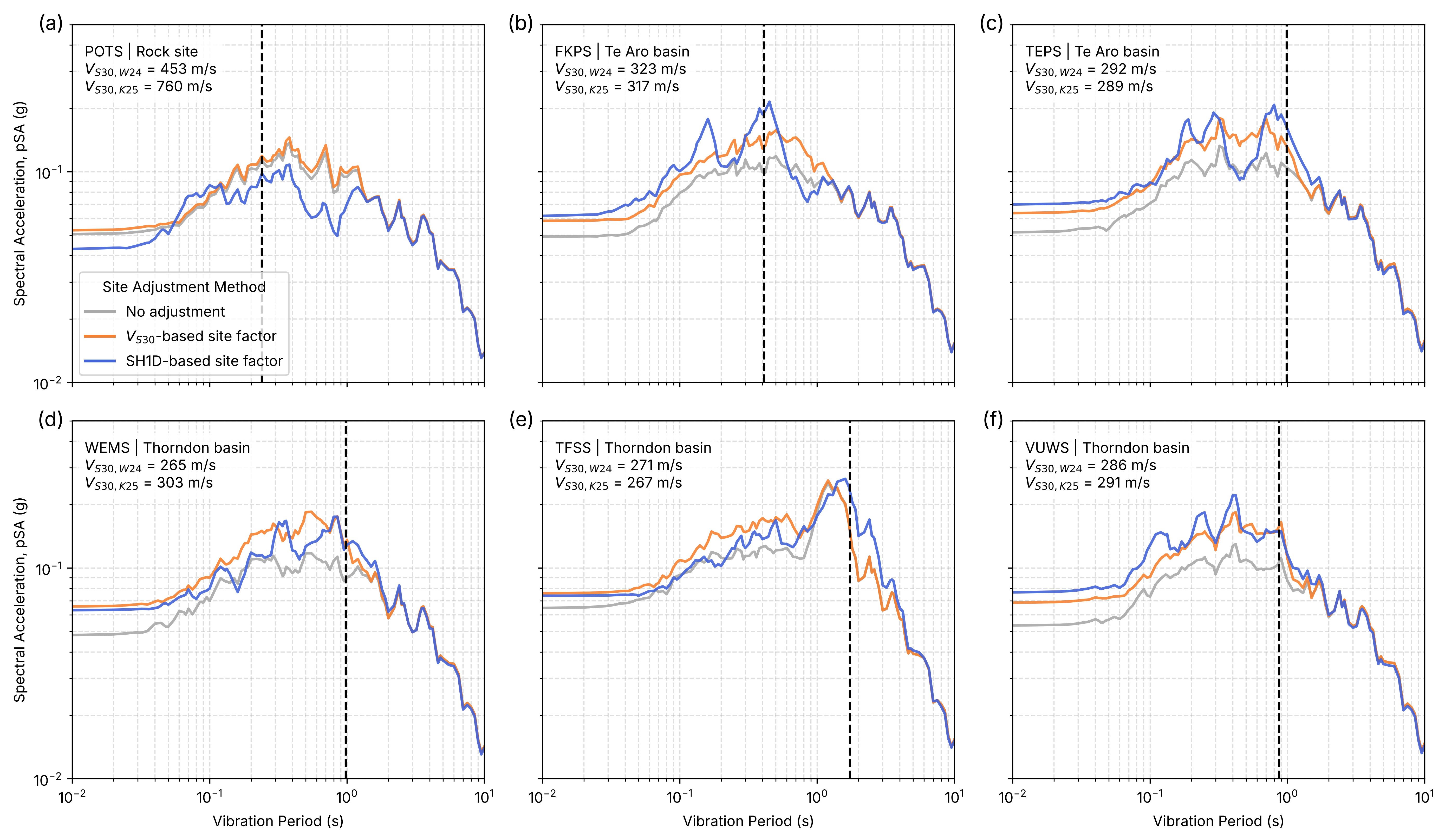}
		\caption{Comparison of simulated spectral accelerations for the Alpine-Hope R1 rupture scenario at six Wellington sites: (a)~POTS, (b)~FKPS, (c)~TEPS, (d)~WEMS, (e)~TFSS, and (f)~VUWS. Results are shown for alternative approaches used to model shallow site effects. $V_{S30,W24}$ denotes the value from~\citet{wotherspoon_development_2024} employed in the $V_{S30}$-based site adjustment. $V_{S30,K25}$ corresponds to the value derived from the site-specific $V_S$ profile adopted for the SH1D-based adjustment \citep{kuncar2025b}. The vertical dashed black line indicates the site period, $T_0$, from~\citet{wotherspoon_development_2024}.}
		\label{fig:response_spectra_6sites_hope_R1}
	\end{figure}

\begin{figure}[h]
		\centering
		\includegraphics[width=\linewidth]{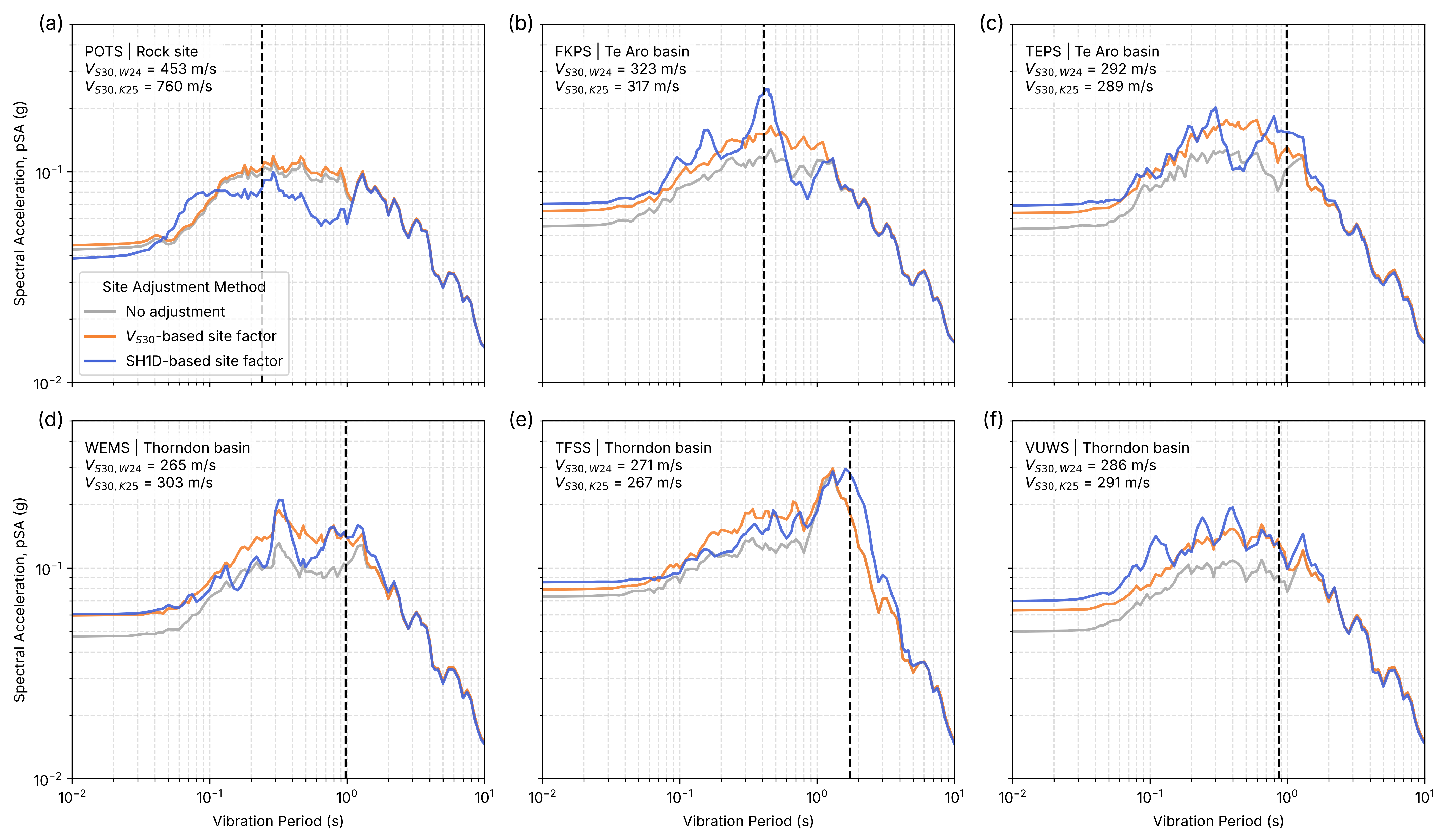}
		\caption{Comparison of simulated spectral accelerations for the Alpine-Hope R2 rupture scenario at six Wellington sites: (a)~POTS, (b)~FKPS, (c)~TEPS, (d)~WEMS, (e)~TFSS, and (f)~VUWS. Results are shown for alternative approaches used to model shallow site effects. $V_{S30,W24}$ denotes the value from~\citet{wotherspoon_development_2024} employed in the $V_{S30}$-based site adjustment. $V_{S30,K25}$ corresponds to the value derived from the site-specific $V_S$ profile adopted for the SH1D-based adjustment \citep{kuncar2025b}. The vertical dashed black line indicates the site period, $T_0$, from~\citet{wotherspoon_development_2024}.}
		\label{fig:response_spectra_6sites_hope_R2}
	\end{figure}

\begin{figure}[h]
		\centering
		\includegraphics[width=\linewidth]{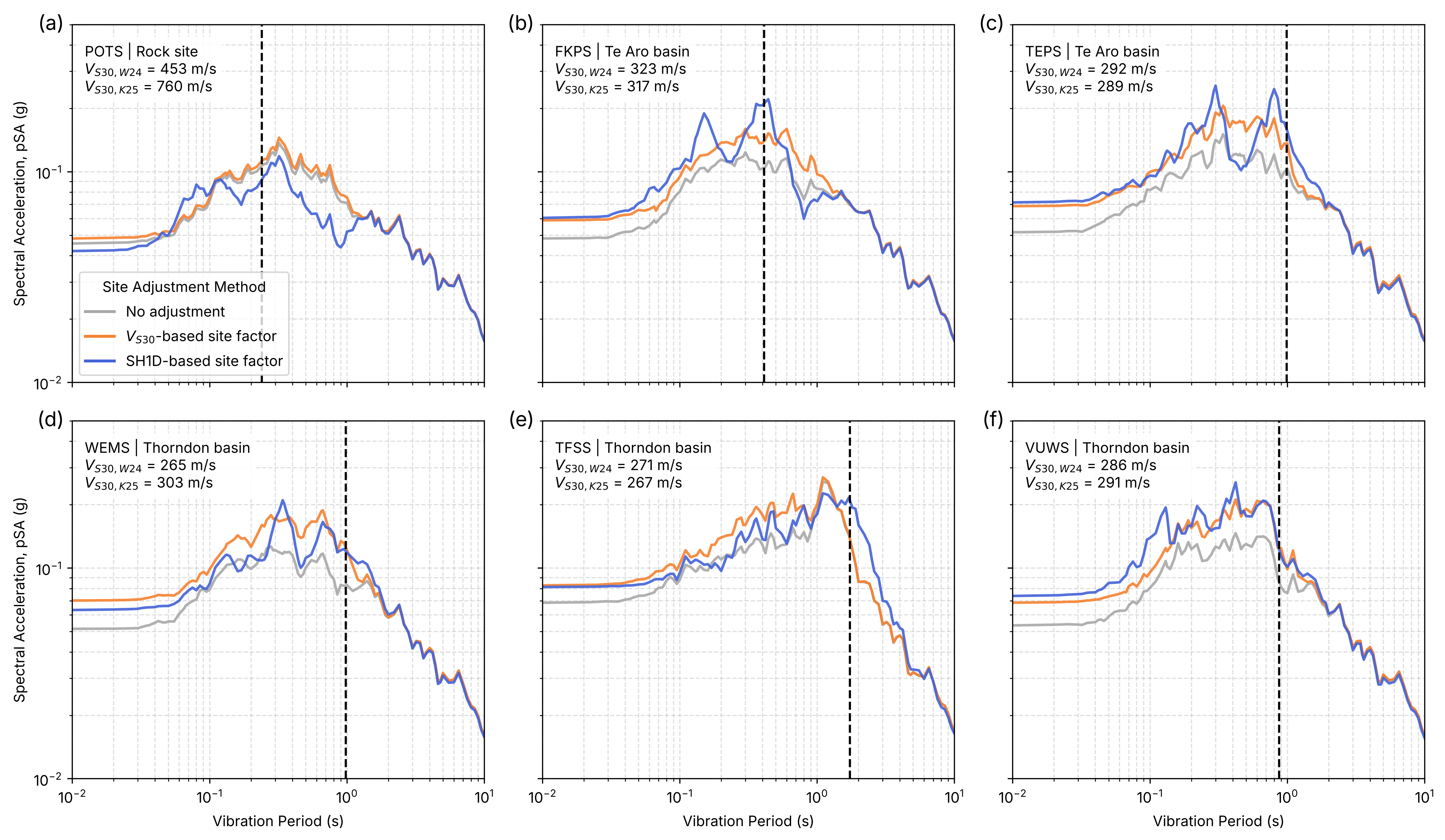}
		\caption{Comparison of simulated spectral accelerations for the Alpine-Hope R3 rupture scenario at six Wellington sites: (a)~POTS, (b)~FKPS, (c)~TEPS, (d)~WEMS, (e)~TFSS, and (f)~VUWS. Results are shown for alternative approaches used to model shallow site effects. $V_{S30,W24}$ denotes the value from~\citet{wotherspoon_development_2024} employed in the $V_{S30}$-based site adjustment. $V_{S30,K25}$ corresponds to the value derived from the site-specific $V_S$ profile adopted for the SH1D-based adjustment \citep{kuncar2025b}. The vertical dashed black line indicates the site period, $T_0$, from~\citet{wotherspoon_development_2024}.}
		\label{fig:response_spectra_6sites_hope_R3}
	\end{figure}

\begin{figure}[h]
		\centering
		\includegraphics[width=\linewidth]{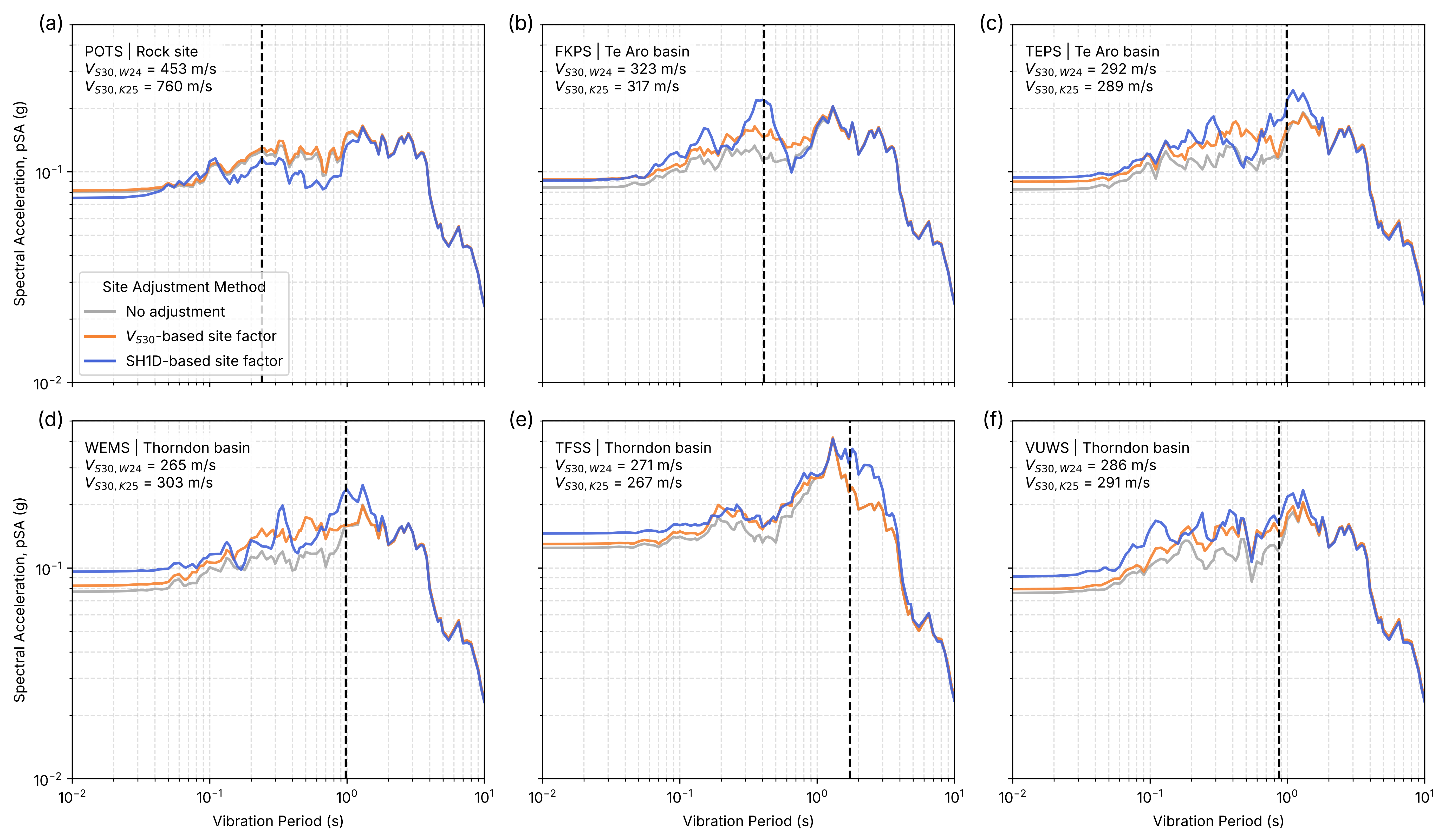}
		\caption{Comparison of simulated spectral accelerations for the Alpine-Wairau R1 rupture scenario at six Wellington sites: (a)~POTS, (b)~FKPS, (c)~TEPS, (d)~WEMS, (e)~TFSS, and (f)~VUWS. Results are shown for alternative approaches used to model shallow site effects. $V_{S30,W24}$ denotes the value from~\citet{wotherspoon_development_2024} employed in the $V_{S30}$-based site adjustment. $V_{S30,K25}$ corresponds to the value derived from the site-specific $V_S$ profile adopted for the SH1D-based adjustment \citep{kuncar2025b}. The vertical dashed black line indicates the site period, $T_0$, from~\citet{wotherspoon_development_2024}.}
		\label{fig:response_spectra_6sites_wairau_R1}
	\end{figure}

\begin{figure}[h]
		\centering
		\includegraphics[width=\linewidth]{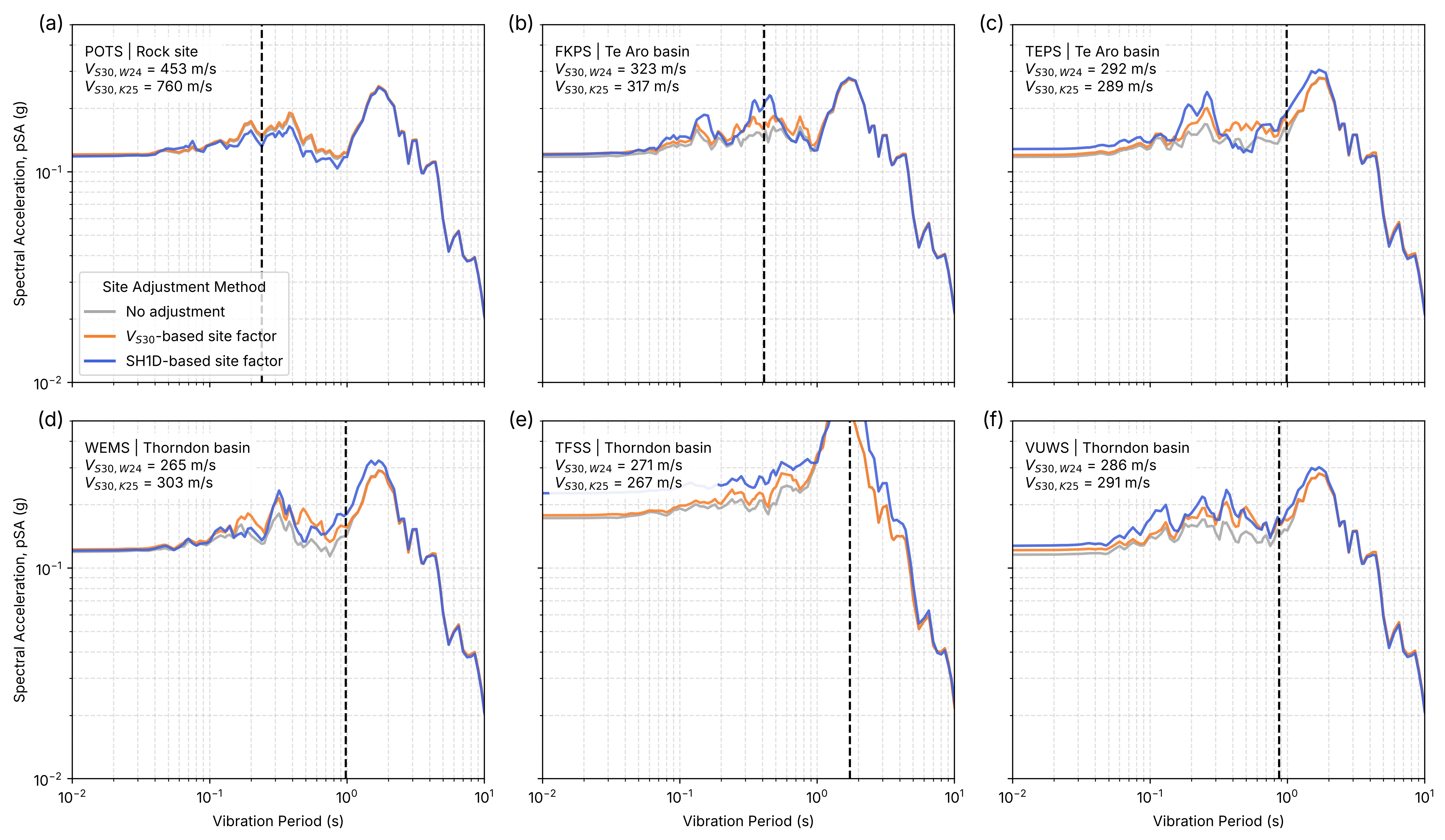}
		\caption{Comparison of simulated spectral accelerations for the Alpine-Wairau R2 rupture scenario at six Wellington sites: (a)~POTS, (b)~FKPS, (c)~TEPS, (d)~WEMS, (e)~TFSS, and (f)~VUWS. Results are shown for alternative approaches used to model shallow site effects. $V_{S30,W24}$ denotes the value from~\citet{wotherspoon_development_2024} employed in the $V_{S30}$-based site adjustment. $V_{S30,K25}$ corresponds to the value derived from the site-specific $V_S$ profile adopted for the SH1D-based adjustment \citep{kuncar2025b}. The vertical dashed black line indicates the site period, $T_0$, from~\citet{wotherspoon_development_2024}.}
		\label{fig:response_spectra_6sites_wairau_R2}
	\end{figure}

\begin{figure}[h]
		\centering
		\includegraphics[width=\linewidth]{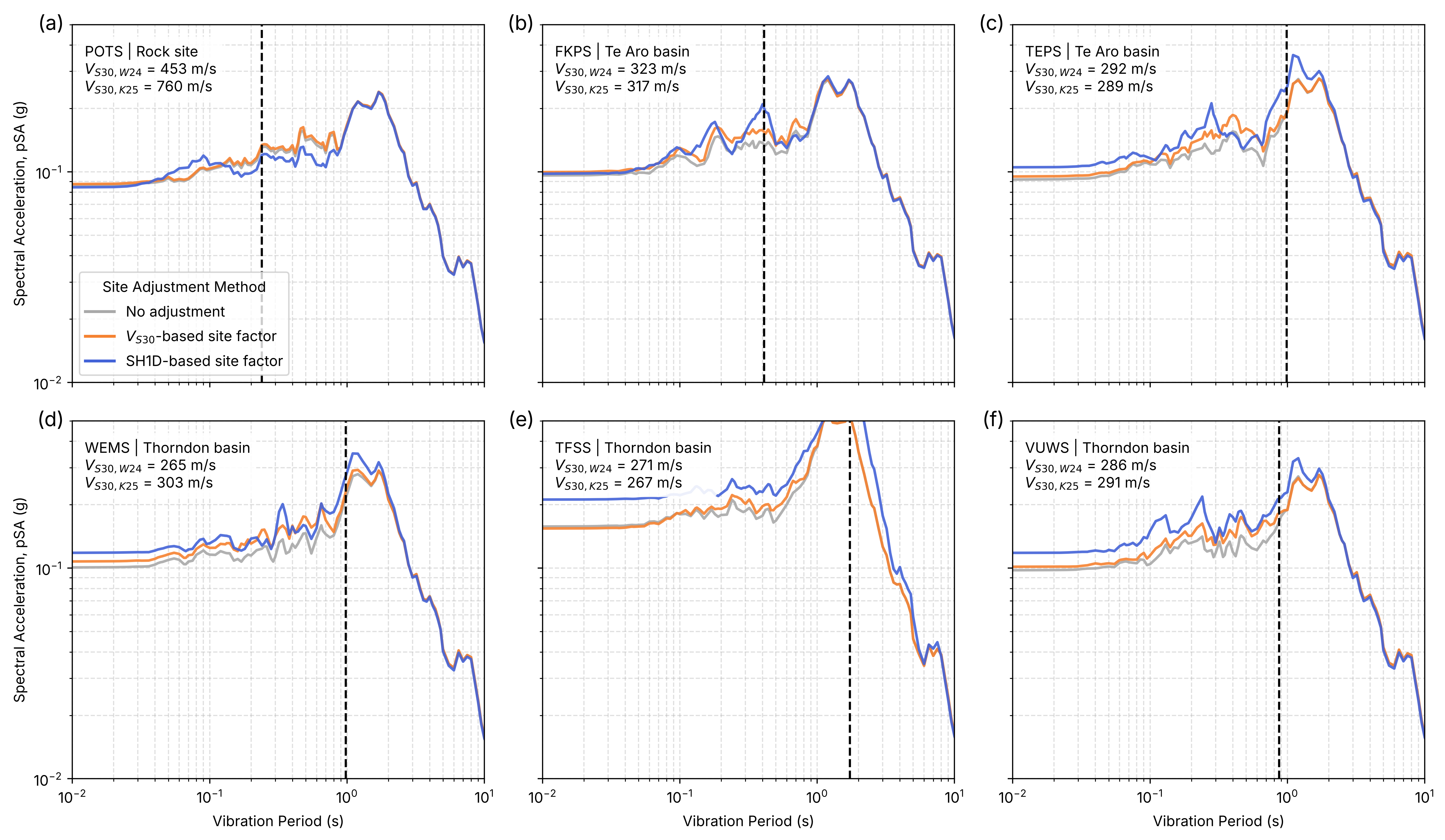}
		\caption{Comparison of simulated spectral accelerations for the Alpine-Wairau R3 rupture scenario at six Wellington sites: (a)~POTS, (b)~FKPS, (c)~TEPS, (d)~WEMS, (e)~TFSS, and (f)~VUWS. Results are shown for alternative approaches used to model shallow site effects. $V_{S30,W24}$ denotes the value from~\citet{wotherspoon_development_2024} employed in the $V_{S30}$-based site adjustment. $V_{S30,K25}$ corresponds to the value derived from the site-specific $V_S$ profile adopted for the SH1D-based adjustment \citep{kuncar2025b}. The vertical dashed black line indicates the site period, $T_0$, from~\citet{wotherspoon_development_2024}.}
		\label{fig:response_spectra_6sites_wairau_R3}
	\end{figure}

\begin{figure}[htbp]
		\centering
		\includegraphics[width=\linewidth]{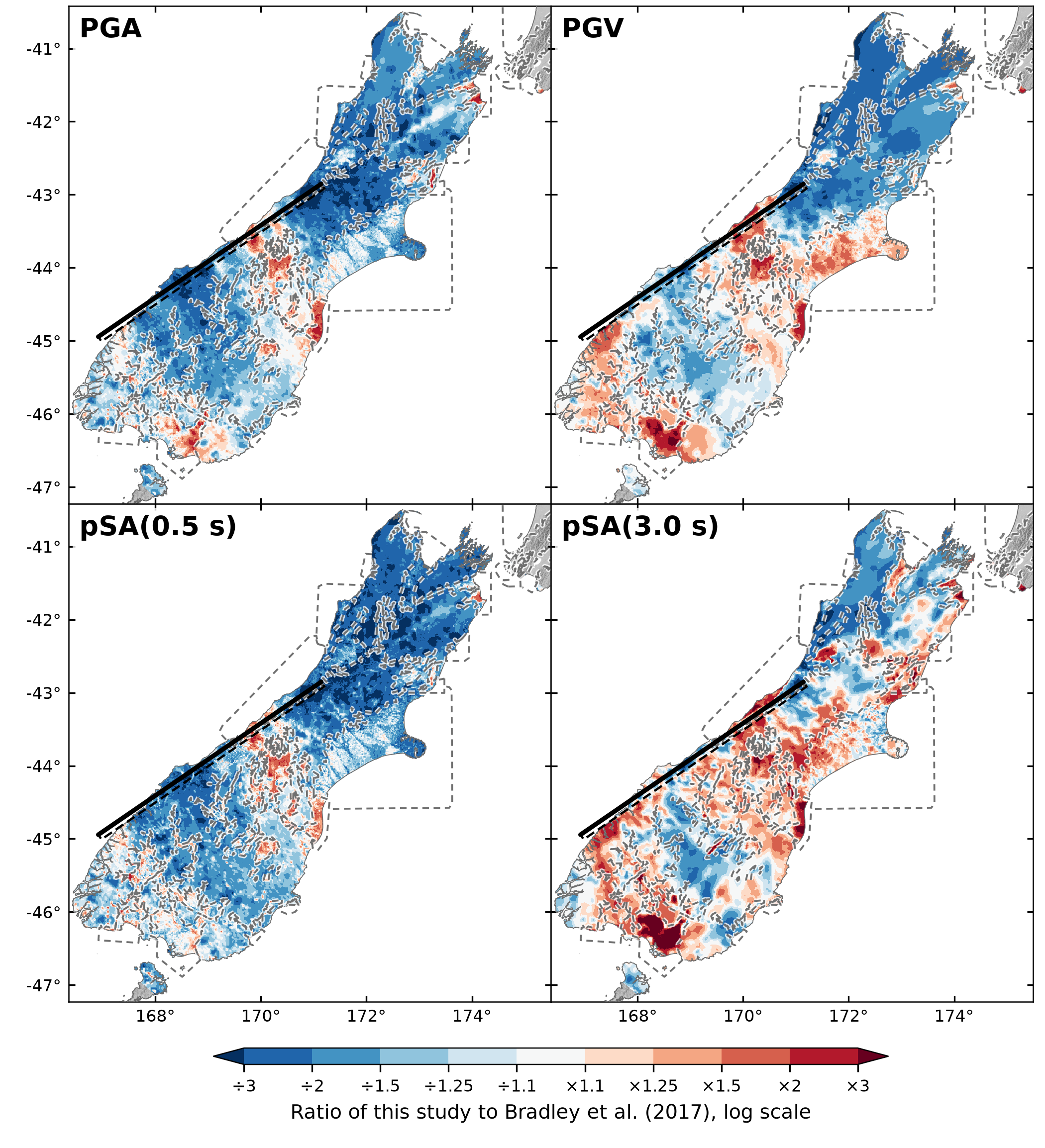}
		\caption{Ratio of this study's base rupture intensity measures to those of the
			AF8 scenario of \citet{bradleyGroundMotionSimulations2017} for PGA, PGV,
			pSA(0.5~s) and pSA(3.0~s), on a logarithmic colour scale, as in
			Figure~\ref{fig:af8-basins} but with all modelled basins outlined (dashed
			grey). The basins are named in Figure~\ref{fig:basins-named}. This study is the
			geometric mean of realisations R1--R3. The AF8 fault surface projection is
			dashed black and its trace bold.}
		\label{fig:af8-ratio}
	\end{figure}

\begin{figure}[htbp]
		\centering
		\includegraphics[width=\linewidth]{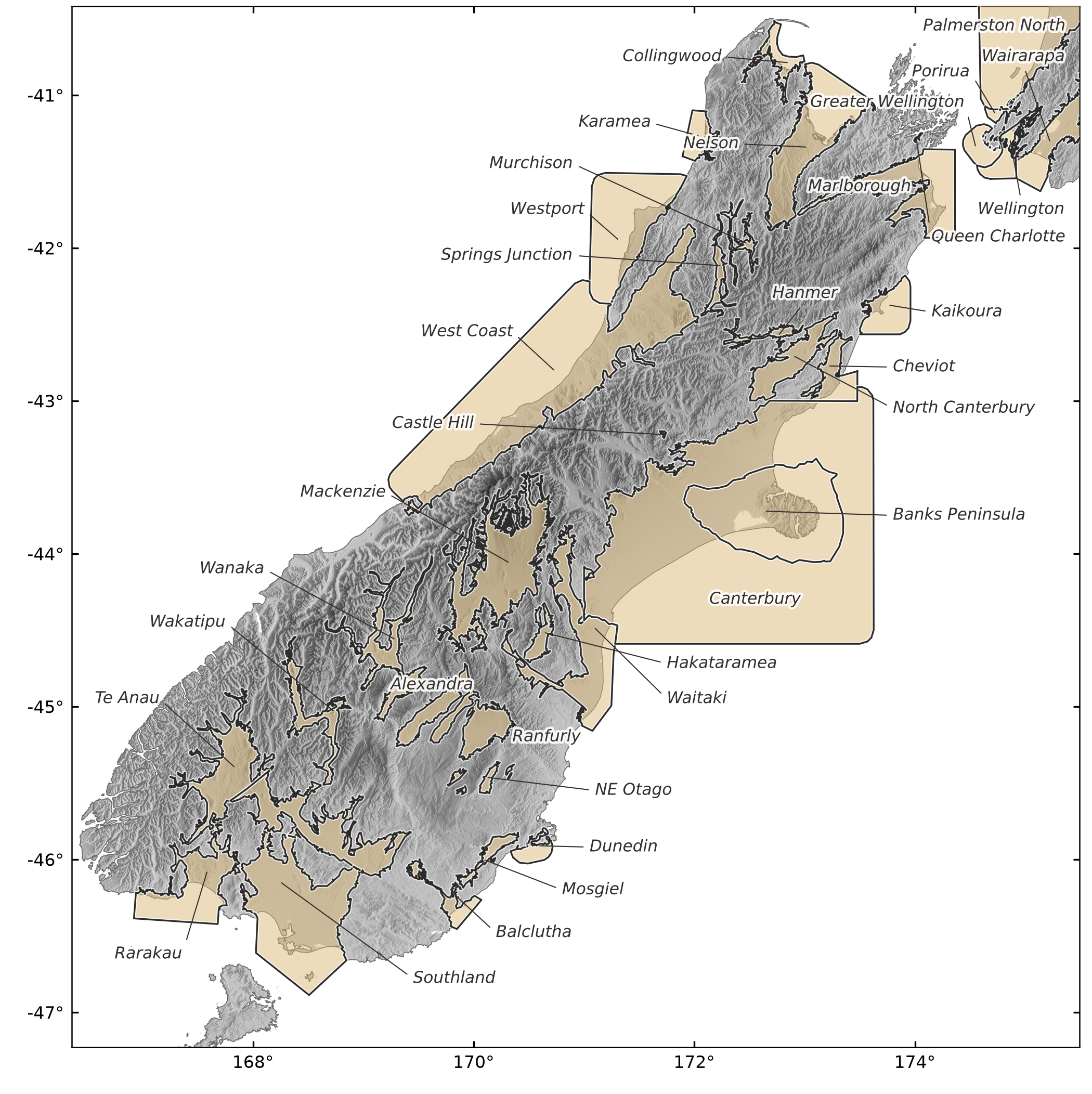}
		\caption{All sedimentary basins explicitly modelled in NZVM v2.09 within the
			region of Figure~\ref{fig:af8-ratio}, outlined and named.}
		\label{fig:basins-named}
	\end{figure}

\begin{figure}[htbp]
		\centering
		\includegraphics[width=\linewidth]{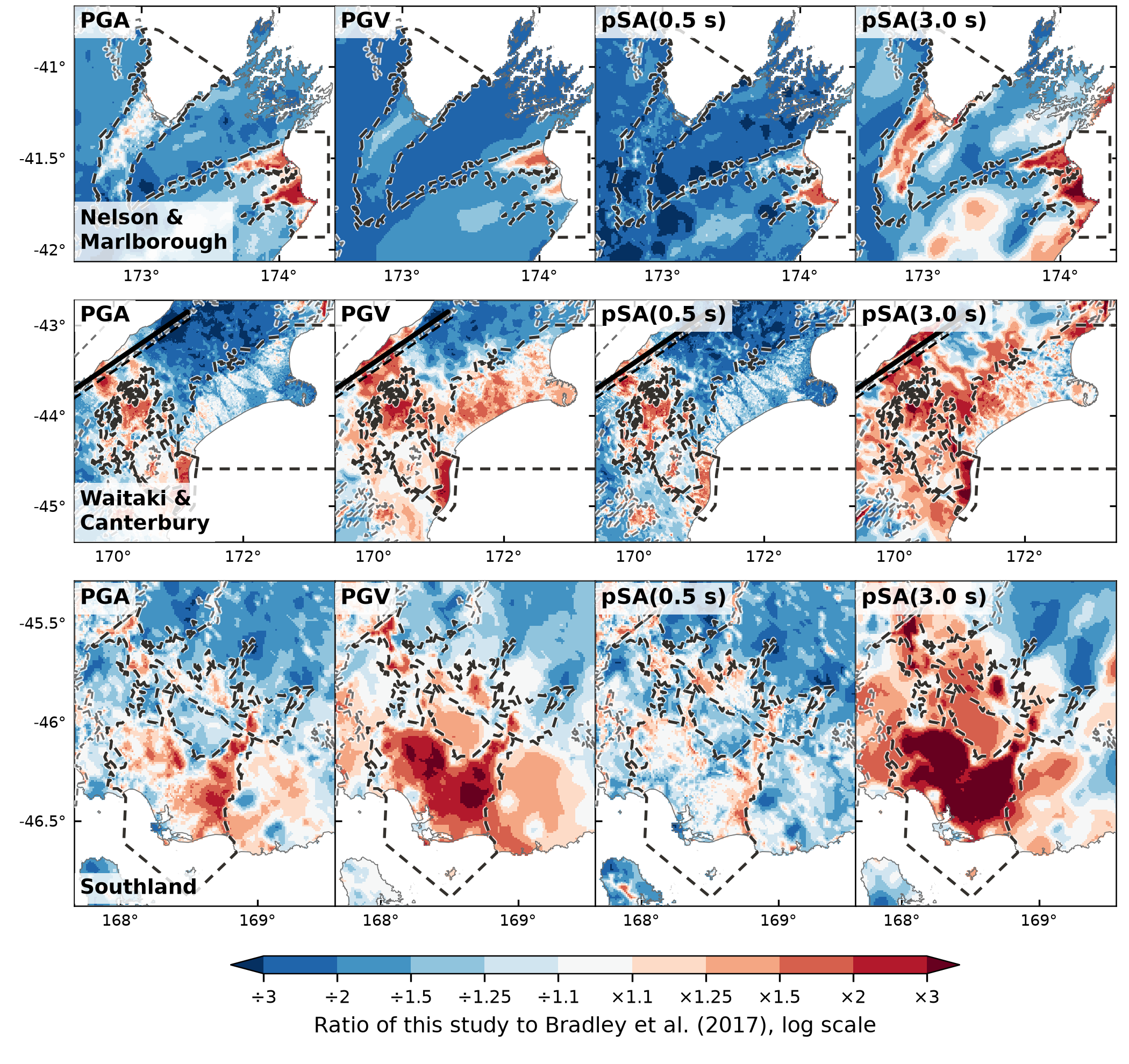}
		\caption{Close-up views of the basins of interest (Nelson and Marlborough;
			Waitaki and Canterbury; Southland) for the ratio of this study's base rupture
			intensity measures to those of the AF8 scenario, on the colour scale of
			Figure~\ref{fig:af8-basins}. Basins of interest are drawn in bold against
			their neighbours.}
		\label{fig:af8-zooms}
	\end{figure}

\begin{figure}[htbp]
		\centering
		\includegraphics[width=\linewidth]{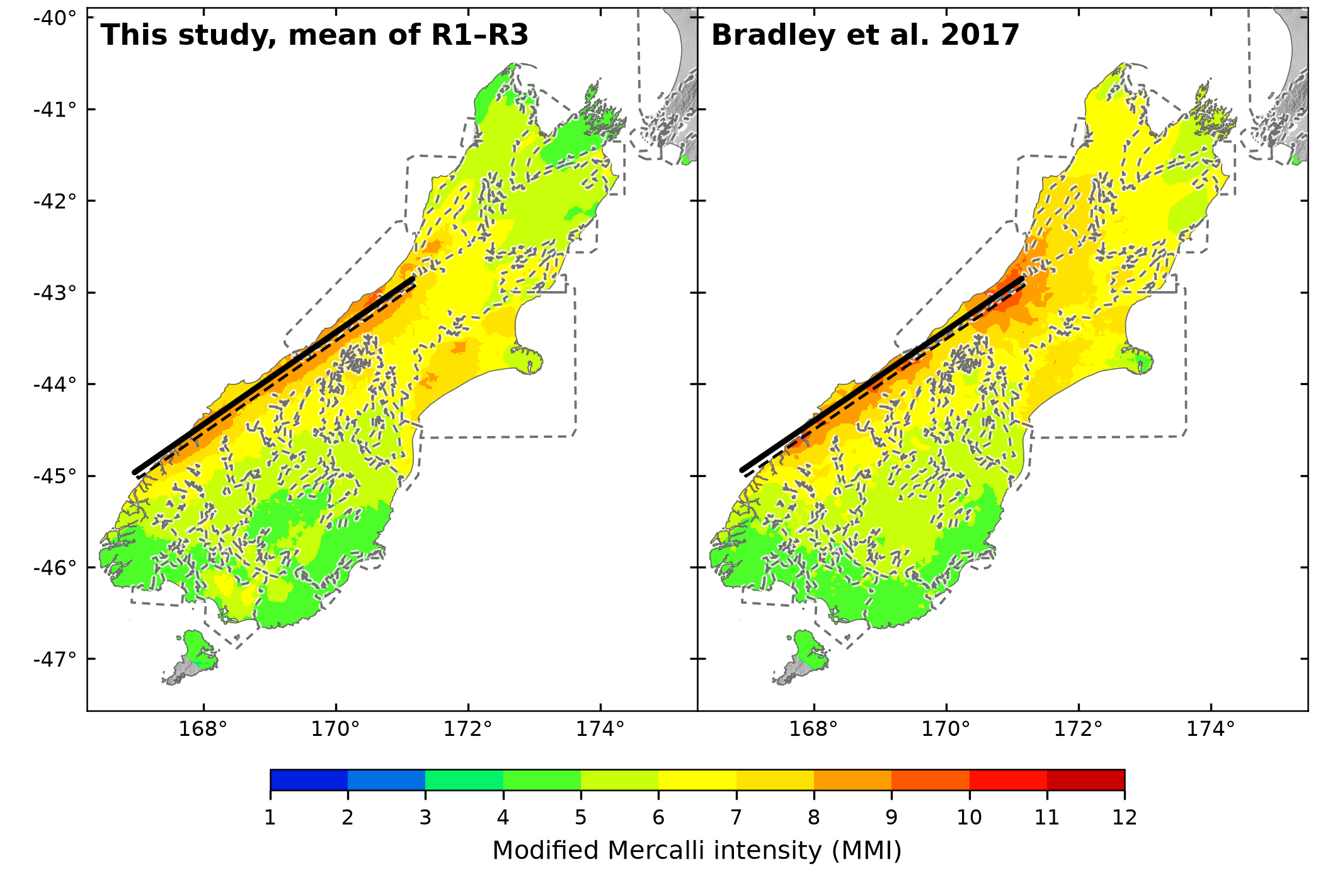}
		\caption{Modified Mercalli intensity, converted from simulated PGV with the
			\citet{wordenProbabilisticRelationshipsGroundMotion2012} ground motion to
			intensity relation, for this study's base rupture (geometric mean of
			R1--R3) and the AF8 scenario of \citet{bradleyGroundMotionSimulations2017}.}
		\label{fig:af8-mmi}
	\end{figure}

	This supplement provides additional results supporting the main text.
	Figures~\ref{fig:response_spectra_6sites_clarence_R2}--\ref{fig:response_spectra_6sites_wairau_R3} extend Figure~\ref{fig:response_spectra_6sites_clarence_R1} to the other rupture scenarios and realizations.
	Figures~\ref{fig:af8-ratio}--\ref{fig:af8-mmi} and Table~\ref{tab:af8-basin-ratios} support the comparison with the AF8 scenario of \citet{bradleyGroundMotionSimulations2017} in Section~\ref{sec:af8-base}.

	\FloatBarrier

	\begin{table}[htbp]
		\centering
		\caption{Geometric-mean ratio of this study's base rupture to the AF8 scenario
			within each highlighted basin, by intensity measure. Values below one mean
			this study gives the lower motion. Taken over the current stations whose
			basin outline contains them, with this study as the geometric mean of realisations
			R1--R3 and the AF8 field interpolated to each station's own position.
			Canterbury is the only one of these basins present in the AF8 velocity model.}
		\label{tab:af8-basin-ratios}
		\begin{tabular}{lrrrrr}
			\toprule
			Basin       & PGA    & PGV    & pSA (0.5\,s) & pSA (3.0\,s) & Stations  \\
			\midrule
			Nelson      & $0.68$ & $0.53$ & $0.45$       & $0.97$       & $1\,871$  \\
			Marlborough & $0.98$ & $0.74$ & $0.73$       & $1.09$       & $1\,397$  \\
			Canterbury  & $0.71$ & $1.13$ & $0.66$       & $1.05$       & $15\,409$ \\
			Mackenzie   & $1.18$ & $1.34$ & $1.27$       & $1.34$       & $4\,006$  \\
			Waitaki     & $1.37$ & $1.51$ & $1.21$       & $1.69$       & $2\,387$  \\
			Southland   & $1.05$ & $1.40$ & $0.94$       & $1.92$       & $6\,519$  \\
			\bottomrule
		\end{tabular}
	\end{table}

	\FloatBarrier
	\section{Electronic Supplement B: Simulation Source and Model Parameters}
	\label{sec:ESupp_B}
	\setcounter{figure}{0}
	\setcounter{table}{0}
	\begin{landscape}

		\begin{table}[p]
			\centering
			\caption{Source geometry and kinematics of every Alpine fault scenario
				realization, alongside \citet{bradleyGroundMotionSimulations2017}.
				Quantities fixed within a scenario are given on its R1 row only. The
				crustal model, wave propagation and output grid are in
				Table~\ref{tab:scenario-model}, the faults each scenario ruptures in
				Table~\ref{tab:scenario-faults}.}
			\label{tab:scenario-source}
			\scriptsize
			\setlength{\tabcolsep}{4pt}
			\begin{tabular}{>{\raggedright\arraybackslash}p{1.55cm}lrrrrrrrrrrrrrrr}
				\toprule
				\multicolumn{17}{@{}l}{\textbf{(a) Source geometry and kinematics}}                                                                                                                                                                                         \\
				\midrule
				Scenario               & Real. & $N_{\mathrm{flt}}$ & $M_w$ & Lat ($^\circ$) & Lon ($^\circ$) & $Z_{\mathrm{hyp}}$ & $L$    & $W$  & $Z_{\mathrm{TOR}}$ & $Z_{\mathrm{BOT}}$ & Area     & Strike     & Dip        & Rake       & $\bar{D}$ & Subfault       \\
				                       &       &                    &       &                &                & (km)               & (km)   & (km) & (km)               & (km)               & (km$^2$) & ($^\circ$) & ($^\circ$) & ($^\circ$) & (m)       & (m)            \\
				\midrule
				Bradley et al.\ (2017) & --    & 1                  & 7.90  & $-44.858$      & $167.219$      & 8.67               & 411.6  & 17.3 & 0.01               & 14.99              & 7121     & 57.0       & 60.0       & 17.2       & 3.78      & $200\times201$ \\
				\addlinespace
				Base                   & R1    & 1                  & 7.90  & $-44.871$      & $167.250$      & 6.23               & 411.0  & 17.0 & 0.00               & 14.72              & 6987     & 56.6       & 60.0       & 15.9       & 4.04      & $100\times100$ \\
				                       & R2    &                    &       &                &                &                    &        &      &                    &                    &          & 56.6       & 60.0       & 15.9       & 4.18      &                \\
				                       & R3    &                    &       &                &                &                    &        &      &                    &                    &          & 56.6       & 60.0       & 15.0       & 4.18      &                \\
				\addlinespace
				Clarence               & R1    & 15                 & 8.31  & $-44.841$      & $167.291$      & 9.14               & 979.1  & 21.8 & 0.00               & 23.76              & 21093    & 57.0       & 66.5       & 170.0      & 5.34      & $100\times100$ \\
				                       & R2    &                    &       &                &                &                    &        &      &                    &                    &          & 57.0       & 66.3       & 169.7      & 5.37      &                \\
				                       & R3    &                    &       &                &                &                    &        &      &                    &                    &          & 56.9       & 66.4       & 169.9      & 5.34      &                \\
				\addlinespace
				Hope                   & R1    & 20                 & 8.31  & $-44.841$      & $167.291$      & 9.14               & 1073.7 & 20.2 & 0.00               & 22.16              & 20497    & 57.0       & 67.0       & 164.7      & 5.65      & $100\times100$ \\
				                       & R2    &                    &       &                &                &                    &        &      &                    &                    &          & 57.0       & 67.0       & 165.1      & 5.70      &                \\
				                       & R3    &                    &       &                &                &                    &        &      &                    &                    &          & 56.9       & 67.2       & 164.9      & 5.66      &                \\
				\addlinespace
				Wairau                 & R1    & 12                 & 8.28  & $-44.841$      & $167.291$      & 9.14               & 894.9  & 22.1 & 0.00               & 24.00              & 19699    & 53.5       & 66.4       & 165.3      & 5.19      & $100\times100$ \\
				                       & R2    &                    &       &                &                &                    &        &      &                    &                    &          & 53.4       & 66.6       & 165.7      & 5.19      &                \\
				                       & R3    &                    &       &                &                &                    &        &      &                    &                    &          & 53.1       & 66.3       & 166.2      & 5.19      &                \\
				\bottomrule
			\end{tabular}
			\begin{flushleft}\scriptsize
				\textit{Notes.} $N_{\mathrm{flt}}$ counts the faults in the realization and
				$M_w$ is the moment sum over them. $\bar{D}$ is the area-weighted mean slip
				over the SRF subfaults. Strike is a potency-weighted axial mean of the
				sub-plane orientations; dip and rake are potency-weighted circular means. $L$ and the area
				are summed over all sub-planes and $W$ is area-weighted.
				The 2017 hypocenter, $\bar{D}$ and subfault size are computed from that
				study's SRF; its other entries are as reported.
			\end{flushleft}
		\end{table}

		\begin{table}[p]
			\centering
			\caption{Crustal model, wave propagation and output grid for every Alpine
				fault scenario realization, alongside
				\citet{bradleyGroundMotionSimulations2017}. None of these settings varies
				between realizations of a scenario, so each is given on the R1 row only, and
				wave propagation is the Graves \& Pitarka hybrid method throughout. The columns ``Mean (km)'', ``Median (km)'', ``< 2km (\%)'' and ``$\geq$ 6 km (\%)'' refer to the separation distances between stations.}
			\label{tab:scenario-model}
			\scriptsize
			\setlength{\tabcolsep}{1.6pt}
			\begin{tabular}{>{\raggedright\arraybackslash}p{1.55cm}lccccccccccccccc}
				\toprule
				\multicolumn{17}{@{}l}{\textbf{(b) Crustal model, wave propagation and output grid}}                                                                                                                                                                       \\
				\midrule
				Scenario                                   & Real. & Geometry              & Rupture   & Slip  & Velocity  & Basins & Topo.    & FD grid & Min $V_S$     & Max LF    & Site                           & Stations  & Mean & Median & $<2$\,km & $\geq6$\,km \\
				                                           &       & source                & generator & real. & model     &        &          & (km)    & (m\,s$^{-1}$) & (Hz)      & response                       &           & (km) & (km)   & (\%)     & (\%)        \\
				\midrule
				\citet{bradleyGroundMotionSimulations2017} & --    & \citet{Stirling_2012} & GP        & 1     & SIVM      & 1      & none     & 0.4     & 500           & 0.25--0.5 & $-$\textsuperscript{$\dagger$} & $18\,962$ & 2.19 & 1.00   & 76       & 13          \\
				\addlinespace
				Base                                       & R1    & NSHM 2022             & GP + BF25 & 3     & NZVM 2.09 & 45     & sq.-tap. & 0.1     & 500           & 1.0       & CB14                           & $92\,297$ & 1.19 & 0.92   & 86       & 0           \\
				                                           & R2    &                       &           &       &           &        &          &         &               &           &                                &           &      &        &          &             \\
				                                           & R3    &                       &           &       &           &        &          &         &               &           &                                &           &      &        &          &             \\
				\addlinespace
				Clarence                                   & R1    & NSHM 2022             & GP + BF25 & 3     & NZVM 2.09 & 45     & sq.-tap. & 0.1     & 500           & 1.0       & CB14                           & $92\,297$ & 1.19 & 0.92   & 86       & 0           \\
				                                           & R2    &                       &           &       &           &        &          &         &               &           &                                &           &      &        &          &             \\
				                                           & R3    &                       &           &       &           &        &          &         &               &           &                                &           &      &        &          &             \\
				\addlinespace
				Hope                                       & R1    & NSHM 2022             & GP + BF25 & 3     & NZVM 2.09 & 45     & sq.-tap. & 0.1     & 500           & 1.0       & CB14                           & $92\,297$ & 1.19 & 0.92   & 86       & 0           \\
				                                           & R2    &                       &           &       &           &        &          &         &               &           &                                &           &      &        &          &             \\
				                                           & R3    &                       &           &       &           &        &          &         &               &           &                                &           &      &        &          &             \\
				\addlinespace
				Wairau                                     & R1    & NSHM 2022             & GP + BF25 & 3     & NZVM 2.09 & 45     & sq.-tap. & 0.1     & 500           & 1.0       & CB14                           & $92\,297$ & 1.19 & 0.92   & 86       & 0           \\
				                                           & R2    &                       &           &       &           &        &          &         &               &           &                                &           &      &        &          &             \\
				                                           & R3    &                       &           &       &           &        &          &         &               &           &                                &           &      &        &          &             \\
				\bottomrule
			\end{tabular}
			\begin{flushleft}\scriptsize
				\textsuperscript{$\dagger$}Site amplification was not applied in \citet{bradleyGroundMotionSimulations2017}.
			\end{flushleft}
		\end{table}

		\begin{table}[p]
			\centering
			\caption{Participating fault systems of each scenario, named as in the NSHM 2022 source model.}
			\label{tab:scenario-faults}
			\scriptsize
			\setlength{\tabcolsep}{4pt}
			\begin{tabular}{>{\raggedright\arraybackslash}p{1.55cm}l>{\raggedright\arraybackslash}p{13.2cm}}
				\toprule
				Scenario                                   & $N_{\mathrm{flt}}$ & Faults                                                                                                                                                                                                                                                                                                                                                                                                                                                                                                                                                                        \\
				\midrule
				\citet{bradleyGroundMotionSimulations2017} & 1                  & Alpine fault as a single segment                                                                                                                                                                                                                                                                                                                                                                                                                                                                                                                                              \\
				\addlinespace
				Base                                       & 1                  & As in \citet{bradleyGroundMotionSimulations2017}.                                                                                                                                                                                                                                                                                                                                                                                                                                                                                                                             \\
				\addlinespace
				Clarence                                   & 15                 & \texttt{Alpine: Caswell}; \texttt{Alpine: Caswell - South George}; \texttt{Alpine: George landward}; \texttt{Alpine: George to Jacksons}; \texttt{Alpine: Jacksons to Kaniere}; \texttt{Alpine: Kaniere to Springs Junction}; \texttt{Alpine: Nancy}; \texttt{Alpine: Resolution - Dagg}; \texttt{Alpine: Resolution - Five Fingers}; \texttt{BooBoo}; \texttt{Clarence: Central}; \texttt{Clarence: Southwest}; \texttt{Fidget}; \texttt{Kekerengu 1}; \texttt{Needles}                                                                                                      \\
				\addlinespace
				Hope                                       & 20                 & \texttt{Alpine: Caswell}; \texttt{Alpine: Caswell - South George}; \texttt{Alpine: George landward}; \texttt{Alpine: George to Jacksons}; \texttt{Alpine: Jacksons to Kaniere}; \texttt{Alpine: Nancy}; \texttt{Alpine: Resolution - Dagg}; \texttt{Alpine: Resolution - Five Fingers}; \texttt{Hanmer}; \texttt{Hope: Conway}; \texttt{Hope: Hanmer NW}; \texttt{Hope: Hope River}; \texttt{Hope: Hurunui}; \texttt{Hope: Kakapo-2-Hamner}; \texttt{Hope: Seaward}; \texttt{Hope: Te Rapa}; \texttt{Kelly}; \texttt{Opouawe - Uruti}; \texttt{Te Rapa 1}; \texttt{Te Rapa 2} \\
				\addlinespace
				Wairau                                     & 12                 & \texttt{Alpine: Caswell}; \texttt{Alpine: Caswell - South George}; \texttt{Alpine: George landward}; \texttt{Alpine: George to Jacksons}; \texttt{Alpine: Jacksons to Kaniere}; \texttt{Alpine: Kaniere to Springs Junction}; \texttt{Alpine: Resolution - Charles}; \texttt{Alpine: Resolution - Dagg}; \texttt{Alpine: Resolution - Five Fingers}; \texttt{Alpine: Springs Junction to Tophouse}; \texttt{Wairau}; \texttt{Wairau 2}                                                                                                                                        \\
				\bottomrule
			\end{tabular}
		\end{table}

	\end{landscape}

	\FloatBarrier
	\section{Electronic Supplement C: Computational Details}
	\label{sec:ESupp_C}
	\setcounter{figure}{0}
	\setcounter{table}{0}

	Low-frequency simulations were conducted on the TACC's Stampede3 (\url{https://tacc.utexas.edu/systems/stampede3/}) cluster, with each job run on 64 "Skylake" (SKX) compute nodes. Every SKX node has two Intel Xeon Platinum 8160 processors, each having 24 physical cores and 192 GiB of RAM. The reproduction of results in \citet{bradleyGroundMotionSimulations2017} were calculated using 16 compute nodes on ESNZ's Cascade cluster; each node provides 2 x AMD EPYC 9654 96-Core processors and 755 GiB of RAM. High-frequency calculations and post-processing were conducted on REANNZ's HPC3 cluster, specifically the AMD 9634 EPYC Genoa compute nodes.
	Core hours (physical cores $\times$ wallclock) are tabulated for low-frequency calculations in Table~\ref{tab:core-hour-table}.

	\begin{table}[h]
		\centering
		\caption{Core hour consumption for low-frequency calculations tabulated by scenario and realisation.}
		\label{tab:core-hour-table}
		\begin{tabular}{llrrr r}
			\toprule
			Scenario                            & Realisation & Nodes       & Physical cores & Wallclock (h) & Core-hours \\
			\midrule
			Base                                & R1          & 64          & 3072           & 28.5          & 87,639.9   \\
			Base                                & R2          & 64          & 3072           & 28.4          & 87,092.9   \\
			Base                                & R3          & 64          & 3072           & 28.2          & 86,551.0   \\
			Clarence                            & R1          & 64          & 3072           & 28.5          & 87,520.4   \\
			Clarence                            & R2          & 64          & 3072           & 28.2          & 86,778.9   \\
			Clarence                            & R3          & 64          & 3072           & 28.1          & 86,409.4   \\
			Hope                                & R1          & 64          & 3072           & 28.4          & 87,142.4   \\
			Hope                                & R2          & 64          & 3072           & 28.4          & 87,205.5   \\
			Hope                                & R3          & 64          & 3072           & 28.4          & 87,302.8   \\
			Wairau                              & R1          & 64          & 3072           & 28.3          & 86,958.9   \\
			Wairau                              & R2          & 64          & 3072           & 28.5          & 87,536.6   \\
			Wairau                              & R3          & 64          & 3072           & 28.4          & 87,268.7   \\
			Alpine 2017                         & R1          & 16          & 6144           & 43.7          & 268,759.0  \\
			\midrule
			\multicolumn{4}{l}{Total (13 runs)} & 384.0       & 1,045,407.4                                               \\
			\bottomrule
		\end{tabular}
	\end{table}

	Simulations were run in a common domain: 314.6 km east-west and 1201.6 km north-south rotated 44.4 degrees in azimuth. The velocity model extends to a depth of 62 km and was discretized at 100 m for a total of $N_x \times N_y \times N_z = 3146 \times 12016 \times 620 = 23,437,448,320$ grid points. The simulations were run for 94,000 time steps with $\Delta t = 0.005\,s$ for a total physical simulation duration of 470 s.

\end{appendices}

\end{document}